\documentclass[12pt]{article}
\usepackage{amsfonts}
\usepackage{amssymb}
\usepackage{graphicx,subcaption}
\usepackage{amsmath}
\usepackage[bottom]{footmisc}
\usepackage{xcolor}
\usepackage[longnamesfirst]{natbib}
\usepackage{rotating}
\usepackage{bm}
\usepackage{array}
\usepackage{graphicx}
\usepackage{booktabs}
\usepackage[normalem]{ulem}

\usepackage[Export]{adjustbox}
\usepackage{tabularx}
\usepackage{cellspace}

\renewcommand{\baselinestretch}{1.3}
\newcommand{\bc}{\begin{center}}
\newcommand{\ec}{\end{center}}
\newcommand{\be}{\begin{equation}}
\newcommand{\ee}{\end{equation}}
\newcommand{\bea}{\begin{eqnarray}}
\newcommand{\eea}{\end{eqnarray}}
\newcommand{\bean}{\begin{eqnarray*}}
\newcommand{\eean}{\end{eqnarray*}}

\newtheorem{assumption}{Assumption}

\newtheorem{theorem}{Theorem}

\newtheorem{definition}{Definition}
\newtheorem{lemma}{Lemma}
\newcounter{pkt}
\newenvironment{tlist}{\begin{list}{(\roman{pkt})}{\usecounter{pkt}\parskip0ex\parsep0ex\itemsep0ex\topsep0ex}}{\end{list}}

\def\EE{\mathbb{E}}
\def\PP{\mathbb{P}}

\newcommand{\tr}{\operatorname*{tr}} 
\newcommand{\argmax}{\operatorname*{argmax}}

\newcommand{\plim}{\operatorname*{plim}}
\newcommand{\vecr}{\operatorname*{vecr}}

\newcounter{saveeqn}

\begin{document}

\title{\vspace*{-0.5 in} 
\textbf{Misspecification-Robust Shrinkage and Selection for VAR Forecasts and IRFs}}

\author{Oriol Gonz\'alez-Casas\'us\thanks{
		\setlength{\baselineskip}{4mm}  Correspondence: O. Gonz\'alez-Casas\'us and F. Schorfheide: Department of Economics, University of Pennsylvania, Philadelphia, PA 19104-6297. Email: oriol.gonzalez@upf.edu (Gonz\'alez-Casas\'us) and schorf@ssc.upenn.edu (Schorfheide). We thank Frank Diebold, Mikkel Plagborg-M\o ller, Christian Wolf, and participants at various seminars and conferences for helpful comments and suggestions.} \\
	{\em \small Universitat Pompeu Fabra \& BSE}
	\and Frank Schorfheide \\
	{\em \small University of Pennsylvania}, \\ {\em \small CEPR, PIER, NBER} }

\date{This Version: \today}
\maketitle


\begin{abstract}
Vector autoregressions (VARs) are vulnerable to dynamic misspecification, forcing researchers to navigate complex bias-variance trade-offs when forecasting or estimating impulse response functions (IRFs). We derive novel, task-specific selection criteria -- PC for forecasting and IRFC for IRF estimation -- based on asymptotically unbiased estimates of frequentist risk under local dynamic misspecification. These criteria give empirical researchers a fully data-driven and misspecification-aware workflow that simultaneously selects among candidate estimators, the degree of  Bayesian shrinkage, and the lag length. IRFC is the first criterion allowing researchers to seamlessly choose between iterated-VAR and local-projection IRF estimators while jointly tuning shrinkage and lag length for the task at hand. (JEL C11, C32, C52, C53)
\end{abstract} 


\thispagestyle{empty}

\clearpage
\setcounter{page}{1}

\section{Introduction}
\label{sec:introduction}

Bayesian vector autoregressions (VARs) have been widely used for macroeconomic forecasting and to analyze the dynamic effects of economic shocks since the early 1980s. They combine the VAR likelihood function with a prior distribution that shrinks the distance between the maximum likelihood estimator (MLE) and the prior mean, thereby reducing the variability of the posterior mean estimator in settings where the number of parameters is large relative to the number of available observations. In a typical implementation, one or more hyperparameters, denoted by $\lambda$ in this paper, control the precision of the prior distribution and hence the relative weight assigned to the prior mean in the construction of the posterior mean. For the posterior mean to deliver accurate forecasts and estimates of impulse response functions (IRFs), a data-driven hyperparameter determination is important. Researchers commonly use a marginal data density (MDD) -- also called marginal likelihood -- to select (e.g., \citealt{DelNegro2004}) or average over (e.g., \citealt{GiannoneLenzaPrimiceri2015}) $\lambda$ and also the number of lags $p$ of the VAR.

The prescription of a Bayesian framework is to estimate a single model (or a single mixture of models in the case of averaging), here a VAR with a specific $(\lambda,p)$, and then derive all decisions and inference, e.g., multi-step forecasts or IRF analyses, from the posterior distribution of this model. The advantage is that across different tasks there is a high degree of coherency. We refer to this as a model-driven workflow. In practice, we observe that many researchers prefer to tailor their estimates and model specifications toward the tasks at hand, which are multi-step forecasting and IRF estimation in this paper. The typical justification for a task-based workflow is a potential misspecification of the workhorse econometric model, here the VAR. This paper provides researchers who favor a task-based approach with tailored information criteria based on estimates of frequentist prediction or IRF estimation risks. These criteria can be used for the joint determination of $(\lambda,p)$ and also to choose between two estimators indexed by $\iota$ and described below. We study their large-sample properties and document their performance in simulations and an empirical application. Throughout the paper, we focus on point prediction and estimation evaluated under quadratic loss.

In a Gaussian VAR setting, the MLE minimizes in-sample one-step-ahead squared forecast errors. If the goal is $h$-step-ahead forecasting of an $n$-dimensional vector time series $y_t$, the model-driven workflow would prescribe iterating the one-step-ahead MLE-based forecasts forward. Alternatively, a task-based approach would use multi-step regressions that project $y_t$ on $y_{t-h}$ and additional lags. We refer to the resulting estimator as the loss-function estimator (LFE), where ``loss'' refers to the $h$-step-ahead forecast error. In the context of IRF analysis, these multi-step regressions are called local projections (LP) and are used to estimate $h$th-order moving average (MA) coefficient matrices; see \cite{Jorda2005}. In the remainder of this paper, we associate the MLE with iterative VAR IRF estimates, and the LFE with LP IRF estimates. Thus, in regard to estimator choice, we consider $\iota \in \{mle,lfe\}$ which leads to the triplet $(\iota,\lambda,p)$ that indexes our estimators and predictors. The model-driven workflow would set $\iota=mle$, determine $(\hat{\lambda},\hat{p})$, and hold them fixed across all tasks. The task-based workflow determines the triplet separately for each horizon and loss function. 

Ultimately, the ranking of the two workflows depends on the degree of model misspecification. For instance, it is well known that the MLE tends to be associated with low variance but a potentially large bias under model misspecification. Conversely, the LFE has higher variance, but lower or no bias under misspecification. The empirical evidence is ambiguous and, depending on the utility assigned to deriving estimates and predictions from a single model, can support either approach. For instance, \cite{MSW2006} undertake a large-scale empirical comparison of MLE versus LFE plug-in predictors using data on more than 150 monthly macroeconomic time series, grouping series randomly so that the degree of VAR misspecification varies across samples. They find that MLE plug-in predictions tend to yield smaller forecast errors in many, but certainly not all, settings. Similarly, \cite{Li2022}, henceforth LPW, conduct a simulation study based on a multitude of estimated VARs to illustrate the bias-variance trade-off in the context of IRF estimation and find that sometimes VAR IRF estimates are sharper and sometimes LP estimates have lower risk. This means that in practice, a researcher who is open to a task-based analysis is well served by having tailored information criteria to assess the bias-variance trade-off, choose between estimators, and select hyperparameters and lag lengths for the task at hand. Our proposed PC (for ``prediction'') and IRFC (for IRF estimation) are such criteria. In particular, IRFC is the first criterion that allows empirical researchers to simultaneously choose between VAR and LP IRF estimates, determine the degree of shrinkage, and select the number of lags. 

The starting point of our analysis is the local misspecification framework in \cite{Schorfheide2005}, henceforth S2005. The paper assumes that the data are  generated by a stationary infinite-order vector moving average (VMA) process that drifts toward a VAR($p_*$) at rate $T^{-1/2}$, where $T$ is the time-series dimension of the sample. The asymptotic lag length $p_*$ is fixed and finite. For any finite sample of size $T$, the wedge between the infinite-order VMA generating the data and the finite-order model used by the econometrician for forecasting or IRF analysis is the source of misspecification. The $T^{-1/2}$ drift in the misspecification balances the bias-variance trade-off between the MLE and the LFE asymptotically. S2005 proposed a prediction criterion $PC_T(\iota,p)$ that provides an asymptotically unbiased estimate of the $h$-step-ahead prediction risk and can be used to select between the predictor $\iota$ and lag length $p$ based on information available at the end-of-sample forecast origin $T$. S2005 restricts attention to the prediction problem using least squares estimates, while we generalize the estimators to a shrinkage class and further address the non-trivial extension to the IRF problem.

The contributions and findings in this paper are as follows. First, we show that the proposed criteria deliver asymptotically unbiased estimates of the prediction risk and IRF estimation risk, respectively, as a function of $(\iota,\lambda,p)$. However, because of the local misspecification, the risk function estimates remain stochastic in the $T \longrightarrow \infty$ limit. Second, we highlight similarities and differences between the forecasting and the IRF estimation applications. On the one hand, they are based on the same multi-step regressions or forward iterations of one-step-ahead estimates. On the other hand, forecasting relies on coefficients that interact with $p$ lags of $y_t$, whereas IRF estimation only utilizes the estimated MA coefficient matrix of order $h$, which is an object that does not grow with the number of lags $p$ in the VAR. Thus, the task-based approach tends to choose different triplets of $(\iota,\lambda,p)$ for forecasting and IRF estimation.  

We calibrate data-generating processes (DGPs) to estimates of the medium and large VARs used for recursive forecasting in \cite{GiannoneLenzaPrimiceri2015} and then pick different dynamic misspecification settings and weight matrix choices for the loss functions. Inspecting the asymptotic risk lines, we find that shrinkage is generally beneficial and that the optimization over $(\lambda,p)$ often narrows the risk differentials between MLE and LFE procedures. Moreover, conditional on $\iota$, adjusting $\lambda$ reduces risk differentials across different lag lengths. Thus, rankings obtained conditional on $\lambda=0$ (no shrinkage) can be quite different from rankings given $\hat{\lambda}$. A Monte Carlo simulation examining the forecast performance shows that the $(\iota,\lambda,p)$ selection generally works well. In the absence of misspecification, the task-based workflow performs only slightly worse than a model-driven workflow that sets $\iota = mle$ and uses the MDD to determine $(\lambda,p)$ as prescribed by the Bayesian paradigm. However, if there is misspecification the adaptivity of the task-based workflow leads to substantially lower prediction risks than the model-driven workflow.  

Finally, we optimize IRFC and PC over $(\iota,\lambda,p)$ on 1,000 empirical seven-variable VARs constructed from the FRED-QD database. The empirical results reveal intricate patterns across the $(\iota,\lambda,p)$ dimensions, and hence provide strong support for a data-driven choice in IRF analyses, exactly in the form that our IRFC is tailored for. Moreover, PC selection of $(\iota,\lambda,p)$ is quite different from IRFC selection.  Based on this finding, our recommendation for practitioners is to use an IRF estimation risk criterion and not a criterion that measures forecast performance, when choosing $(\iota,\lambda,p)$ for an IRF estimator.

Our paper is connected to three broad strands of the econometrics literature: multi-step forecasting, IRF estimation, and model selection. The currently most active branch is the one on IRFs, partly due to the increasing popularity of LPs. Our LP IRF estimator is similar to the Bayesian LP estimator proposed by \cite{Miranda-AgrippinoRicco2021}. They use a quasi-MDD for hyperparameter determination (averaging rather than selection), whereas we propose to use a criterion that directly targets IRF estimation risk. \cite{PlagborgMoeller2021} show that LPs and VARs estimate the same IRFs in population if the number of lags is unrestricted. In finite samples, there is however a bias-variance trade-off that is illustrated in the aforementioned large-scale simulation study by LPW. Our IRFC is designed to resolve this trade-off. 


An important insight in the LP literature developed in the papers by \cite{MontielOleaPlagborgMoller2021} and \cite{MontielOleaEtAl2024}, henceforth MPQW, is that the use of ``additional'' lags in LPs, called lag augmentation, can alleviate inference problems caused by serial correlation in LP regression errors. In particular, using the same drifting DGP framework of S2005, MPQW show that under lag augmentation, which in the S2005 framework means that the number of lags in the VAR exceeds the asymptotic lag order of the DGP, the LP estimator is correctly centered up to order $O(T^{-1/2})$. MPQW exploit this centering property to construct IRF confidence intervals that have better coverage properties than the corresponding VAR confidence intervals and are hence robust to misspecification. For us, the correct centering eliminates misspecification bias that contributes to the mean squared error (MSE) of the IRF estimator. Lag length selection with IRFC accounts for the fact that additional lags can induce correct centering of the LP estimator. 

\cite{Ludwig2024} derives a finite-sample equivalence result between VAR and LP regressions, showing based on the Frisch-Waugh-Lovell Theorem that an iteration of increasingly larger order VARs can exactly replicate an LP, and similarly a multi-step VAR prediction can be replicated by LPs of equal and lower orders. He makes the case that a fair comparison of the two types of IRF estimators should not use the same number of lags for VAR and LP, but instead equalize model size by accounting for the result that a collection of VARs can replicate an LP and vice versa. Our setting adds a further dimension of complexity by allowing for shrinkage. We illustrate the intricate interplay between the three dimensions considered, $(\iota,\lambda,p)$, in an asymptotic experiment as well as in finite samples. Our IRFC model determination resolves the critique raised by \cite{Ludwig2024} by optimizing jointly over estimator and lag length.

In the multi-step forecasting literature, LFEs (also called multi-step or direct estimators) have been studied, among others, by \cite{Findley1983},  \cite{Weiss1991}, \cite{Bhansali1997}, \cite{ClementsHendry1998}, \cite{Ing2003}, S2005, and \cite{MSW2006}. Much of the theoretical work is based on the assumption that the DGP is fixed and the class of misspecified candidate forecasting models is increasing with sample size, e.g., \cite{Shibata1980}, \cite{SpeedYu1993}, \cite{Bhansali1996}, \cite{IngWei2003}. Thus, the discrepancy between the best estimated forecasting model and the DGP vanishes asymptotically. Building on S2005, we follow the opposite approach by keeping the class of forecasting models fixed and letting the degree of misspecification asymptotically vanish. In our setup, the degree of misspecification is ``too small'' to be consistently estimable. Hence, PC and IRFC provide only an asymptotically unbiased, but not consistent, estimate of the final prediction or IRF estimation risk.

With regard to the literature on model determination and information criteria, PC is a modification of \citeauthor{Shibata1980}'s (\citeyear{Shibata1980}) final prediction error criterion. In the empirical Bayes literature, the use of an unbiased risk estimate as an objective function for hyperparameter selection dates back to \cite{Stein1981}. In the recent Bayesian time series literature, hyperparameter determination based on MDDs dominates. In this regard, \cite{GiannoneLenzaPrimiceri2015} has been a very influential paper, albeit not the first to propose the use of MDDs. There is other work proposing objective functions that target the estimation risk of VAR coefficient estimators and transformations thereof, e.g., IRFs. Examples of such work include \cite{hansen2016stein} and \cite{Lohmeyer2018}. However, assumptions about model misspecification are different from ours, which leads to different risk estimates. Rather than using a model determination criterion, one could use a hold-out sample for $(\iota,\lambda,p)$ determination. This is straightforward in a forecasting application, because forecasts generated based on the estimation sample can be evaluated with the hold-out sample. Pseudo-out-of-sample forecasting performance was used for hyperparameter determination, for instance, in early implementations of the Minnesota prior, e.g., \cite{Todd1984}, \cite{Litterman1986}, \cite{DoanLittermanSims1984}, and the DSGE model prior in \cite{IngramWhiteman1994}. However, in the IRF estimation application it is less clear how to evaluate IRF estimates from the estimation sample using information from the hold-out sample. 

The remainder of this paper is organized as follows. Section~\ref{sec:VAR1} describes the shrinkage estimators, the drifting DGP, and derives the prediction risk for the forecasting application and the estimation risk for the IRF application. All of this is done in the context of a VAR(1), which then is generalized to a VAR($p$) setting in Section~\ref{sec:VARp} using companion form notation. The proposed PC and IRFC criterion functions for the choice of estimator, hyperparameter, and lag length are developed in Section~\ref{sec:modeldetermination}. Section~\ref{sec:numerics} provides a numerical illustration of the asymptotic formulas derived previously, Section~\ref{sec:MC} presents results from Monte Carlo experiments, and an empirical application is conducted in Section~\ref{sec:empirics}. Finally, Section~\ref{sec:conclusion} concludes. Proofs, derivations, and additional simulation results are relegated to the Online Supplement.

\section{VAR(1)-Based Multi-step Regressions}
\label{sec:VAR1}

Suppose that an econometrician considers a potentially misspecified VAR of the form 
\be
y_t = \Phi y_{t-1} + u_t, \quad u_t \sim {\cal N}(0,\Sigma_{uu}),
\label{eq:var}
\ee
where $y_t$ is an $n \times 1$ vector, to estimate a prediction function or IRFs for an infinite-order VMA process. The estimators are posterior means that combine an MLE or LFE with a prior mean using  hierarchical models; see Section~\ref{subsec:VAR1.estimators}. The distributional assumptions therein are only used to define the class of estimators considered, and no further reference is made in our theory to any of the distributional assumptions imposed by the hierarchical model. The degree of shrinkage is determined by a hyperparameter that controls the weight on the prior information. Setting this hyperparameter to zero leads to the least squares estimator that has been considered for forecasting in S2005.  The DGP is described in Section~\ref{subsec:VAR1.dgp}. It takes the form of a VAR but the innovations are distorted by an infinite-dimensional linear process that vanishes at rate $T^{-1/2}$. In Section~\ref{subsec:VAR1.limitdistribution} we derive the limit distribution of the MLE and LFE. The resulting frequentist prediction and IRF estimation risks (expected losses) are presented in Sections~\ref{subsec:VAR1.forecasting} and~\ref{subsec:VAR1.IRF}. To keep the exposition relatively simple, we first analyze forecasts and IRFs from a locally misspecified VAR(1). The extension to multiple lags and an unknown lag order $p$ is provided in Section~\ref{sec:VARp}. 

\subsection{MLE and LFE Shrinkage Estimators} 
\label{subsec:VAR1.estimators}

Under the assumption that forecasts are evaluated under a quadratic loss function of the form
\be
L(y_{T+h},\hat{y}_{T+h}) = \mbox{tr} \big[ W(y_{T+h}-\hat{y}_{T+h})(y_{T+h}-\hat{y}_{T+h})' \big] = \| y_{T+h}-\hat{y}_{T+h} \|_W^2,
\label{eq:loss}
\ee
where $W$ is a symmetric and positive-definite $n \times n$ weight matrix, the VAR in (\ref{eq:var}) implies that  the optimal $h$-step-ahead point predictor at forecast origin $T$ is
\be
 \hat{y}_{T+h} = \Phi^h y_T, \quad h=1,\ldots,H.  \label{eq:VAR1.yhat} 
\ee
Moreover, if one is willing to add more structure to the problem and assumes that the one-step-ahead forecast errors $u_t$ can be expressed as
\be
   u_t = \Sigma_{uu}^{tr} \Omega e_t,
   \label{eq:VAR.ID}
\ee
where $\Sigma_{uu}^{tr}$ is the lower-triangular Cholesky factor of $\Sigma_{uu}$, $\Omega$ is an orthogonal matrix, and $e_t \sim (0,I)$ is a vector of structural shocks, then the  IRF can be expressed as
\be
  \frac{\partial  y_{T+h}}{\partial e_T'} = \Phi^h \Sigma_{uu}^{tr} \Omega, \quad h=0,\ldots,H. \label{eq:VAR1.IRF}
\ee 
The reason for studying the prediction problem and the IRF estimation problem side by side is that both rely on an estimate of $\Phi^h$. Such an estimate can be generated iteratively, by estimating $\Phi$ using a regression of $y_t$ on $y_{t-1}$, which we refer to as MLE, or it can be obtained directly, using a multi-step-ahead regression of $y_t$ on $y_{t-h}$. We refer to the latter estimator as LFE because in the forecasting context it is obtained by using the loss function under which the forecasts are evaluated also for parameter estimation. Rather than using these two estimators directly, we combine their estimation objective functions with a prior distribution to obtain a posterior mean estimator that can be interpreted as a shrinkage estimator. The degree of shrinkage is controlled by a hyperparameter that we determine in Section~\ref{sec:modeldetermination}.

\noindent {\bf MLE Shrinkage Estimator.}
Define $S_{T,kl} = \sum_{t=1}^T y_{t-k} y_{t-l}'$. The MLE can be expressed as
\be
\hat{\Phi}_T(mle) = S_{T,01} S_{T,11}^{-1}.
\ee
The likelihood-based shrinkage estimator of $\Phi$ is defined as the posterior mean obtained by combining the likelihood function associated with (\ref{eq:var}) with the following prior (the covariance matrix conforms with column vectorization $\mbox{vec}(\Phi)$):
\be
\Phi | \Sigma_{uu} \sim \mathcal{N} \big( \underline{\Phi}_T, \;  (\tilde{\lambda}_T \underline{P}_\Phi)^{-1} \otimes \Sigma_{uu} \big).
\label{eq:prior.mle}
\ee
The prior is indexed by the hyperparameter $\tilde{\lambda}_T$ that controls the degree of shrinkage. The mean and the hyperparameter of the prior distribution are indexed by the sample size $T$ for a reason that will be explained below.   Using standard calculations, the posterior mean can be expressed as the matrix-weighted average of the prior mean and the MLE:
\be
\bar{\Phi}_T(mle,\tilde{\lambda}_T) = \big[ \tilde{\lambda}_T  \underline{\Phi}_T \underline{P}_\Phi  + \hat{\Phi}_T(mle) S_{T,11} \big]\bar{P}_\Phi^{-1}(\tilde{\lambda}_T), \quad \bar{P}_\Phi(\tilde{\lambda}_T) = \tilde{\lambda}_T \underline{P}_\Phi + S_{T,11}.
\label{eq:barphi.mle.tildelambda}
\ee
Note that for $\tilde{\lambda}_T=0$ we obtain that $\bar{\Phi}_T(mle,\tilde{\lambda}_T) = \hat{\Phi}_T(mle)$. Moreover, $\bar{\Phi}_T(mle,\tilde{\lambda}_T) = \underline{\Phi}_T$ if $\tilde{\lambda}_T\to\infty$. Let $\Psi = \Phi^h$. Then we can define the likelihood-based (plug-in) shrinkage estimator of $\Phi^h$ as\footnote{We are using a plug-in estimator $\bar{\Phi}^h$ rather than the posterior mean of $\Phi^h$ which would also depend on higher-order moments the posterior, which are negligible in our asymptotic analysis.}
\be
\bar{\Psi}_T(mle,\tilde{\lambda}_T) = \bar{\Phi}^h_T(mle,\tilde{\lambda}_T).
\label{eq:estimator.mle.plugin}
\ee

\noindent {\bf LFE Shrinkage Estimator.} The loss-function-based predictor is based on the multi-step regression
\be
y_t = \Psi y_{t-h} + v_t, \quad v_t \sim {\cal N}(0,\Sigma_{vv}),
\label{eq:multistep.regression}
\ee
ignoring the serial correlation in $v_t$ implied by the VAR(1) in (\ref{eq:var}). The rationale behind this estimator is that it directly targets the $h$-step-ahead forecast error covariance matrix. Define
\be
\hat{\Psi}_T(lfe) = S_{T,0h} S_{T,hh}^{-1}.
\ee
Using the prior 
\be
\Psi |\Sigma_{vv} \sim {\cal N} \big( \underline{\Psi}_T, \;  (\tilde{\lambda}_T \underline{P}_\Psi)^{-1} \otimes \Sigma_{vv} \big),
\label{eq:multistep.prior}
\ee
we obtain the quasi-posterior mean
\be
\bar{\Psi}_T(lfe,\tilde{\lambda}_T) = \big[ \tilde{\lambda}_T  \underline{\Psi}_T \underline{P}_\Psi  + \hat{\Psi}_T(lfe) S_{T,hh} \big]\bar{P}^{-1}_\Psi(\tilde{\lambda}_T), \quad \bar{P}_\Psi(\tilde{\lambda}_T) = \tilde{\lambda}_T \underline{P}_\Psi + S_{T,hh}.
\label{eq:multistep.posterior}
\ee

\subsection{Drifting DGP and Prior}
\label{subsec:VAR1.dgp}

We assume that the sample has been generated from a covariance stationary DGP with an infinite-dimensional VMA representation. While the sample size $T$ is fixed in practice, we use $T \longrightarrow \infty$ asymptotics to approximate the frequentist risk. If the DGP and the lag length of the misspecified forecasting model are fixed then the variance of the estimators of $\Phi^h$ vanishes at rate $T^{-1}$ whereas the misspecification bias does not disappear. Thus, eventually, the loss-function-based predictor dominates the likelihood-based predictor along this asymptote, even if the misspecification is small. 

To generate asymptotics that better reflect the finite-sample trade-offs one has two choices: either increase the dimensionality of the forecasting model with sample size or let the DGP drift toward the forecasting model. As in S2005, we pursue the latter approach and assume that the DGP takes the form of a drifting VMA process that is local to the VAR in (\ref{eq:var}):
\be
y_t = F y_{t-1} + \epsilon_t + \frac{\alpha}{\sqrt{T}} \left(\sum_{j=1}^\infty A_j L^j \right) \epsilon_{t}, \quad \epsilon_t \sim (0,\Sigma_{\epsilon \epsilon} ),
\label{eq:dgp}
\ee
where $L$ is the lag operator. This means that the misspecification bias of the MLE of $\Phi$ in (\ref{eq:var}) relative to the ``true'' $F$ in (\ref{eq:dgp}) is of order $O(T^{-1/2})$. The contribution of parameter estimation to prediction loss can be represented as the sum of a squared bias and a variance term. The $O(T^{-1/2})$ drift guarantees that these two terms are asymptotically of the same order. 

Posterior means combine information from the prior and likelihood function. If the eigenvalues of $F$ are less than one in absolute value and the $A_j$s satisfy a summability condition that will be stated more formally below, the information in the likelihood function accumulates at rate $T$ and the MLE behaves asymptotically as 
\be\label{eq:mlerep}
\hat{\Phi}_T(mle) - F = T^{-1/2} \xi_T +  O_p(T^{-1}), 
\ee
where $\xi_T$ is an $O_p(1)$ random variable. To arrive at a framework that leads to non-trivial shrinkage decisions we rewrite the prior in (\ref{eq:prior.mle}) as follows:
\be \label{eq:driftpriorrep}
(\Phi - F) | \Sigma_{uu} \sim {\cal N} \big( (\underline{\Phi}_T-F), \;  (\tilde{\lambda}_T \underline{P}_\Phi)^{-1} \otimes \Sigma_{uu} \big) \equiv  T^{-1/2} {\cal N} \big( \underline{\phi}, \;  (\lambda \underline{P}_\Phi)^{-1} \otimes \Sigma_{uu} \big),
\ee
where $\equiv$ denotes distributional equivalence.
We ensure that the information accumulates at rate $T$ by setting the precision as: 
\be
	\tilde{\lambda}_T = \lambda T. \label{eq:prior.precision.drift}
\ee
In addition, we let the prior mean drift toward $F$ at rate $T^{-1/2}$: 
\be
\underline{\Phi}_T  = F + T^{-1/2} \underline{\phi}.
\label{eq:prior.mean.drift}
\ee
Using (\ref{eq:barphi.mle.tildelambda}) to combine the assumptions about the prior in \eqref{eq:driftpriorrep} with the MLE representation \eqref{eq:mlerep}, we deduce that the posterior mean has the form 
\be
\bar{\Phi}_T(mle,\lambda) -F= \; T^{-1/2} \cdot \big[ \lambda  \underline{\phi} \underline{P}_\Phi + \xi_T (S_{T,11}/T) \big] \big( \lambda \underline{P}_\Phi + S_{T,11}/T \big)^{-1} + O_p(T^{-1}),
\ee 
where $S_{T,11}/T = O_p(1)$. 

Note that from now on we use the local degree of shrinkage $\lambda$ instead of $\tilde{\lambda}_T$ as an argument of the $\Phi_T(\cdot)$ and $\Psi_T(\cdot)$ functions. By construction, the relative weights on MLE and prior mean are no longer sample size dependent. This captures the fact that in practice prior distributions play an important role in regularizing VAR parameter estimates to obtain good empirical performance. Our asymptotic frequentist risk calculations will focus on the $O_p(T^{-1/2})$ term in the expansion of the posterior estimator. Analogously, we rewrite the prior for the LFE in \eqref{eq:multistep.prior} as
\be 
(\Psi - F^h) | \Sigma_{vv} \sim   T^{-1/2} {\cal N} \big( \underline{\psi}, \;  (\lambda \underline{P}_\Psi)^{-1} \otimes \Sigma_{vv} \big).
\ee

\subsection{Limit Distribution of MLE and LFE Shrinkage Estimators} 
\label{subsec:VAR1.limitdistribution}

We proceed by deriving the limit distributions for $\bar{\Psi}_T(mle,\lambda)$ and  $\bar{\Psi}_T(lfe,\lambda)$. To do so, we state some regularity conditions:

\begin{assumption} \hspace*{1cm}\\[-3ex]
	\label{a:dgp}
	\begin{tlist}
		\item All eigenvalues of $F$ have modulus less than one.
		\item The sequence of $n \times n$ matrices $\{A_j\}_{j=0}^\infty$ satisfies the following
		summability condition: $\sum_{j=0}^\infty j^2 \|A_j\| < \infty$.
		\item $\{ \epsilon_t \}$ is a sequence of independent, $n$-dimensional, mean
		zero random variates with $\EE[ \epsilon_t \epsilon_t'] = \Sigma_{\epsilon \epsilon}$. 
		\item The $\epsilon_t$'s are uniformly Lipschitz over all directions, that is,
		there exist $K > 0$, $\delta > 0$, and $\nu > 0$ such that for all
		$0 \le w-u \le \delta$,
		\[
		\sup_{a'a = 1 } \; \PP \{ u < a' \epsilon_t < w \} \le K (w-u)^\nu.
		\]
		\item There exists an $\eta > 0 $ such that 
		\[
		\EE \bigg[ \| \epsilon_t \|^{6H + \eta}  \bigg] < \infty,
		\]
		where $H$ is the maximum horizon in (\ref{eq:VAR1.yhat}) and \eqref{eq:VAR1.IRF}.
	\end{tlist}
\end{assumption}

Assumptions~\ref{a:dgp}(i) and~(ii) guarantee that for any fixed $T$ the DGP is stationary. Assumption~(iii) assumes that the innovations are homoskedastic. This condition is a natural benchmark that simplifies the presentation of the analytical results. In the Online Supplement we discuss how the objects in the main text generalize to heteroskedasticity of certain forms, including stationary stochastic volatility, Markov switching and deterministic breaks. Assumptions~(iv) and~(v) ensure that the finite sample moments of the estimators eventually exist, which will allow us to calculate expected losses based on the moments of the limit distribution. These conditions were also imposed in S2005 and are taken from \cite{FindleyWei2002}. All elliptical distributed $\epsilon_t$'s with bounded densities have the uniform Lipschitz property, including Gaussian and $t$-distributed random variables. The following theorem characterizes the limit distribution of the MLE and LFE shrinkage estimators for a fixed $\lambda$. We use $\Longrightarrow$ to denote convergence in distribution.

\begin{theorem}\label{thm:limitdis}
	Suppose that the DGP satisfies Assumptions~\ref{a:dgp}~(i) to (iii). Then, for $\iota\in\{lfe,mle\}$ and $\lambda \ge 0$: 
	\begin{equation}
		\Bar{\Psi}_T(\iota,\lambda)=F^h+T^{-1/2}[\delta(\iota,\lambda)+\alpha\mu(\iota,\lambda)+\zeta_T(\iota,\lambda)],
	\end{equation} 
	where $\zeta_T(\iota,\lambda) \Longrightarrow \zeta(\iota,\lambda) \sim {\cal N}(0,V(\iota,\lambda))$.
\end{theorem}

The formulas for the bias terms $\delta(\iota,\lambda)$ and $\mu(\iota,\lambda)$ and the asymptotic covariance matrix $V(\iota,\lambda)$ are provided in the proof of Theorem~\ref{thm:limitdis} in the Online Supplement. 
$\bar{\Psi}_T(\iota,\lambda)$ converges to $F^h$ in probability as $T \longrightarrow \infty$. The important terms for the subsequent frequentist risk calculations are those premultiplied by $T^{-1/2}$. Consider the case $\lambda=0$, then the weight on the prior mean is zero and $\delta(\iota,\lambda)=0$. The bias term $\mu(\iota,\lambda)$ arises from the covariance between $\Sigma_{j=1}^\infty A_j \epsilon_{t-j}$ and $y_{t-1}$ (or $y_{t-h}$).  For $\lambda > 0$, there is a second bias term, $\delta(\iota,\lambda)$, which captures the effect of the prior distribution. At $\lambda = \infty$, $\delta(\iota,\lambda)$ equals the local prior mean and $\mu(\iota,\lambda) = 0$. Finally, $\zeta(\iota,\lambda)$ is a mean-zero normal random variable. 

It is well known that $V(mle,0) < V(lfe,0)$ for $h>1$; see, for instance, S2005. The LFE is inefficient, because it ignores the serial correlation of $h$-step-ahead forecast errors in its estimation objective function. Shrinkage, i.e., $\lambda >0$, does not just affect the asymptotic bias, but also reduces the variance of the estimators. The larger the precision, the smaller the sampling variance of the shrinkage estimator. 

\subsection{Forecasting Application: Prediction Risk}
\label{subsec:VAR1.forecasting}

\noindent {\bf Risk and Optimal Prediction.} The MLE and LFE shrinkage predictors are defined as 
\be
\hat{y}_{T+h}(\iota,\lambda) = \bar{\Psi}_T(\iota,\lambda) y_T, \quad \iota \in \{mle, \, lfe\}.
\label{eq:predictor.mle}
\ee
As is common in the literature, to streamline the theoretical derivations we assume that there are two independent processes, $\{y_t\}$ and $\{ \tilde{y}_t\}$, both generated from the DGP in (\ref{eq:dgp}); see, for instance, \cite{Baillie1979}, \cite{Reinsel1980}, \cite{Shibata1980}, and  \cite{LewisReinsel1985,LewisReinsel1988}. The former is used for parameter estimation and the latter is the process to be forecast. This assumption removes the (asymptotically negligible) correlation between the parameter estimates and the lagged values at the forecast origin. The optimal predictor of a future observation $\tilde{y}_{T+h}$ generated from the
DGP is the conditional mean
\be
\hat{y}_{T+h}^{opt} = \mathbb{E}_T[\tilde{y}_{T+h}],
\ee
where the expectation is taken conditional on the (infinite) history
of the process up to time $T$ and the parameters $\alpha$,
$F$ and $A(L)$. The expected loss of $\hat{y}_{T+h}^{opt}$
provides a lower bound for the frequentist risk of any estimator.
We normalize the prediction risk ${\cal R}(\hat{y}_{T+h})$
of a predictor $\hat{y}_{T+h}$ as follows
\be
{\cal R}(\hat{y}_{T+h})
= \mathbb{E} \bigg[ \| \tilde{y}_{T+h} - \hat{y}_{T+h} \|_W^2 \bigg]
- \mathbb{E} \bigg[ \| \tilde{y}_{T+h} - \hat{y}_{T+h}^{opt} \|_W^2 \bigg]  =  \mathbb{E} \bigg[ \| \hat{y}_{T+h} - \hat{y}_{T+h}^{opt} \|_W^2 \bigg].
\label{eq:normrisk}
\ee
This expression can be interpreted as an expected regret. Under the quadratic loss function, it is the squared discrepancy between the actual predictor and the conditional mean.

\noindent {\bf Pseudo-optimal Value.} To characterize the pseudo-optimal value for $\Psi$ in the VAR(1)-based prediction function $\Psi \tilde{y}_T$ we define $A_0=0$ and $A(L) = \sum_{j=0}^\infty A_j L^j$, and let 
\[
z_t = A(L) \epsilon_t, \quad \Gamma_{yy,h} = \sum_{j=0}^\infty F^{j+h} \Sigma_{\epsilon \epsilon} F^{j'} , \quad  \Gamma_{zy,h} = \sum_{j=0}^\infty A_{j+h} \Sigma_{\epsilon \epsilon} F^{j'}.
\]
It was shown in S2005 that the pseudo-optimal value takes the form 
\be
\tilde{\Psi}_{*,T}(prd) = F^h +  \alpha T^{-1/2} \mu_*(prd)  + \alpha O(T^{-1}), \quad 
\mu_*(prd) =  \sum_{j=0}^{h-1} F^j \Gamma_{zy,h-j} \Gamma_{yy,0}^{-1}, 
\label{eq:Psi.pov}
\ee
where $prd$ here is shorthand for prediction. Comparing the formula for $\mu_*(prd)$ to the formula for $\mu(lfe,0)$ given in the Online Supplement as part of the proof of Theorem~\ref{thm:limitdis}, we deduce that in the absence of shrinkage, i.e., $\lambda=0$, the limit distribution of the LFE is centered at the pseudo-optimal value. The result is stated in the following lemma:
	
\begin{lemma} \label{lem:mulfe.mulfe}
For any horizon $h$, $\mu(lfe,0)= \mu_*(prd)$.
\end{lemma} 

\noindent\textbf{Prediction Risk.} The next theorem characterizes the limit of the finite-sample prediction risk 
of the MLE and LFE shrinkage predictors based on their limit distribution. Because of the normalization in (\ref{eq:normrisk}), the prediction risk is determined by the bias and variance of the estimators of $\Psi$. Assumptions~\ref{a:dgp} (iii) to (v) ensure that the finite-sample moments of the estimators are eventually finite and converge to the moments of the limit distribution.

To express the asymptotic prediction (and, later on, IRF estimation) risks compactly, we introduce a weighted norm for matrices. For a matrix $A$ and symmetric weight matrices $W$ and $\Gamma$ that conform to the row and column dimensions of $A$, respectively, we write\footnote{The operator vecr denotes row-wise vectorization, i.e., $\vecr(A)=\textup{vec}(A')$.}
\be
\left\Vert A \right\Vert_{W\otimes\Gamma}^2 = \tr\left\{ (W\otimes\Gamma)\, \vecr(A)\, \vecr(A)' \right\} = \tr\left\{ W A \Gamma A' \right\},
\label{eq:matrixnorm}
\ee
which extends the vector norm in~(\ref{eq:loss}) to matrix arguments.

\begin{theorem}
	\label{thm:prediction.risk}
	Suppose Assumption~\ref{a:dgp} is satisfied. Then, for $\iota\in\{mle,lfe\}$ and $\lambda \ge 0$:
	\be
	\lim_{T \longrightarrow \infty} T {\cal R}(\hat{y}_{T+h}(\iota,\lambda) ) = \Bar{\mathcal{R}}_{B}(\iota,\lambda)+\Bar{\mathcal{R}}_{V}(\iota,\lambda)+C, \label{eq:asymp.pred.risk}
	\ee
	where    
	\begin{eqnarray}
		\Bar{\mathcal{R}}_{B}(\iota,\lambda)&=&\left\Vert \delta(\iota,\lambda)+\alpha(\mu(\iota,\lambda)-\mu_*(prd))\right\Vert_{W\otimes\Gamma_{yy,0}}^2 \, , \nonumber \\
		\Bar{\mathcal{R}}_{V}(\iota,\lambda) &=&\tr \bigg\{ ( W \otimes \Gamma_{yy,0} ) V( \iota,\lambda) \bigg\} \, , \nonumber 
	\end{eqnarray}
	and $C$ is a constant characterized in the Online Supplement.
\end{theorem}

A proof of Theorem~\ref{thm:prediction.risk} is provided in the Online Supplement. The proof eventually replaces the finite-sample second moments of $\zeta_T(\iota,\lambda)$ by the moments of the limit random variable $\zeta(\iota,\lambda)$ in Theorem~\ref{thm:limitdis}. This requires that the sequence $\zeta_T(\iota,\lambda)$ is uniformly integrable. A formal proof using Assumptions~\ref{a:dgp} (iv) and (v) for $\lambda=0$ was provided in S2005. The proof can be extended to $\lambda>0$ by noting that the Bayes estimators considered in this paper are weighted averages of the least squares estimators in S2005 and (non-stochastic) prior means.

Recall that in (\ref{eq:normrisk}) we defined an expected regret that equals the discrepancy between a predictor $\hat{y}_{T+h}$ and the conditional mean $\hat{y}_{T+h}^{opt}$ associated with the infinite-dimensional DGP in (\ref{eq:dgp}). According to the calculations underlying Theorem~\ref{thm:prediction.risk}, the conditional mean $\hat{y}_{T+h}^{opt}$ can be replaced, without any consequences for hyperparameter selection, by a pseudo-optimal predictor from the VAR model $\tilde{\Psi}_{*,T}(prd)$; see (\ref{eq:Psi.pov}). The constant $C$ arises from this replacement, and since it does not depend on the predictor $(\iota,\lambda)$, it is irrelevant for rankings of predictors because it cancels out when computing risk differentials.

The prediction risk is decomposed into a bias term $\bar{\mathcal{R}}_B(\iota,\lambda)$ and a variance term $\bar{\mathcal{R}}_V(\iota,\lambda)$. Consider the LFE, which corresponds to $\iota=lfe$. Recall that for $\lambda=0$ the prior-induced bias $\delta(lfe,\lambda)=0$, and the regression-induced bias term $\mu(lfe,\lambda) = \mu_*(prd)$. Thus, $\bar{\mathcal{R}}_B(lfe,\lambda)=0$, but the variance term $\bar{\mathcal{R}}_V(lfe,\lambda)$ is large. Raising $\lambda$ generates some bias, but also reduces the variance contribution to the prediction risk. The same logic applies to the MLE shrinkage predictor, i.e., $\iota = mle$, except that $\mu_*(prd) - \mu(mle,0) \not=0$. For $\lambda > 0$, the $\delta(\iota,\lambda)$ term generated by the prior could either increase or decrease the estimation bias component. The smaller the misspecification $\alpha$, the less important is the bias term, and the more important the variance component of the risk, $\bar{\mathcal{R}}_V(\iota,\lambda)$ becomes, when choosing between the MLE and LFE shrinkage predictors. 

\subsection{IRF Application: Estimation Risk}
\label{subsec:VAR1.IRF}

\noindent {\bf Population IRFs.}  The DGP in (\ref{eq:dgp}) can be rewritten as an MA($\infty$) process of the form
\be
y_t = \sum_{s=0}^\infty F^s \epsilon_{t-s} + \frac{\alpha}{\sqrt{T}}\left( \sum_{s=0}^\infty F^s L^s \right)\left( \sum_{j=1}^\infty A_j L^j \right)  \epsilon_t.
\label{eq:dgp.irf}
\ee
Connecting the innovations $\epsilon_t$ to a vector of economic shocks $e^*_t \sim (0,I)$ as in (\ref{eq:VAR.ID}), we write 
\be
   \epsilon_t = \Sigma_{\epsilon \epsilon}^{tr} \Omega_* e^*_t,
   \label{eq:dgp.ID}
\ee 
where $\Sigma_{\epsilon \epsilon}^{tr}$ is the lower-triangular Cholesky factor of $\Sigma_{\epsilon \epsilon}$ and $\Omega_*$ is an orthogonal matrix. The true effect of a shock $e_{T}^*$ on $y_{T+h}$ is given by the MA coefficient matrix 
\be\label{eq:IRFp1}
\frac{\partial y_{T+h}}{\partial e_{T}^{*'}} = \left( F^h+
\frac{\alpha}{\sqrt{T}} \mu_*(irf) \right)\Sigma_{\epsilon \epsilon}^{tr} \Omega_* , \quad \mu_*(irf) = \sum_{j=0}^{h-1} F^j A_{h-j},  \quad h\geq 1. 
\ee
To simplify the notation, we dropped the IRF horizon $h$ from the $\mu(\cdot)$ argument. 

\noindent {\bf IRF Estimation Risk.} The IRF estimates can be obtained from $\bar{\Psi}_T(\iota,\lambda)$, where $\iota = mle$ corresponds to the VAR-based estimate and $\iota = lfe$ yields the LP based estimate. The posterior mean estimate $\bar{\Psi}_T(\iota,\lambda)$ needs to be multiplied by the matrix that connects the one-step-ahead forecast errors to the structural shocks; see (\ref{eq:dgp.ID}). In the subsequent risk calculation, we ignore the discrepancy between an estimate of $\Sigma_{uu}^{tr} \Omega$ and the population analogue $\Sigma_{\epsilon \epsilon}^{tr} \Omega_*$ because it is unrelated to the choice between VAR-based IRF estimation and LPs.\footnote{
	The selection problem considered in this paper is orthogonal to the structural shock identification problem for schemes that solely rely on an estimate of the reduced form error variance, since in that case the effect of misspecification is of order $O(1/T)$. For other identification schemes, such as sign restrictions or external proxies, dynamic misspecification may interfere with the recovery of structural shocks. However, as stressed by \cite{PlagborgMoeller2021}, the population problem of structural identification is logically distinct from the choice of finite-sample estimation approach. Therefore, our selection criteria focus on the latter while holding the identification scheme fixed.
} In many applications, the object of interest is the response to a subset of the structural shocks, rather than all structural shocks. Without loss of generality, we assume that the $n_{sh}$ shocks of interest are ordered first. We denote the first $n_{sh}$ columns of the matrix $\Sigma_{\epsilon \epsilon}^{tr}\Omega_*$ by the $n \times n_{sh}$ matrix $\Xi$ and simply treat $\Xi$ as known. The asymptotic behavior of the IRF estimation risk 
\be
		{\cal R}_{IRF}(\iota,\lambda) =  \mathbb{E}\left[\left\|\bar{\Psi}_T(\iota,\lambda)  -\left[F^h+\frac{\alpha}{\sqrt{T}}\mu_*(irf)\right]  \right\|^2_{ W \otimes \Xi\Xi'} \right] \, , 
\ee
is summarized in the following theorem, which mirrors Theorem~\ref{thm:prediction.risk} above.

\begin{theorem}
	\label{thm:IRF.risk}
	Suppose Assumption~\ref{a:dgp} is satisfied. Then, for $\iota\in\{mle,lfe\}$ and $\lambda \ge 0$:
	\be
	\lim_{T \longrightarrow \infty}\; T {\cal R}_{IRF}(\iota,\lambda) = \Bar{\mathcal{R}}_{IRF,B}(\iota,\lambda) + \Bar{\mathcal{R}}_{IRF,V}(\iota,\lambda), \label{eq:IRFrisklim} 
	\ee
	where
	\begin{eqnarray} 
		\Bar{\mathcal{R}}_{IRF,B}(\iota,\lambda) &=&  
		\bigg\| \delta(\iota,\lambda) +\alpha(\mu(\iota,\lambda)-\mu_*(irf))  \bigg\|^2_{ W \otimes \Xi\Xi'} \, ,\nonumber \\
		\Bar{\mathcal{R}}_{IRF,V}(\iota,\lambda)
		&=& \tr\bigg\{\left[W\otimes (\Xi \Xi')\right]V(\iota,\lambda)\bigg\} \, . \nonumber 
	\end{eqnarray}
\end{theorem}

The $n \times n$ weight matrix $W$ in Theorem~\ref{thm:IRF.risk} reweights the IRF estimation errors across the response variables. The only two differences between the IRF estimation risk and the prediction risk are the weight matrices, $W \otimes \Xi\Xi'$ versus $W \otimes \Gamma_{yy,0}$, and the desired biases, $\mu_*(irf)$ versus $\mu_*(prd)$.

Consider the no-shrinkage case, $\lambda=0$. The variance formulas in Theorem~\ref{thm:limitdis} imply that the LP estimate always has higher variance than the VAR-based estimate; see S2005. Moreover, there is no clear ranking of $\Bar{\mathcal{R}}_{IRF,B}(lfe,0)$ and $\Bar{\mathcal{R}}_{IRF,B}(mle,0)$, which means that even under large misspecification $\alpha$, the VAR-based IRF estimator may dominate the LP estimator. This is broadly consistent with the simulation experiments reported by LPW (Figures 2 and 3), where the authors illustrate that their LP estimates dominate the VAR estimates in terms of bias but the VAR estimates have lower variance, leading to ambiguous rankings. LPW emphasize that the preferred estimator largely depends on how much one trades off variance and bias and conclude that, for researchers who also care about precision, VAR methods are the most attractive.  We will come back to this point when we consider a numerical illustration of the asymptotic risk formulas.

\section{Multiple Lags and Unknown $p_*$}
\label{sec:VARp}

So far, we assumed that the predictors and IRF estimators are derived from a VAR with a single lag and that the DGP is local to a VAR(1). In order to capture the more realistic case of unknown lag order, we  now extend the analysis to the case in which the econometrician considers VARs of lag orders $p=1,\ldots,q$ and the DGP is local to a VAR($p_*$), where $p_* < q$. Going forward, all $(\iota,\lambda)$ arguments in the formulas in Section~\ref{sec:VAR1} will be replaced by $(\iota,\lambda,p)$ to highlight the dependence on the number of lags.

\noindent {\bf Companion Form.}  To cleanly generalize the formulas derived in Section~\ref{sec:VAR1}, we write the VAR in $q$th-order companion form and then restrict the lag length to $p \le q$. Thus, $q$ can be thought of as the maximum number of lags considered by the researcher. We provide an overview here in the main text and further details in the Online Supplement. The $q$-companion form is given by
\be
Y_t  = \Phi  Y _{t-1} + U _t, \quad \Sigma_{UU} = M \Sigma_{uu} M',
\label{eq:var.companionform}
\ee
where
\[
\underbrace{Y_t}_{nq \times 1} = \left[ \begin{array}{c} y_t \\ y_{t-1} \\ \vdots \\ y_{t-q+1} \end{array} \right], \quad
\underbrace{\Phi}_{nq \times nq} = \left[ \begin{array}{ccccccc} \phi_1  & \cdots & \phi_{q-1} & \phi_{q} \\
	I_n    & \cdots & 0_n      & 0_n \\
	\vdots &  \ddots &  \vdots         & \vdots \\
	0_n    & \cdots & I_n      & 0_n 
\end{array} \right], \quad 
\underbrace{U_t}_{nq \times 1} = \left[ \begin{array}{c} u_t \\ 0_{n \times 1} \\ \vdots \\ 0_{n \times 1} \end{array} \right], \quad 
\underbrace{M}_{nq \times n}  = \left[ \begin{array}{c} I_n \\ 0_n \\ \vdots \\ 0_n \end{array} \right].
\]
Here $M$ is a selection matrix such that $y_t = M'Y_t$. The $q$-companion form looks identical to (\ref{eq:var}), except that we replaced lower-case by upper-case variables.

\noindent {\bf Imposing Lag-length Restrictions.} To impose the restriction that the coefficient matrices on lags $p+1,\ldots,q$ are equal to zero, we define the selection matrices 
\be
\underbrace{R_p}_{nq \times n(q-p)}  = \begin{bmatrix} 0_{np \times n(q-p)}\\ I_{n(q-p)} \end{bmatrix}, \quad
\underbrace{R_{p\perp}}_{nq \times np} = \begin{bmatrix} I_{np} \\ 0_{ n(q-p) \times np} \end{bmatrix} . \label{eq:def.R}
\ee
Then, the lag length restriction can be expressed as
\be
M'\Phi R_p = 0. 
\label{eq:M.Phi.Rp}
\ee
Moreover, the $p$-companion form coefficient matrix of the VAR($p$) is given by $R_{p \perp}' \Phi R_{p \perp}$.

\noindent {\bf Companion-form DGP.} The DGP in $q$-companion form is given by:
\be
Y_t = F Y_{t-1} + M \epsilon_t + \frac{\alpha}{\sqrt{T}} M \left(\sum_{j=1}^{\infty} A_j L^j \right)  \epsilon_{t}, \quad Z_t =  M \left(\sum_{j=1}^{\infty} A_j L^j \right)  \epsilon_{t}  \label{eq:dgp.companionform},
\ee
where $F$ is an $nq \times nq$ companion form matrix and the $n\times n$ matrices $A_j$ are identical to those in (\ref{eq:dgp}). We assume that the asymptotic lag order is $p_* < q$, which implies that $M' F R_p = 0$ if and only if $p \ge p_*$. In the prediction problem, the pseudo-optimal value generated from a VAR($p$) model generalizes to $\mu_*(prd,p)$, where the notation explicitly inherits the dimension of the estimated model $p$. The effect of dynamic misspecification on IRFs in a local-to-VAR($p_*$) DGP is given by 
\be
\label{eq:IRF.pov}
\mu_*(irf) = \sum_{j=0}^{h-1}  F^j M A_{h-j}M',
\ee 
which generalizes the object found in \eqref{eq:IRFp1} for $p_*=1$ to any $p_*\geq 1$. In contrast to the pseudo-true centering in the prediction problem, $\mu_*(irf)$ does not change with the order of the estimated model. We let $\Sigma_{EE} = M\Sigma_{\epsilon \epsilon} M'$ and use $\Gamma_{YY,h}$ to denote companion form autocovariances.

\noindent {\bf Limit Distribution and Risks.} First, Theorems~\ref{thm:limitdis}, \ref{thm:prediction.risk}, and \ref{thm:IRF.risk} have straightforward, albeit tedious, generalizations to the case of $p>1$. For instance, the LFE can be written as
\begin{eqnarray}
	\sqrt{T} \big( \bar{\Psi}_T(lfe,\lambda,p)  - F^h \big)  
	&=& - \sqrt{T} F^hR_p(R_p'\bar{S}_{T,hh}^{-1}R_p)^{-1}R_p'\bar{S}_{T,hh}^{-1}  \label{eq:LFE.limit}  \\
		&& + \delta(lfe,\lambda,p) + \alpha \mu(lfe,\lambda,p) + \zeta_T(lfe,\lambda,p), \nonumber
\end{eqnarray}	
The terms $\delta(\cdot)$, $\mu(\cdot)$, and $\zeta_T(\cdot)$, now also with $p$ as an argument, are the companion form generalizations of the terms in Theorem~\ref{thm:limitdis}. We provide formulas for the bias terms and the asymptotic covariance matrices in the Online Supplement. The following lemma is useful for illustrating the consequences of setting $p < p_*$:

\begin{lemma}
	\label{lem:MFhRp} For horizons $h \ge 1$, we have $M' F^h R_p = 0$ if and only if $p \ge p_*$. 
\end{lemma}

According to (\ref{eq:LFE.limit}), multiplying $\sqrt{T} \big( \bar{\Psi}_T(lfe,\lambda,p)  - F^h \big)$ by the matrix $M'$ to select the first $n$ rows -- which are the ones used for predictions and IRFs for $y_{T+h}$ -- generates the term $-\sqrt{T} M'F^hR_p(R_p'\bar{S}_{T,hh}^{-1}R_p)^{-1}R_p'\bar{S}_{T,hh}^{-1}$. Lemma~\ref{lem:MFhRp}  implies that this term will diverge whenever $p < p_*$. Therefore, it is crucial to use a lag length selection criterion with the property that the probability of $\hat{p} < p_*$ goes to zero as $T \longrightarrow \infty$.

\noindent {\bf Differences Between Prediction and IRF Estimation.} There are two important differences between prediction and IRF estimation. First, the former is based on the $n \times np$ matrix $M'\bar{\Psi}_T(\iota,\lambda,p)$, whereas the latter is based only on the $n\times n$ $h$-order MA coefficient matrix $M'\bar{\Psi}_T(\iota,\lambda,p)M$. Second, in the prediction problem the LFE is (pseudo)-optimally centered, i.e., $M'\mu(lfe,0,p) = M'\mu_*(prd,p)$, whereas we found that for the IRF estimation the centering of the LFE is suboptimal: $M'\mu(lfe,0,p)M \not= M'\mu_*(irf)M$ for $p=p_*=1$. 
However, by increasing $p$ beyond $p_*$ there is scope for improving the centering of the LFE in the IRF application. The fact that LPs use the higher-order lags only as control variables, but do not utilize the coefficients associated with them, leads to the following important relationship between $\mu(lfe,0,p)$ and $\mu_*(irf)$: 

\begin{lemma}
	\label{thm:mulfe.eq.muirf} $M' \mu(lfe,0,p) M = M'\mu_*(irf)M$ if and only if $p \ge p_*+1$.
\end{lemma}

The benefit of using additional lags as controls, termed lag-augmentation, has been previously stressed in the work by \cite{MontielOleaPlagborgMoller2021} and  MPQW. Lemma~\ref{thm:mulfe.eq.muirf} reproduces the benefit of lag-augmentation in the context of the drifting coefficient DGP considered in this paper: in the absence of shrinkage, the LP IRF estimate is optimally centered for any $p>p_*$. In addition, the variance remains constant. For $\lambda=0$, Proposition 3.1 in MPQW applies also to our setting and implies that the LP estimation risk is identical for all $p>p_*$. Moreover, according to Corollary~3.2 of MPQW, the VAR-based IRF estimation risk does not depend on $p$ and is identical to the LP risk for $\lambda=0$ and horizons $h \le p-p_*$. These two properties of LP IRF estimates are quite different from the properties of LFE predictors. In the forecasting application, additional lags always weakly reduce the bias relative to the true conditional expectation,\footnote{Notice that as $p$ is increased, the constant $C$ in Theorem~\ref{thm:prediction.risk} weakly decreases because the true conditional expectation of $\tilde{y}_{T+h}$ is projected onto a larger space.} but also increase the variance of the predictor.

\section{Model Determination and Estimator Selection}
\label{sec:modeldetermination}

The researcher has to determine the triplet $(\iota,\lambda,p)$. Thus, we now derive two criteria that are based on asymptotically unbiased risk estimates (URE) for the prediction risk and the IRF estimation risk, respectively, and can be used for task-based selection.\footnote{The analysis in this section is based on a generalization of Theorems~\ref{thm:limitdis}, \ref{thm:prediction.risk} and~\ref{thm:IRF.risk} to $p>1$ in $q$-companion form, building on (\ref{eq:LFE.limit}).} The URE objective function for the forecasting problem in Section~\ref{subsec:modeldetermination.forecasting} generalizes the PC criterion proposed in S2005 to cover hyperparameter $\lambda$ determination. Section~\ref{subsec:modeldetermination.IRF} derives a criterion that targets the IRF estimation risk. A step-by-step implementation guide is provided in Section~\ref{subsec:modeldetermination.cookbook}. Because in the Bayesian VAR literature it is common to select prior hyperparameters using an MDD, we present an MDD criterion for the hyperparameter selection in Section~\ref{subsec:modeldetermination.MDD}. Finally, we offer some brief remarks about optimality in Section~\ref{subsec:modeldetermination.optimality}.

\subsection{Forecasting} 
\label{subsec:modeldetermination.forecasting}

\noindent {\bf In-sample Prediction Loss.} The in-sample $h$-step ahead mean squared forecast error matrix can be written as
\be
MSE(\iota,\lambda,p) = \frac{1}{T}  \sum_{t=1}^T (y_t - M'\Bar{\Psi}_{T}(\iota,\lambda,p) Y_{t-h})
(y_t - M'\Bar{\Psi}_{T}(\iota,\lambda,p) Y_{t-h})' ,
\label{eq:MSE.iota.lambda.p}
\ee
where $y_t = M'Y_t$. We normalize the forecast error by the MSE of the unshrunk loss-function predictor with the maximum number of lags $q$, which gives the smallest in-sample MSE, and define the loss differential
\be
\Delta_{L,T} (\iota,\lambda,p)
= T \left( \tr\left\{ W \cdot MSE(\iota,\lambda,p)\right\} - \tr\left\{W \cdot MSE(lfe,0,q)\right\} \right) \ge 0.
\label{eq:mse.differential}
\ee
The asymptotic behavior of the risk differential can be characterized as follows:

\begin{theorem}
	\label{thm:ure}
	Suppose that Assumption~\ref{a:dgp} is satisfied. Then, for $\iota\in\{mle,lfe\}$, $\lambda \ge 0$, and $p_* \le p \le q$:
	\begin{tlist}
		\item  The in-sample
		forecast error loss differential has the following limit distribution
		\begin{eqnarray*}
			\Delta_{L,T} (\iota,\lambda,p)
			&\Longrightarrow& \left\Vert M'\delta(\iota,\lambda,p)\right\Vert_{W\otimes\Gamma_{YY,0}}^2+\alpha^2\left\Vert M'\left(\mu_*(prd,q)-\mu(\iota,\lambda,p)\right)\right\Vert_{W\otimes\Gamma_{YY,0}}^2\\
			&&+\left\Vert M'\left(\zeta(lfe,0,q)-\zeta(\iota,\lambda,p)\right)\right\Vert_{W\otimes\Gamma_{YY,0}}^2\\
			&&+2\alpha \tr\left\{W M'\left[\mu_*(prd,q)-\mu(\iota,\lambda,p)\right]\Gamma_{YY,0}\left[\zeta(lfe,0,q)-\zeta(\iota,\lambda,p)\right]'M\right\}\\
			&&-2\alpha \tr\left\{W M'\delta(\iota,\lambda,p)\Gamma_{YY,0}
			\left[\mu_*(prd,q)-\mu(\iota,\lambda,p)\right]'M\right\}\\
			&&-2\tr\left\{W M'\delta(\iota,\lambda,p)\Gamma_{YY,0}
			\left[\zeta(lfe,0,q)-\zeta(\iota,\lambda,p)\right]'M\right\}.
		\end{eqnarray*}

		\item The expected in-sample forecast error differential converges to
		\begin{eqnarray*}
			\mathbb{E}\left[\Delta_{L,T} (\iota,\lambda,p)\right] 			&\longrightarrow&
			\bar{\cal R}_B(\iota,\lambda,p) +\bar{\cal R}_V(\iota,\lambda,p) - \big(\bar{\cal R}_B(lfe,0,q) +\bar{\cal R}_V(lfe,0,q) \big) \\
			&& + 2\bar{\cal R}_V(lfe,0,q) - 2 \tr\left\{\left((MWM')\otimes\Gamma_{YY,0}\right)Cov(lfe,0,q;\iota,\lambda,p)\right\}.
		\end{eqnarray*}
	\end{tlist}
\end{theorem}

The distribution of the random variable $\zeta(\iota,\lambda,p)$ corresponds to the limit distribution of $\zeta_T(\iota,\lambda,p)$, defined in (\ref{eq:LFE.limit}) and its MLE analogue. The convergence (in distribution) result was stated in Theorem~\ref{thm:limitdis} for the special case of $p=1$ and has a straightforward generalization to $p>1$.  It is important to note that, in our local analysis, the loss differential $\Delta_{L,T}(\iota,\lambda,p)$ converges in distribution to a random variable and not a constant. The analytical expression for the covariance term can be found in the Online Supplement.

\noindent {\bf From In-Sample to Out-of-Sample Prediction Risk.} We now construct a selection criterion whose expected value, for large $T$, matches the asymptotic prediction risk 
up to a term that differences out across model specifications. Because $\bar{\cal R}_V(lfe,0,q)$ is irrelevant for comparisons across different $(\iota,\lambda,p)$, the formula for the limit of the expected loss suggests correcting the MSE by twice the covariance component of the asymptotic risk.

\begin{definition} \label{def:pc}
	Define the $PC_T(\iota,\lambda,p)$ criterion for the joint selection of estimator type, degree of shrinkage, and lag length as
	\[
	PC_T(\iota,\lambda,p) = T \tr[W \cdot MSE(\iota,\lambda,p)] + 2 \hat{\cal R}_{Cov}(lfe,0,q;\iota,\lambda,p),
	\]
	where $\hat{\cal R}_{Cov}(lfe,0,q;\iota,\lambda,p)$ has the property that
	\[
	\mathbb{E}[\hat{\cal R}_{Cov}(lfe,0,q;\iota,\lambda,p)] \longrightarrow \tr\left\{\left((MWM')\otimes\Gamma_{YY,0}\right)Cov(lfe,0,q;\iota,\lambda,p)\right\}.
	\]
\end{definition}

The term $2\hat{\cal R}_{Cov}(lfe,0,q;\iota,\lambda,p)$ in the definition of the PC criterion can be viewed as a penalty term that turns the in-sample fit measured by $MSE(\iota,\lambda,p)$ into a measure of out-of-sample fit. It must fulfill an asymptotic unbiasedness condition to guarantee that PC provides a URE for the asymptotic prediction risk. After combining Definition~\ref{def:pc} of the selection criterion with the MSE differential formula in (\ref{eq:mse.differential}) and Theorem~\ref{thm:ure}, we can deduce that
\begin{eqnarray}
	\lefteqn{\EE \big[ PC_T(\iota,\lambda,p) - PC_T(\iota',\lambda',p') \big] } \label{eq:E.PC.limit}\\
	&=& \EE \big[ \Delta_{L,T}(\iota,\lambda,p) - \Delta_{L,T}(\iota',\lambda',p') \big] + 2  \mathbb{E} \big[\hat{\cal R}_{Cov}(lfe,0,q;\iota,\lambda,p) - \hat{\cal R}_{Cov}(lfe,0,q;\iota',\lambda',p') \big] \nonumber \\
	&\longrightarrow & \Big(\bar{\mathcal{R}}_B(\hat{y}_{T+h}(\iota,\lambda,p)) + \bar{\mathcal{R}}_V(\hat{y}_{T+h}(\iota,\lambda,p)) \Big) - \Big(  \bar{\mathcal{R}}_B(\hat{y}_{T+h}(\iota',\lambda',p')) +  \bar{\mathcal{R}}_V(\hat{y}_{T+h}(\iota',\lambda',p')) \Big) \nonumber
\end{eqnarray}
as $T \longrightarrow \infty$. $PC_T(\iota,\lambda,p)$ can be used to choose between MLE and LFE, to select the shrinkage hyperparameter $\lambda$, and to determine the lag length $p$. Note that in our local misspecification framework it is not possible to consistently estimate the end-of-sample $h$-step-ahead prediction risk due to the rescaling in \eqref{eq:asymp.pred.risk}. One can only construct estimates that remain noisy in the limit. PC provides such an estimate and has the property that it is unbiased for the risk. In Section~\ref{subsec:MC.riskdifferentials.pc} we provide a numerical illustration that compares draws from the distribution of $PC_T(\iota,\lambda,p)$ to the asymptotic risk function.

\noindent {\bf An Alternative Selection Criterion.} One could define an alternative selection criterion by replacing the goodness-of-fit term $T \tr[W \cdot MSE(\iota,\lambda,p)]$ with the difference between $\bar{\Psi}_T(\iota,\lambda,p)$ and $\bar{\Psi}_T(lfe,0,q)$, which can be expanded as follows:
\begin{eqnarray*}
	\lefteqn{\sqrt{T}\big( \bar{\Psi}_T(\iota,\lambda,p)-\bar{\Psi}_T(lfe,0,q) \big) } \\
	&=& \sqrt{T} \big( \bar{\Psi}_T(\iota,\lambda,p) - F^h - \alpha \mu_*(prd,q) \big) - \sqrt{T} \big( \bar{\Psi}_T(lfe,0,q) - F^h - \alpha \mu_*(prd,q) \big).
\end{eqnarray*}
Define the scaled and weighted differential between the two posterior mean estimators as 
\be
   \Delta_{\Psi,T}(\iota,\lambda,p) = T\left\Vert M'\left(\bar{\Psi}_T(\iota,\lambda,p)-\bar{\Psi}_T(lfe,0,q)\right)\right\Vert_{W\otimes \Gamma_{YY,0}}^2.
\ee 
In turn, we can construct a slightly modified selection criterion:

\begin{definition} \label{def:pc.star}
	Define the $PC_T^*(\iota,\lambda,p)$ criterion for the joint selection of estimator type, shrinkage, and lag length as
	\[
	PC_T^*(\iota,\lambda,p) =  \Delta_{\Psi,T}(\iota,\lambda,p)+ 2 \hat{\cal R}_{Cov}(lfe,0,q;\iota,\lambda,p).
	\]
\end{definition}

The calculations underlying the proof of Theorem~\ref{thm:ure} can be modified to establish that
\begin{eqnarray}
	\mathbb{E} \big[\Delta_{\Psi,T}(\iota,\lambda,p)  \big]  
	&\longrightarrow & \bar{\mathcal{R}}_B(\iota,\lambda,p) + \bar{\mathcal{R}}_V(\iota,\lambda,p)
	+ \bar{\mathcal{R}}_V(lfe,0,q) \label{eq:expected.psibar.differential} \\
	&&- 2 \tr\left\{\left((MWM')\otimes\Gamma_{YY,0}\right)Cov(lfe,0,q;\iota,\lambda,p)\right\} \nonumber
\end{eqnarray}
as $T \longrightarrow \infty$. Thus, $PC_T^*(\iota,\lambda,p)$ also provides, up to the constant $\bar{\mathcal{R}}_V(lfe,0,q)$, an asymptotically unbiased risk estimate. The criterion $IRFC_T(\iota,\lambda,p)$ developed in the next section will be closely connected to $PC_T^*(\iota,\lambda,p)$. In unreported simulations, we find that both $PC_T$ and $PC_T^*$ perform equally well in multi-step forecasting applications.

\subsection{IRF Estimation}
\label{subsec:modeldetermination.IRF}

Given that LP estimates do not uniformly (in terms of misspecification) dominate VAR-based IRF estimates, it is natural to employ a method like the PC criteria in Definitions~\ref{def:pc} and~\ref{def:pc.star} to find the preferred IRF estimator, akin to what has been proposed above for multi-step point prediction. We can now modify the derivations of $PC_T^*(\iota,\lambda,p)$ to obtain a selection criterion for IRF estimation. Lemma~\ref{thm:mulfe.eq.muirf}, in combination with the assumption that $p_*<q$, lets us replace $M' \mu(lfe,0,q) M$ by $M'\mu_*(irf)M$. Define the differential of IRF estimators as 
\be
  \Delta_{IRF,T}(\iota,\lambda,p) = T \left\|M'\left(\bar{\Psi}_T(\iota,\lambda,p)-\bar{\Psi}_T(lfe,0,q)\right)\right\|^2_{W\otimes(M\Xi \Xi'M')}.
\ee 
Then we can turn (\ref{eq:expected.psibar.differential}) into 
\begin{eqnarray}\label{eq:Eirfnormdif}
	\mathbb{E}\big[\Delta_{IRF,T}(\iota,\lambda,p) \big] 
	&\longrightarrow&
	\Bar{\mathcal{R}}_{IRF,B}(\iota,\lambda,p) +  \Bar{\mathcal{R}}_{IRF,V}(\iota,\lambda,p) + \Bar{\mathcal{R}}_{IRF,V}(lfe,0,q) \\
	&& - 2 \mbox{tr}\bigg\{\left[(MWM')\otimes (M\Xi \Xi'M')\right]Cov(lfe,0,q;\iota,\lambda,p)\bigg\}, \nonumber 
\end{eqnarray}
as $T \longrightarrow \infty$. Here $\Bar{\mathcal{R}}_{IRF,B}(\iota,\lambda,p)$ and $\Bar{\mathcal{R}}_{IRF,V}(\iota,\lambda,p)$ are companion form generalizations of $\Bar{\mathcal{R}}_{IRF,B}(\iota,\lambda)$ and $\Bar{\mathcal{R}}_{IRF,V}(\iota,\lambda)$ in Theorem~\ref{thm:IRF.risk}, respectively. The only major difference between multi-step prediction and IRF estimation is the weighting applied to the $\bar{\Psi}_T(\cdot)$ differential. A more subtle difference is the need for a strict inequality $p_*<q$ rather than $p_*\le q$. By construction, the squared bias term in (\ref{eq:Eirfnormdif}) is identical to that in (\ref{eq:IRFrisklim}). The term $\Bar{\mathcal{R}}_{IRF,V}(lfe,0,q)$ is independent of $(\iota,\lambda,p)$ and does not affect the ranking of estimators. Thus, we deduce that the following criterion provides, up to a constant that does not depend on $(\iota,\lambda,p)$,  an asymptotically unbiased estimate of the large sample risk:

\begin{definition} \label{def:pcIRF}
	Define the $IRFC_T(\iota,\lambda,p)$ criterion for the joint selection of IRF estimator, shrinkage, and lag length, with impact matrix $\Xi$, as
	\[
	IRFC_T(\iota,\lambda,p) = T\left\Vert M'\left(\bar{\Psi}_T(\iota,\lambda,p)-\bar{\Psi}_T(lfe,0,q)\right)\right\Vert^2_{W\otimes(M\Xi\Xi'M')}+ 2 \hat{\cal R}_{CovIRF}(lfe,0,q;\iota,\lambda,p),
	\]
	where $\hat{\cal R}_{CovIRF}(lfe,0,q;\iota,\lambda,p)$ has the property that 
	\[
	\mathbb{E}[\hat{\cal R}_{CovIRF}(lfe,0,q;\iota,\lambda,p)] \longrightarrow \tr\left\{\left(MWM'\otimes M\Xi\Xi'M'\right)Cov(lfe,0,q;\iota,\lambda,p)\right\}.
	\]
\end{definition}

If $p < p_*$, then $IRFC_T$ diverges. Intuitively, one would expect that $IRFC_T$ will on average decrease if $p$ is raised to $p_*+1$ because the asymptotic bias is reduced. A further increase in the number of lags cannot improve the asymptotic bias component. We explore these trade-offs numerically by evaluating the asymptotic estimation risk formulas for various configurations of the DGP.

\subsection{Step-By-Step Implementation and Practical Considerations}
\label{subsec:modeldetermination.cookbook}

Our model determination approach is task-based, meaning that for every horizon $h$ a different specification can be selected. The scope of the method proposed in this paper is to construct point forecasts or point IRF estimates. While we use the $q$-companion form representation in Sections~\ref{sec:VARp} and~\ref{sec:modeldetermination} to present the formulas in a clean way, the expressions can also be evaluated directly using VAR($p$) formulas, which might be faster. 
	
\noindent {\bf Preliminary Implementation Steps Common to Both Applications.}
\begin{enumerate}
	\item Specify the weight matrices $W$, which might depend on horizon or application type.
    \item Specify a maximum number of lags $q$ (analogously as for AIC/BIC lag length selection), and a hyperparameter grid $\Lambda$, e.g., equally-spaced for $\ln \lambda$ with a zero appended. 
    \item Specify the means and covariance matrices of the prior distributions given by the companion-form generalizations of (\ref{eq:prior.mle}) and (\ref{eq:multistep.prior}):
    \begin{enumerate} 
	    \item Set the unrestricted elements in $\underline{\phi}_T$ and $\underline{\psi}_T$ in the companion form prior mean matrices $\underline{\Phi}_T$ and $\underline{\Psi}_T$. Examples include: (i) $\underline{\phi}_T = \underline{\psi}_T = 0$; (ii) prior mean matrices derived from univariate unit-root representations (Minnesota prior). (Approximate) alignment of prior means across MLE and LFE can be achieved using the first-order Taylor expansion of $\Phi^h - F^h$ (see Section~\ref{sec:numerics} for an application)
	    \be
	    \Phi^h-F^h \approx \sum_{j=0}^{h-1}F^j(\Phi-F)F^{h-1-j}. \label{eq:Phih.Phi.Taylor}
	    \ee
	    \item To specify the $nq \times nq$ companion-form analogues of the precision matrices $\underline{P}_{ \Phi}$ and $\underline{P}_{\Psi}$ it suffices to set the $np \times np$ submatrix that corresponds to the precision of the coefficients for lags 1 through $p$. Following the Minnesota prior literature, a benchmark choice is a diagonal precision matrix, potentially scaled by the sample variance of each $y_{i,t-j}$ element; see, for instance, \cite{Sims1998}. 	    
	    \end{enumerate} 
\end{enumerate}

\noindent {\bf Forecasting Implementation Steps:} For each forecast horizon $h$:
	\begin{enumerate}
		\item Evaluate $PC_T(\iota,\lambda,p)$ in Definition~\ref{def:pc} for $\iota \in \{ mle,lfe \}$, $\lambda \in \Lambda$, and $0 \le p \le q$. To do so, use (\ref{eq:MSE.iota.lambda.p}) to evaluate $MSE(\iota,\lambda,p)$, and use the correction term estimate (demean $Y_t$ as needed for the autocovariance estimates):
		\begin{eqnarray}
			\hat{\cal R}_{Cov}(lfe,0,q;\iota,\lambda,p) &=& \tr\left\{\left((MWM')\otimes\widehat{\Gamma}_{YY,0}\right)\widehat{Cov}(lfe,0,q;\iota,\lambda,p)\right\} \label{eq:Rhatcov} \\
			\widehat{Cov}(lfe,0,q;\iota,\lambda,p) &=& \mbox{use \eqref{appeq:cov.companionform} with population objects replaced by}
            \nonumber \\
			\widehat{\Gamma}_{YY,0} &=& \frac{1}{T} \sum_{t=1}^T Y_t Y_t', \quad \widehat{F} = \bar{\Phi}_T(mle,0,q),  \nonumber \\
			\widehat{\Sigma}_{EE} &=& M \left(\frac{1}{T} \sum_{t=1}^T \big( y_t - M'\widehat{F} Y_{t-1}\big)\big( y_t - M'\widehat{F} Y_{t-1} \big)'\right) M'. \nonumber 
		\end{eqnarray}		 
		\item Find the global argmin $(\hat{\iota},\hat{\lambda},\hat{p})$ and generate the point forecast  $M'\bar{\Psi}_T(\hat{\iota},\hat{\lambda},\hat{p})Y_T$. 
	\end{enumerate} 

\noindent {\bf IRF Estimation Implementation Steps:} Specify the impact matrix $\Xi$ that connects the structural shocks $e_t$ to the reduced-form innovations $\epsilon_t$. Recall that $\Xi$ was defined as the first $n_{sh}$ columns of $\Sigma_{\epsilon \epsilon}^{tr} \Omega_*$ in (\ref{eq:dgp.ID}). Then, for each forecast horizon $h$:
	\begin{enumerate}
		\item Evaluate $IRFC_T(\iota,\lambda,p)$ in Definition~\ref{def:pcIRF} for $\iota \in  \{ mle,lfe \}$, $\lambda \in \Lambda$, and $0 \le p \le q$, where $\hat{\cal R}_{CovIRF}(lfe,0,q;\iota,\lambda,p)$ can be obtained by modifying the weighting in (\ref{eq:Rhatcov}). 
		\item Find the global argmin $(\hat{\iota},\hat{\lambda},\hat{p})$ and generate IRF point estimate  $M'\bar{\Psi}_T(\hat{\iota},\hat{\lambda},\hat{p})M\Xi$.
	\end{enumerate} 

\noindent {\bf Multi-dimensional Hyperparameters.} In the Bayesian VAR literature, it is common to use multiple hyperparameters; see, for instance, \cite{DoanLittermanSims1984}, \cite{Sims1998}, and \cite{GiannoneLenzaPrimiceri2015}. Our prior distribution could be augmented by additional hyperparameters that control other features of the parameter distribution. For instance, one could incorporate a variance decay rate $\tau$ for lags of order $j=2,\ldots,p$ by multiplying the corresponding entries in the precision matrices $\underline{P}_\Phi$ and $\underline{P}_\Psi$ by $j^\tau$, and replace $\underline{P}_{\cdot}$ by $\underline{P}_{\cdot}(\tau)$ to highlight the dependence. Because this modification preserves the Kronecker structure of the prior and the paper makes no claims about a uniform convergence of the $PC_T(\cdot)$ and $IRFC_T(\cdot)$ objective functions to limit objective functions, the analysis in the paper remains valid and no adjustments to the selection criteria are necessary to jointly determine  $(\iota,\lambda,\tau,p)$. 

\noindent {\bf Heteroskedasticity.} Under heteroskedasticity, the error variance would depend on time, $\Sigma_{t,\epsilon\epsilon}$. Under the quadratic loss function, the predictor $\hat{y}_{T+h}(\iota,\lambda,p)$ does not directly depend on the innovation covariance matrix $\Sigma_{T,\epsilon \epsilon}$ at the end-of-sample forecast origin. The IRF estimation might indirectly depend on $\Sigma_{T,\epsilon \epsilon}$ through the construction of the shock impact matrix $\Xi$. If it does, then an estimate of $\Sigma_{T,\epsilon \epsilon}$ is required. Important for the generalization of PC and IRFC to a heteroskedastic setting is an estimate of the asymptotic covariance matrix of $\zeta_T(\iota,\lambda,p)$. We sketch in the Online Supplement how a generic adjustment can be constructed for a stochastic $\{\Sigma_{t,\epsilon \epsilon}\}$ process. If the heteroskedasticity takes the form of a structural break, e.g., a drop in volatility after 1984 (Great Moderation) or a volatility spike during the COVID pandemic, the adjustment could be targeted toward the structural breaks by essentially estimating different $\Sigma_{t,\epsilon \epsilon}$ matrices over the subsamples in which the volatility is assumed to be constant.

\noindent {\bf Model Selection Versus Averaging.} Our model determination approach emphasizes selection over averaging. In a Bayesian setting that combines a likelihood $p(Y|\Phi,\Sigma)$ with a hierarchical prior $p(\Phi,\Sigma|\lambda)$, it is natural to specify a prior $p(\lambda)$ over the hyperparameter $\lambda$. A model selection approach can be interpreted as setting $p(\lambda) \propto c$, where $\propto$ denotes proportionality, and making inference conditional on the $\hat{\lambda}$ that maximizes the MDD $p(Y|\lambda)$:
\be
   p(\Phi,\Sigma|Y,\hat{\lambda}), \quad \hat{\lambda} = \argmax_{\lambda \in \Lambda} \; p(Y|\lambda), \quad p(Y|\lambda) = \int p(Y|\Phi,\Sigma) p(\Phi,\Sigma|\lambda) d(\Phi,\Sigma).
\ee
Alternatively, a full Bayesian approach averages over $\lambda$:
\be
  p(\Phi,\Sigma|Y) = \int p(\Phi,\Sigma|Y,\lambda) p(Y|\lambda) \frac{p(\lambda)}{\int p(Y|\lambda) p(\lambda)d\lambda} d \lambda.  
  \label{eq:p.Phi.Sigma.Y} 
\ee
If $p(\Phi,\Sigma|Y,\lambda)$ is insensitive to $\lambda$ for values of $\lambda$ that are associated with a high MDD $p(Y|\lambda)$, then both approaches lead to similar point estimates and predictions. 

Two modifications of our approach would be possible but are not further explored in this paper. First, for a grid $\Lambda$ with $L$ grid points, one could define an average estimator 
\[
  \bar{\Psi}^{ave}_T\big(\iota,\{w_l\}_{l=1}^L,p \big) = \sum_{l=1}^L w_l \bar{\Psi}_T(\iota,\lambda_l,p),
\]
derive a URE for $\bar{\Psi}^{ave}_T\big(\iota,\{w_l\}_{l=1}^L,p \big)$ and use it to determine the weights $\{w_l\}_{l=1}^L$.\footnote{One could, in addition, also average across $\iota$ and $p$. {An example of constructing a URE for averaging across $p$ in a similar misspecified environment is \cite{Lohmeyer2018}.}} Second, one could use, say, $PC_T(\iota,\lambda,p)$ to generate averaging weights to replace $p(Y|\lambda)$ in (\ref{eq:p.Phi.Sigma.Y}):
\[
   \tilde{p}(Y|\lambda) \propto \exp \left\{ - PC_T(\iota,\lambda,p) \right\}.
\]

\noindent {\bf Point Versus Interval Prediction and IRF Estimation.} PC and IRFC are selection criteria that target quadratic prediction and IRF estimation losses. While we do not have any formal results that connect our criteria to the task of generating interval predictions or error bands for IRFs, it strikes us as reasonable to condition coverage intervals on $(\hat\iota,\hat\lambda,\hat p)$ chosen by our criteria; in the same way it is reasonable to condition coverage intervals derived from autoregressive models on the lag length that has been determined by the AIC, which targets the Kullback-Leibler distance to a ``true'', possibly infinite-dimensional DGP. However, the construction of coverage intervals does raise conceptual issues. 

For frequentist inference, the difficulty is the asymptotic bias introduced by the prior distribution and misspecification. Once $\hat{\lambda} > 0$, any symmetric coverage interval around our point prediction or IRF estimate does not have the desired frequentist coverage probability. This problem arises more generally, e.g., also under the MDD-based hyperparameter determination discussed in Section~\ref{subsec:modeldetermination.MDD}, when researchers try to interpret finite-sample Bayesian VAR credible intervals obtained under an informative prior, e.g., the Minnesota prior, as frequentist confidence intervals. The solution proposed in MPQW for IRF estimation is not to shrink ($\lambda=0$) and to use the LFE, so that the IRF estimator is correctly centered. In this case our IRFC can still provide a targeted lag length choice.

From a Bayesian perspective, the key challenge is the presence of unmodelled misspecification, which implies that the likelihood does not cover the ``true'' distribution of the data. Our presumption is that the researcher is aware of the potential misspecification but does not want to estimate a more complicated model. The object of posterior inference (simplifying the notation) would be $F^h + \alpha T^{-1/2} \mu_*$. A limited-information posterior that acknowledges misspecification could be obtained by ``inverting'' the sampling distribution of $\bar{\Psi}_T(\iota,\lambda,p)$ to obtain a quasi-posterior distribution for $F^h + \alpha T^{-1/2}\mu_*$.\footnote{Use Bayes Theorem with a prior $p(\theta) \propto c$ to argue that $p(\theta|\hat{\theta}) \propto p(\hat{\theta}|\theta)$, where $\propto$ denotes proportionality.} The term $\delta$ would reflect the initial prior distribution for $\Psi$, and $\alpha(\mu-\mu_*)$ is a nuisance term that would have to be integrated out using beliefs about the severity of the misspecification. \cite{Muller2013} develops this idea formally in a setting (without a nuisance term of the form $\alpha(\mu-\mu_*)$) in which the misspecification is generated by unmodeled heteroskedasticity that leads to a sandwich covariance matrix in the sampling distribution of a baseline estimator. It is beyond the scope of this paper to develop a similar theory for our problem.

\subsection{MDD Based Hyperparameter Selection}
\label{subsec:modeldetermination.MDD}

In the VAR literature, hyperparameters are often selected using the MDD for a VAR($p$). The VAR($p$) can be written in matrix form as 
\be
Y = X \Phi' + U.
\ee
Here $Y$ and $U$ are $T \times n$ matrices with rows $y_t'$ and $u_t'$, and $X$ is the matrix of stacked regressors. With a single lag, $X$ has rows $y_{t-1}'$ and is $T \times n$; more generally, $X$ stacks the $p$ lags entertained by the researcher, so that $X$ is $T \times np$ and $\Phi$ is $n \times np$. Because the MDD formula given below involves $X$ only through $X'X$ and the associated posterior objects, it is valid for any $p$. The dimension of the coefficient matrix $\Phi'$ also depends on $p$, but we omit the dependence from the notation. We combine the conditional prior for $\Phi|\Sigma_{uu}$ in \eqref{eq:prior.mle}  with a marginal distribution for $\Sigma_{uu}$:
\be
\Sigma_{uu} \sim IW(\underline{\nu},\underline{S}).
\ee
Let (we provide a closed-form formula in the Online Supplement) 
\be
p(Y|\lambda) = \int \int p(Y|\Phi,\Sigma_{uu})p(\Phi,\Sigma_{uu}|\lambda) d\Phi d \Sigma_{uu}.
\ee 
It is convenient to take log differentials. Because the log density is not defined for $\lambda = 0$, we consider deviations from $\lambda = \infty$. We also multiply the differential by $-1$, so that the hyperparameter determination is based on the minimization of the differential. This leads to the criterion (we now make the dependence on the lag length $p$ explicit):
\begin{eqnarray}
	MDD_T(\lambda,p) &=&
	2 \big[ \ln p(Y|\infty) - \ln p(Y|\lambda) \big] \label{eq:mdd.differential} \\
	&=& \bar{\nu} \big\{ \ln  |\bar{S}_T(\lambda)|-\ln |\bar{S}_T(\infty)|  \big\} + n \big\{  \ln |\lambda \underline{P}_\Phi + X'X/T| - \ln |\lambda \underline{P}_\Phi|   \big\}, \nonumber
\end{eqnarray}
where 
\[
\bar{S}(\lambda) = \underline{S} +  (\lambda T) \underline{\Phi}_T \underline{P}_\Phi \underline{\Phi}_T' + Y'Y - \bar{\Phi}_T \bar{P}_\Phi \bar{\Phi}_T', \quad
\bar{\nu} = \underline{\nu} + T.
\]
The first term in the second line of (\ref{eq:mdd.differential}) is an in-sample goodness of fit differential, which is scaled so that it can be shown to converge in distribution to a stochastic process indexed by $\lambda$. Thus, just as $PC_T(\iota,\lambda,p)$, the MDD function remains stochastic in the $T \longrightarrow \infty$ limit. The second term is a penalty differential that is $\infty$ for $\lambda=0$, and 0 for $\lambda= \infty$. For values of $\lambda >0$, it converges to a non-stochastic function of $\lambda$. Hyperparameter and lag length selection are based on the minimization of $MDD_T(\lambda,p)$ with respect to $(\lambda,p)$. We provide some comparisons of MDD versus PC selection in Section~\ref{sec:MC}.

\subsection{A Word on Optimality}
\label{subsec:modeldetermination.optimality}

Because we are considering frequentist risk in a setting in which the model that is estimated is misspecified and we are using criteria that rely on risk estimates, one might think that a natural framework for studying optimality is the one that underlies \cite{Stein1981}'s unbiased risk estimation approach and the literature that builds on it. Key features of this framework are: (i) there is a hyperparameter $\lambda$ that needs to be determined; (ii) an unbiased estimate of the risk as a function of $\lambda$ can be constructed from the available data; (iii) one can envision an oracle that is equipped with knowledge of the ``true'' parameters and can construct a $\lambda_*$ that minimizes estimation or prediction loss; (iv) there is a growing cross-sectional dimension $n$ that ensures that a procedure based on the minimizer $\hat{\lambda}$ of the estimated risk function attains the same risk as a procedure that uses $\lambda_*$. 

In our setting we have (i) and (ii), and for (iii) we also could imagine an oracle that constructs $\lambda_*$. However, (iv) is absent because the cross-sectional dimension $n$ is fixed and an increasing time dimension does not suffice for convergence of PC and IRFC to an oracle objective function because of the $T^{-1/2}$ drift in the DGP. PC and IRFC (and the MDD) remain stochastic in the $T \longrightarrow \infty$ limit, which leaves a gap between what would be $\lambda_*$ in our setting and the $\hat{\lambda}$s generated by our (and other) criteria. We showed formally in (\ref{eq:E.PC.limit}), (\ref{eq:expected.psibar.differential}), and (\ref{eq:Eirfnormdif}) that the average of our criterion across repeated sampling converges pointwise to the asymptotic risk function, but in every sample there is a deviation. This distribution of objective functions will be illustrated in Figure~\ref{fig:RiskvsPC_alpha2_h4_p1} in Section~\ref{sec:MC} below. In the simulations, the shape of our stochastic criterion functions does resemble the shape of the asymptotic risk function, but the distribution of criterion functions never becomes more concentrated as $T \longrightarrow \infty$. We do think that this setting is representative of the challenges faced by researchers in finite samples where oracle objective functions remain elusive.

\section{Numerical Illustration of Asymptotic Risks}
\label{sec:numerics}

We now provide a numerical illustration of the asymptotic prediction and IRF estimation risk formulas derived above. We first specify a DGP in Section~\ref{subsec:numerics.dgp}, then discuss the asymptotic forecasting risk in Section~\ref{subsec:numerics.forecasting}, and finally examine the IRF estimation risk in Section~\ref{subsec:numerics.irfs}.

\subsection{DGP and Prior} 
\label{subsec:numerics.dgp}

\noindent {\bf Data Generating Processes.} To induce realistic dynamics, we use OLS estimates constructed from U.S. data to generate two parameterizations of the DGP in (\ref{eq:dgp}) for $p_*=1$. The calibrated DGPs resemble the medium ($n=7$) and large ($n=20$) VARs recursively estimated in \cite{GiannoneLenzaPrimiceri2015}.\footnote{For the large model, we omitted 2 of the 22 variables that are highly collinear with the other 20 variables.} Because our theory relies on stationarity, we apply the \cite{Hamilton2018} filter to each of the time series to remove low frequency variation. We also standardize the filtered time series to have mean zero and unit variance. The estimation sample ranges from 1962:Q4 to 2019:Q4. Further details are provided in the Online Supplement.

The drift terms $\{A_j\}_{j=1}^{10}$ in (\ref{eq:dgp}) are drawn once from a normal distribution with mean zero and standard deviation $\rho^{j}$. We calibrate $\rho=0.8$. This represents an agnostic nature of misspecification and ensures that distant lags have weaker effects so that the misspecification is non-negligible, yet difficult to detect by standard tools, in the sense that there remains uncertainty as to whether to increase the lag length beyond $p_*$. Simulation designs labeled A$[k]$ use the raw sequence $\{A_j\}_{j=1}^{10}$ and designs labeled B$[k]$ replace the standard deviations $\rho^j$ by one for lags $j \in \{4,8\}$, capturing putative annual dynamics if the period $t$ is interpreted as a quarter. We consider three choices $k=1,2,3$ for the weight matrix. The first option is agnostic and sets $W=\textup{I}$. The second option makes the loss function approximately look like the log determinant of the one-step ahead forecast error covariance matrix and sets $W=\Sigma_{\epsilon \epsilon }^{-1}$; see (\ref{eq:dgp}). Finally, the third choice selects a subset of the variables included in $y_t$: $W$ is a diagonal matrix with ones in the first three elements and 1/100 in the remaining ones; this effectively picks only the first three variables in the VAR, namely real GDP, a price index, and the federal funds rate. A summary is provided in Table~\ref{tab:MCDesigns}.

\begin{table}[t!]
	\caption{\sc Monte Carlo Designs}
	\label{tab:MCDesigns}
	\vspace*{-0.5cm}
	\begin{center}
		\begin{tabular}{lll} \hline \hline 
			Misspecification & A & $\{ A_1,\ldots, A_{10} \}$ \\
			&  B & $\{ A_1,A_2, A_3, A_4/\rho^4, A_5, A_6, A_7, A_8/\rho^8, A_9, A_{10} \}$ \\
			Weight Matrices  & 1 & $W=\textup{I}$ \\
			& 2 & $W=\Sigma_{\epsilon \epsilon}^{-1}$ \\
			& 3 & $W = \begin{bmatrix} \textup{I}_3 & 0 \\ 0 & \textup{I}_{n-3}/100 \end{bmatrix}$  \\ \hline \hline  
		\end{tabular}
	\end{center}
\end{table}

We consider three sample sizes $T$: 130, 250, and 500. In the paper we focus on results for $T=250$ (large enough sample size to estimate a VAR with 20 variables and up to $q=6$ lags) and $T=500$ (further along the $T$ asymptote). In the Online Supplement we also report results for the medium model with $T=130$ and $q=4$. We maintain the drifting structure of the DGP as we vary the sample size $T$. The expected loss calculations are frequentist: we keep the parameters of the DGP fixed as we repeatedly generate data and evaluate the loss associated with various prediction procedures. We consider two choices of $\alpha$ for the six designs: $\alpha = 0$ (no misspecification) and $\alpha=2$ (misspecification). For $\alpha=0$, Designs A$[k]$ and B$[k]$ are identical because the $A_j$ coefficients do not matter.  All results are based on 5,000 Monte Carlo replications. 

\noindent {\bf Prior Distributions.} As is common in the Bayesian VAR literature, we set the VAR prior mean $\underline{\Phi}_T$ for $T=130$ such that it generates univariate unit-root representations (``Minnesota'' prior); see point 3(a) in Section \ref{subsec:modeldetermination.cookbook}. For sample sizes other than $T=130$, we proceed in three steps. First, we convert $\underline{\Phi}_{T=130}$ into the local prior mean $\underline{\phi}$ using (\ref{eq:prior.mean.drift}). Second, we use the Taylor series expansion (\ref{eq:Phih.Phi.Taylor}) to convert $\underline{\phi}$ into $\underline{\psi}$, which is $h$-dependent. Third, we use (\ref{eq:prior.mean.drift}) again to compute $\underline{\Phi}_T$ and $\underline{\Psi}_T$ based on $\underline{\phi}$ and $\underline{\psi}$ from the second step.

\subsection{Multi-Step-Ahead Forecasting}
\label{subsec:numerics.forecasting}

\begin{figure}[t!]
	\caption{\sc Asymptotic Forecasting Risk Differentials, Medium Model, $h=6$}
	\label{fig:asymptotic.risk}
	\vspace*{-0.5cm}
	\setlength\tabcolsep{6pt}
	\adjustboxset{width=\linewidth,valign=c}
	\begin{center}
		\begin{tabular}{lccc}
			& \multicolumn{1}{c}{No Misspecification}
			& \multicolumn{2}{c}{Misspecification} \\
			& \multicolumn{1}{c}{\footnotesize Designs A/B1}
			& \multicolumn{1}{c}{\footnotesize Design A1}
			& \multicolumn{1}{c}{\footnotesize Design B1} \\
			\rotatebox[origin=c]{90}{\footnotesize LFE}
			& \includegraphics[scale=.3]{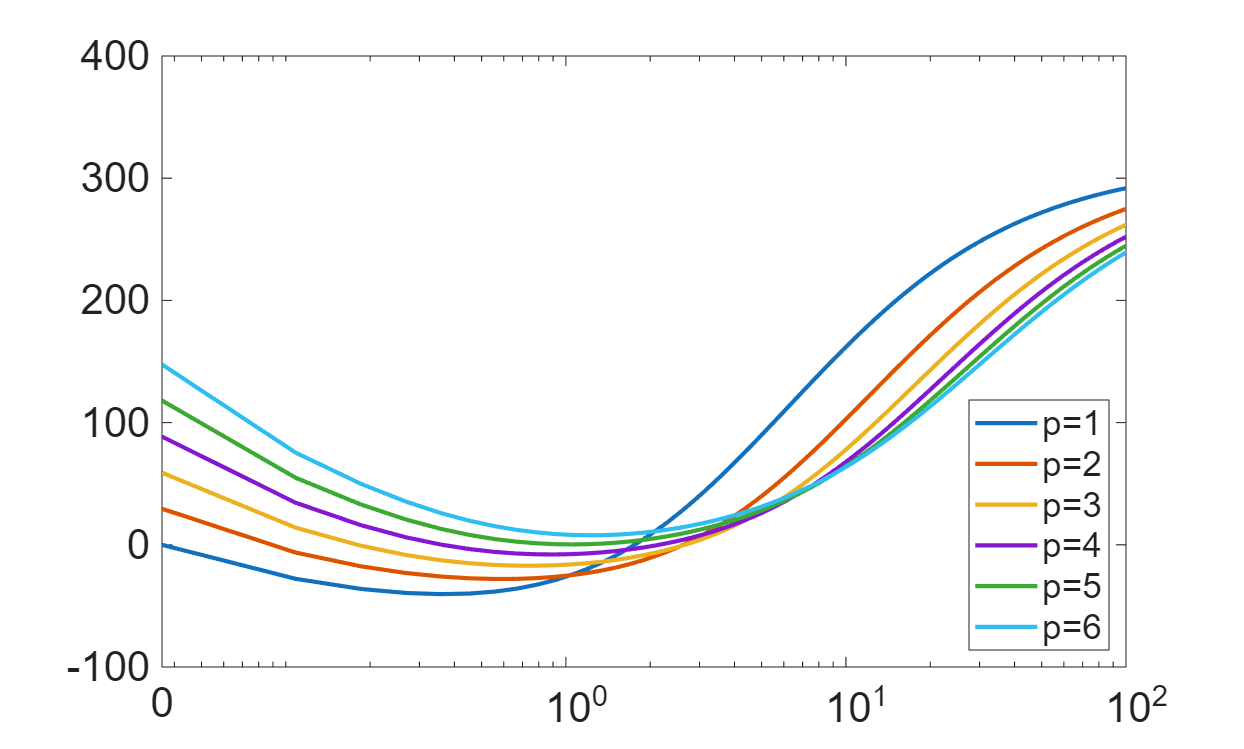}
			& \includegraphics[scale=.3]{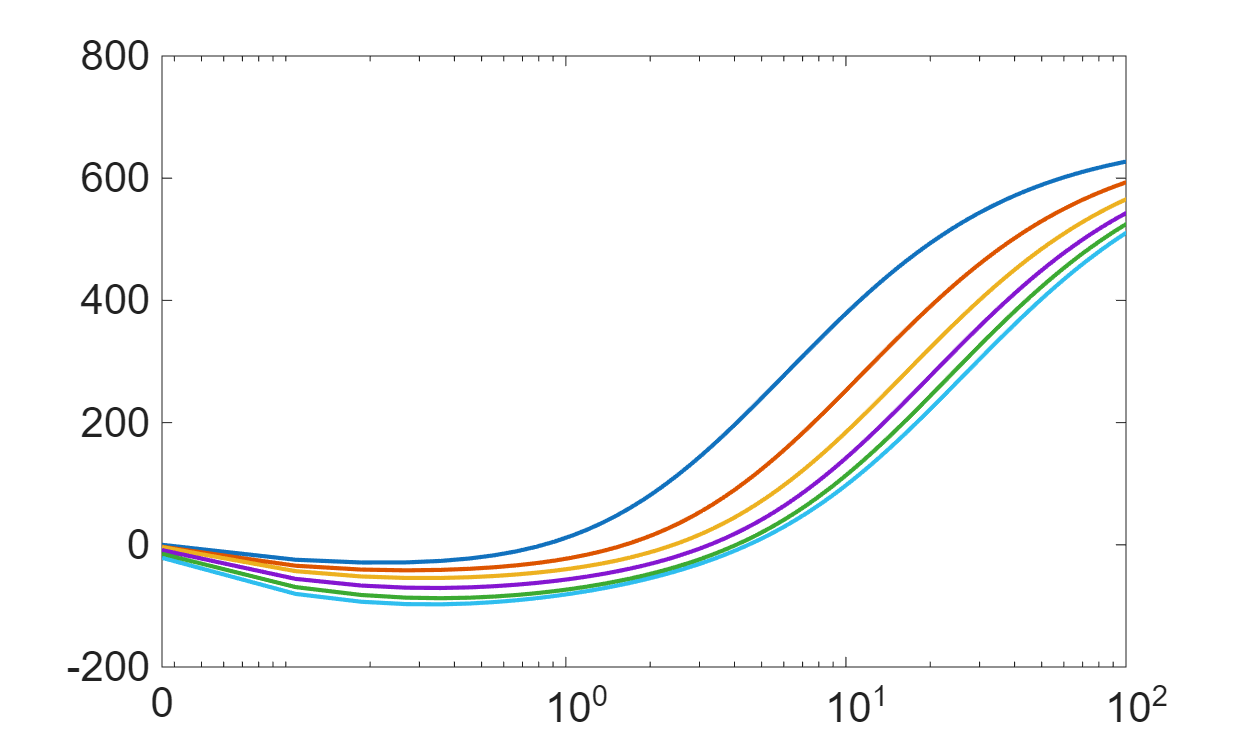}
			& \includegraphics[scale=.3]{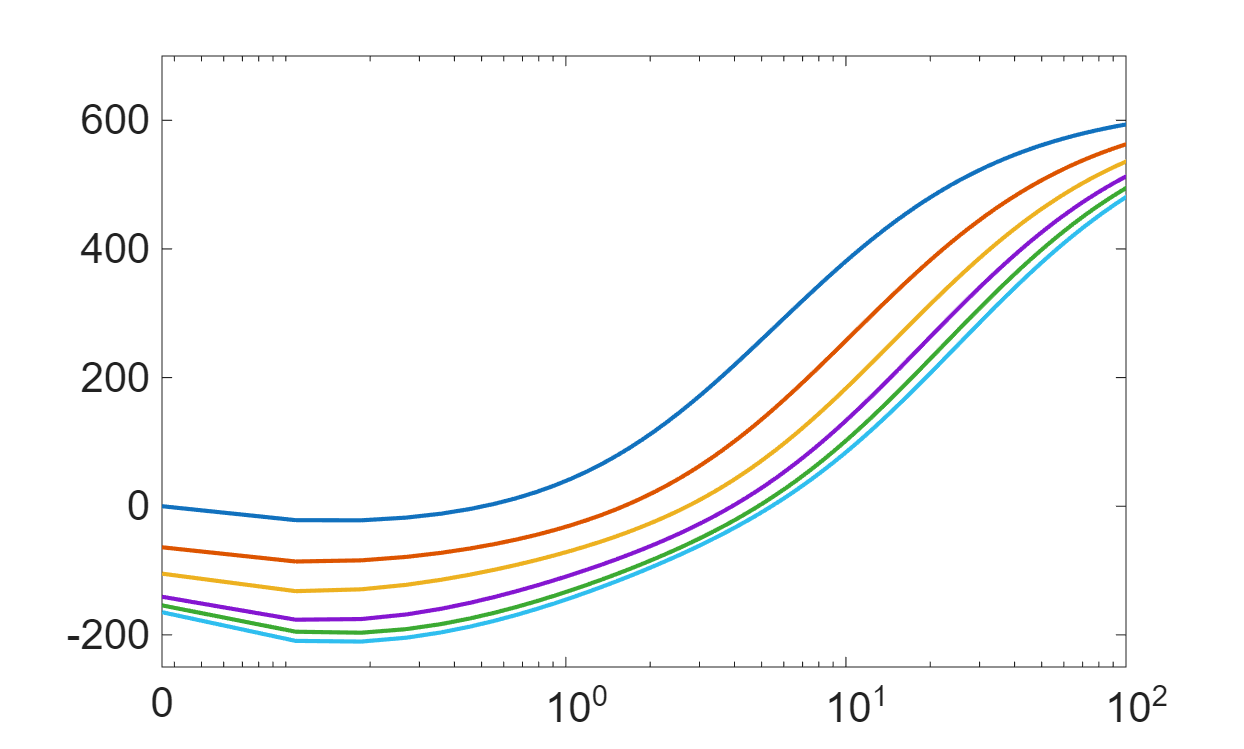} \\
			\rotatebox[origin=c]{90}{\footnotesize MLE}
			& \includegraphics[scale=.3]{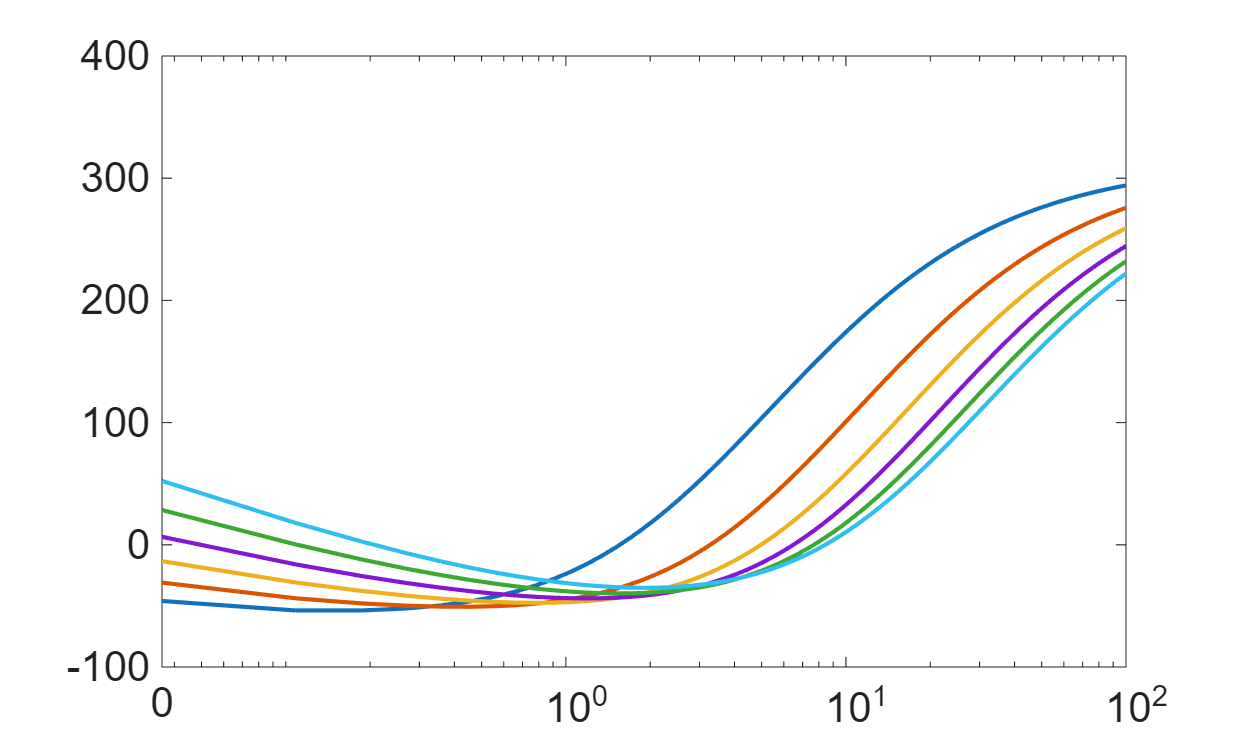}
			& \includegraphics[scale=.3]{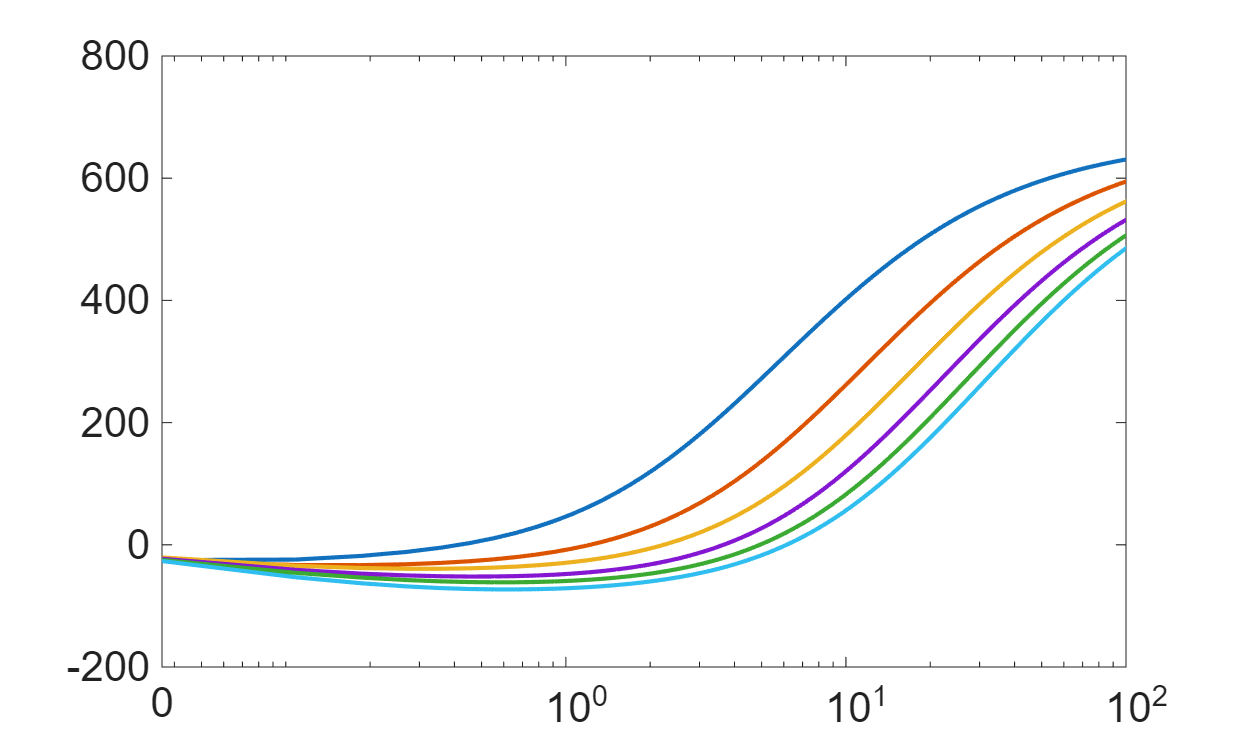}
			& \includegraphics[scale=.3]{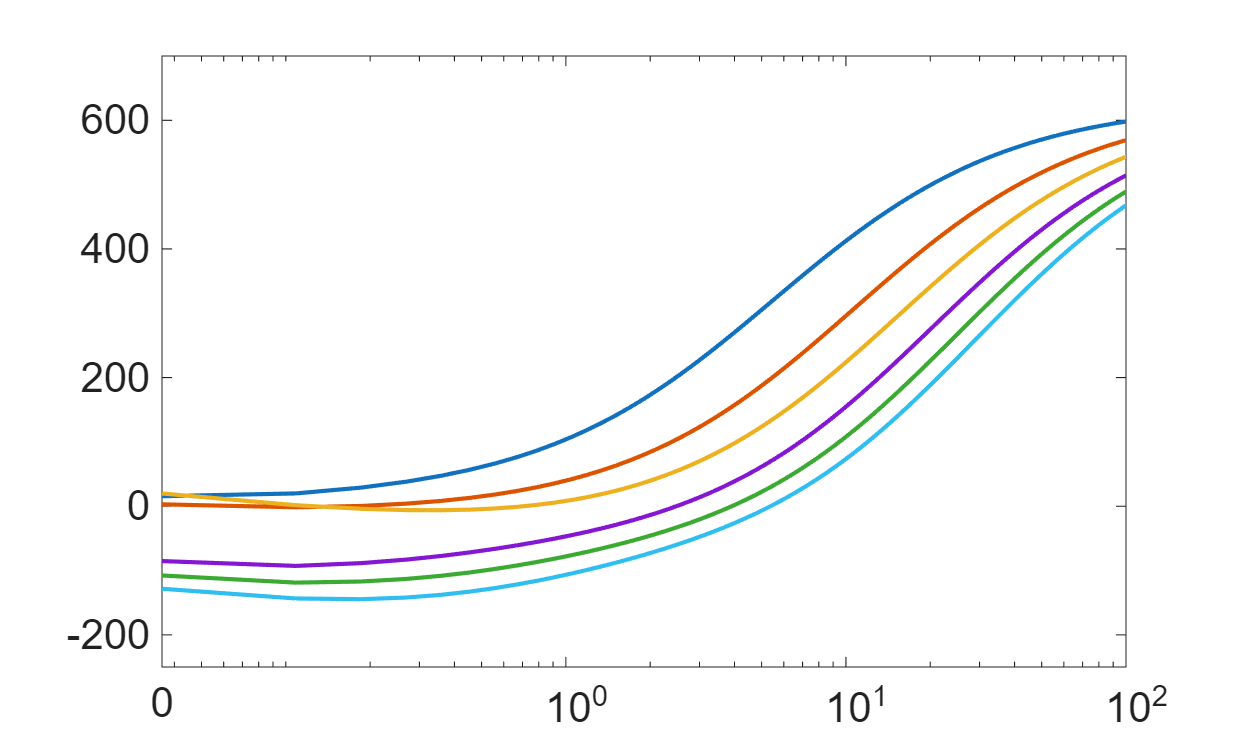} \\[-1ex]
			& {\tiny Hyperparameter $\lambda$} & {\tiny Hyperparameter $\lambda$} & {\tiny Hyperparameter $\lambda$}
		\end{tabular}
	\end{center}
	{\footnotesize {\em Notes}: On the $y$-axis we plot risk differentials between $\hat{y}_{T+h}(\iota,\lambda,p)$ and $\hat{y}_{T+h}(lfe,0,1)$. The $x$-axis is on a logarithmic scale with zero as the left endpoint. Different lines correspond to different lag orders $p$.}\setlength{\baselineskip}{4mm}
\end{figure}

Without loss of generality, we take prediction risk differences with respect to $\hat{y}_{T+h}(lfe,0,p_*=1)$ when reporting results. This eliminates the constant $C$ in Theorem~\ref{thm:prediction.risk}. Figure~\ref{fig:asymptotic.risk} depicts the asymptotic risk differentials for $\hat{y}_{T+h}(\iota,\lambda,p)$ 
as a function of $\lambda$ for $\iota \in \{lfe,mle\}$ and various choices of $p$. For $\lambda = \infty$, MLE and LFE posterior means are equal to the prior means. Because of the alignment of prior means across MLE and LFE, the resulting predictors are equivalent up to first order and have identical risks. Moreover, we keep the means fixed as we increase the lag length. Thus, the risk of any given predictor converges to the same value across different $p$ as $\lambda\to\infty$. This convergence is not visible in the figure because we truncate the $x$ axis at $\lambda = 100$.

\graphicspath{{ResultsMC/Figures/PriorControl/LargeSystem/GLP/Medium/AsymptoticForecRisk/}}

The first column of Figure~\ref{fig:asymptotic.risk} contains results for the no-misspecification case $\alpha=0$. The MLE dominates the LFE conditional on $(\lambda,p)$ because it is more efficient and there is no misspecification bias. For all $(\iota,p)$ combinations, it is optimal to apply some degree of shrinkage. The last two columns in Figure~\ref{fig:asymptotic.risk} contain the results for the $\alpha=2$ case. Under misspecification, shrinkage remains desirable. For Design A1, misspecification is still sufficiently small so that at $\lambda=0$ the MLE based predictors are preferable. However, this changes for Design B1, which scales up the misspecification coefficients $A_4$ and $A_8$. Now the LFE (slightly) dominates the MLE. 

Under the two designs in Figure~\ref{fig:asymptotic.risk}, the optimal level of shrinkage is  non-trivial in the sense that the minimum risk is obtained for an interior value of $\lambda$. The benefit of shrinkage is dependent on the lag order, but substantial throughout.  For $\alpha=0$, the optimal degree of shrinkage increases with lag $p$. Increasing the lag order is undesirable for both predictors in the absence of misspecification, as the lowest prediction risk is obtained for $p=p_*=1$. On the other hand, if the VAR is misspecified with $\alpha=2$, increasing the lag order above $p_*=1$ reduces the misspecification bias and hence is desirable even at low levels of shrinkage.

The asymptotic risk lines in Figure~\ref{fig:asymptotic.risk} also emphasize the intricate interplay between $(\iota,\lambda,p)$, as marginal risk improvements can operate along any dimension. Take, for instance, the LFE under no misspecification. Under no shrinkage, $\lambda=0$, estimating the model with $p=1$ is preferred to $p=2$. However, keeping $p=2$ while adding a bit of shrinkage in turn dominates the unshrunk $p=1$ estimator. The patterns become increasingly convoluted as we vary $\iota$ too, i.e., compare across the rows in Figure~\ref{fig:asymptotic.risk}. This argument has been recently stressed for standard least squares estimates by~\cite{Ludwig2024}, and our PC criterion crucially permits to resolve this joint complex selection.

\subsection{IRF Estimation}
\label{subsec:numerics.irfs}

\begin{figure}[t!]
	\caption{\sc Asymptotic IRF Estimation Risk, Medium Model, $h=4$}
	\label{fig:IRFrisk}
	\vspace*{-0.5cm}
	\setlength\tabcolsep{6pt}
	\adjustboxset{width=\linewidth,valign=c}
	\begin{center}
		\begin{tabular}{lccc}
			& \multicolumn{1}{c}{No Misspecification}
			& \multicolumn{2}{c}{Misspecification $(\alpha=2)$} \\
			& \multicolumn{1}{c}{\footnotesize Designs A/B1}
			& \multicolumn{1}{c}{\footnotesize Design A1}
			& \multicolumn{1}{c}{\footnotesize Design B1} \\
			\rotatebox[origin=c]{90}{\footnotesize LP IRF}
			& \includegraphics[scale=.3]{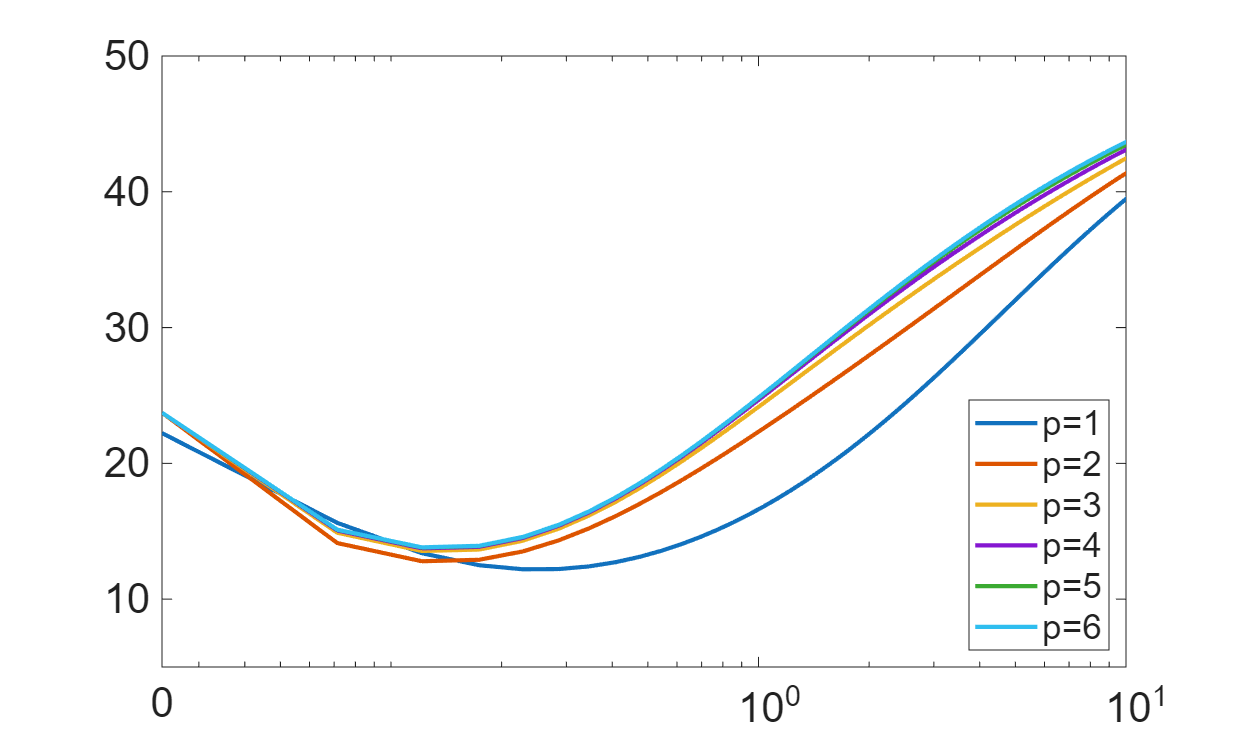}
			& \includegraphics[scale=.3]{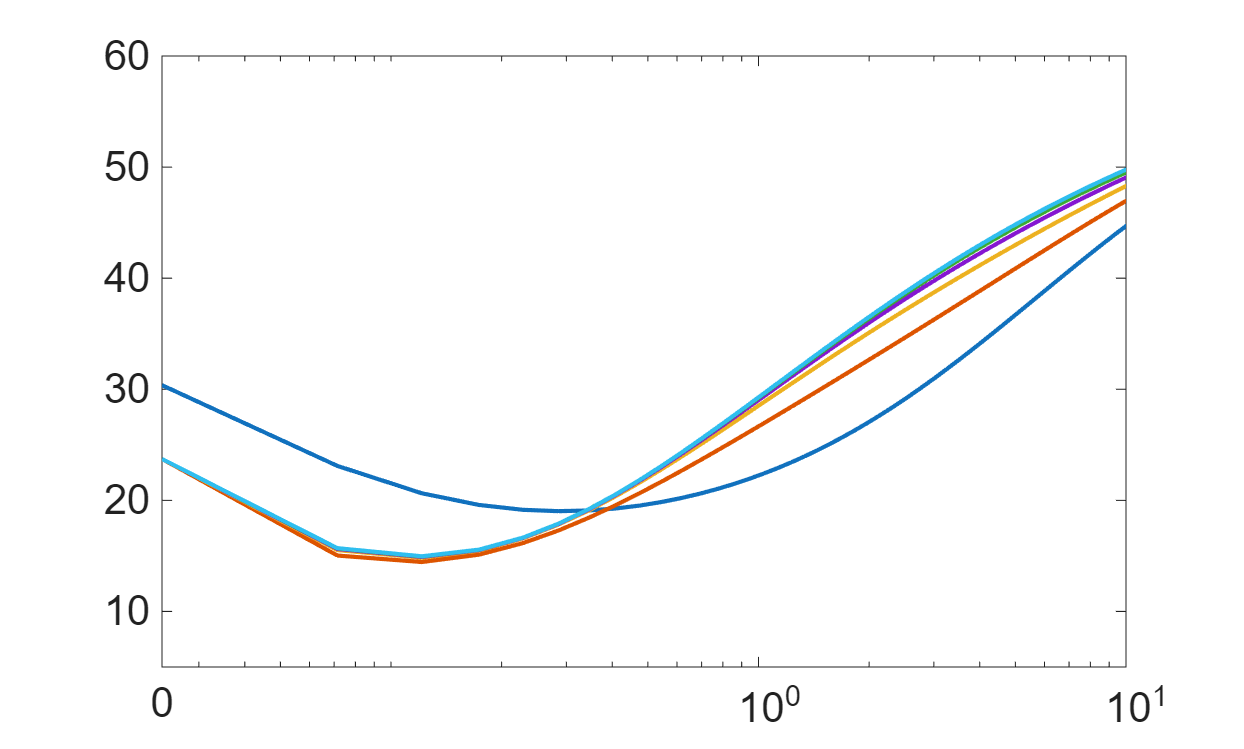}
			& \includegraphics[scale=.3]{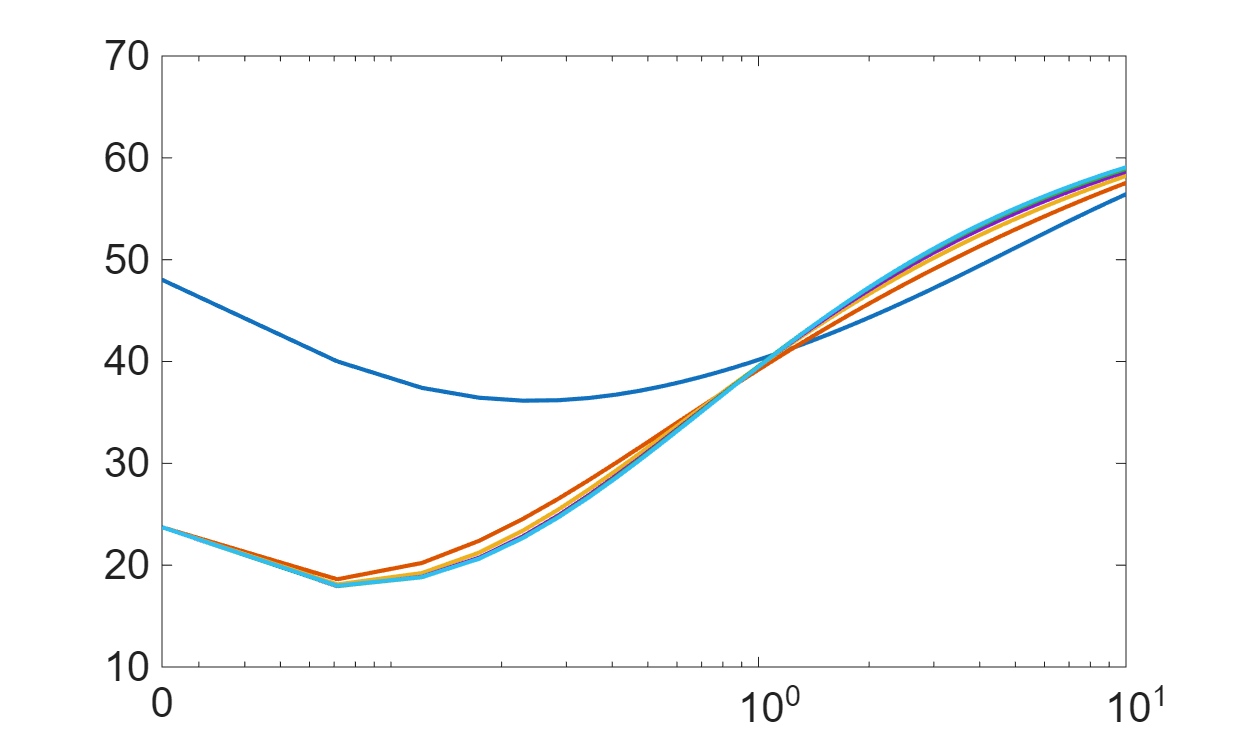} \\
			\rotatebox[origin=c]{90}{\footnotesize VAR IRF}
			& \includegraphics[scale=.3]{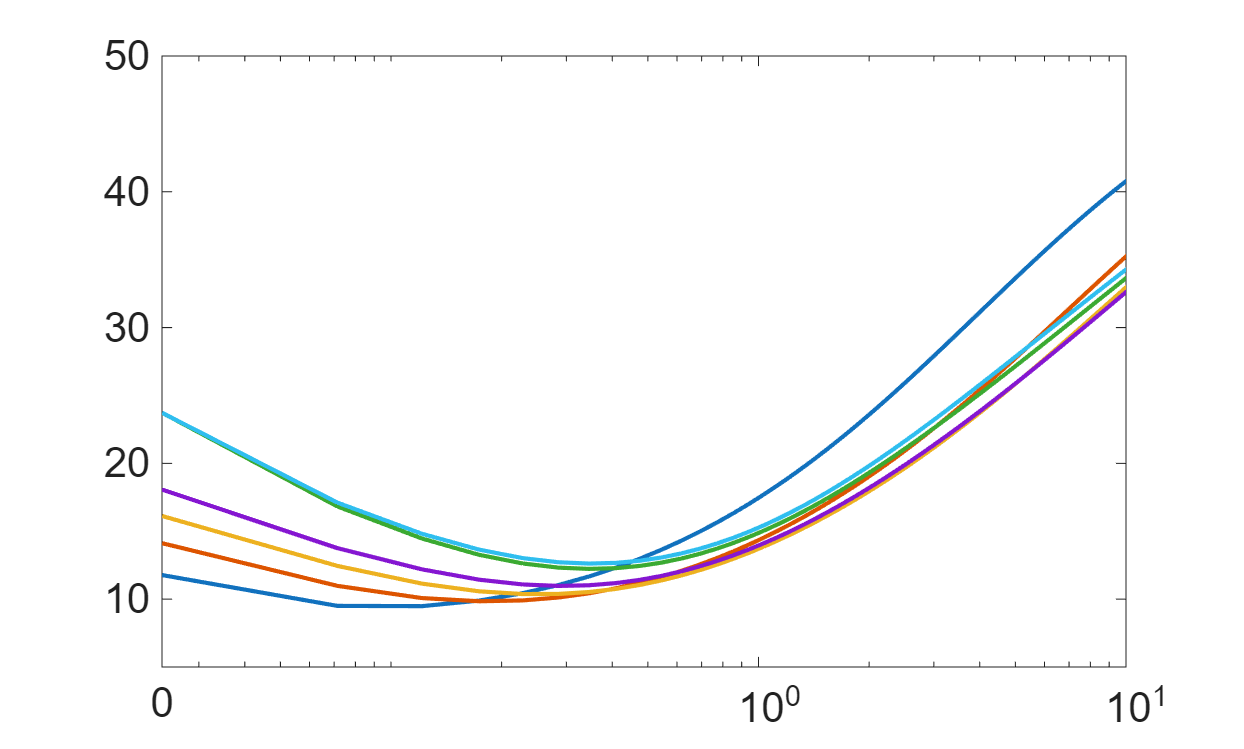}
			& \includegraphics[scale=.3]{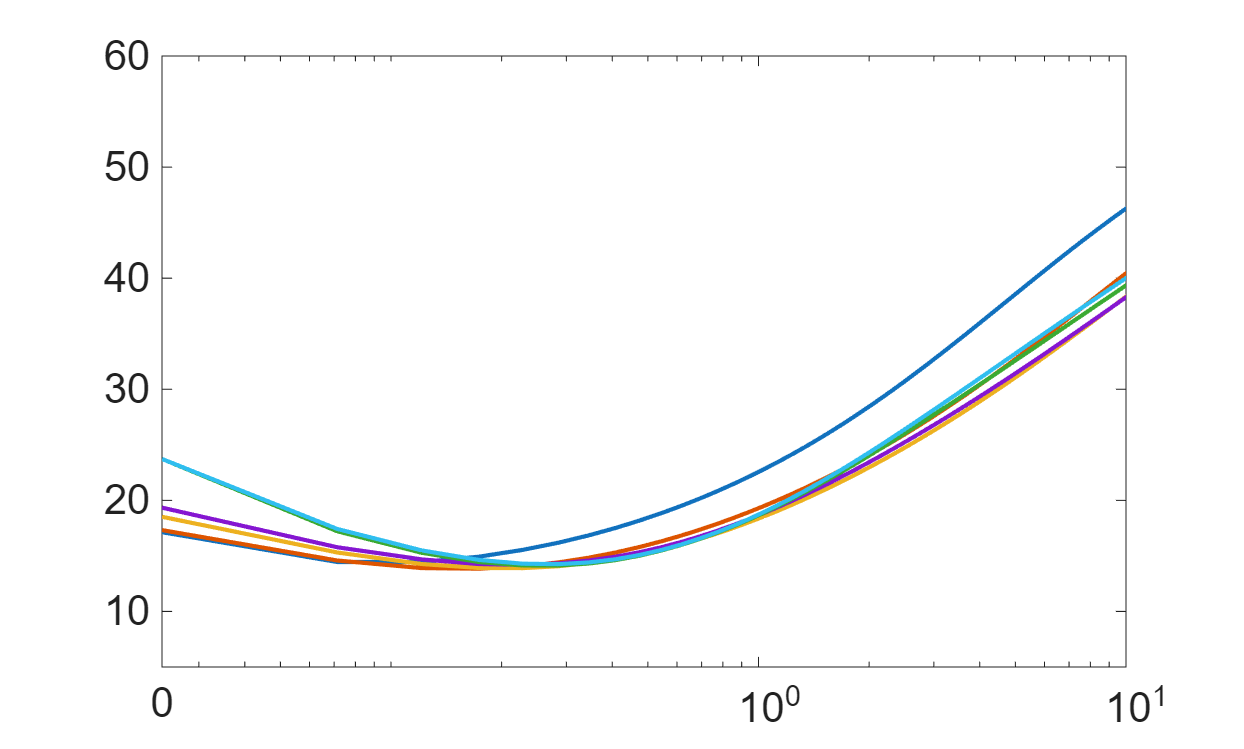}
			& \includegraphics[scale=.3]{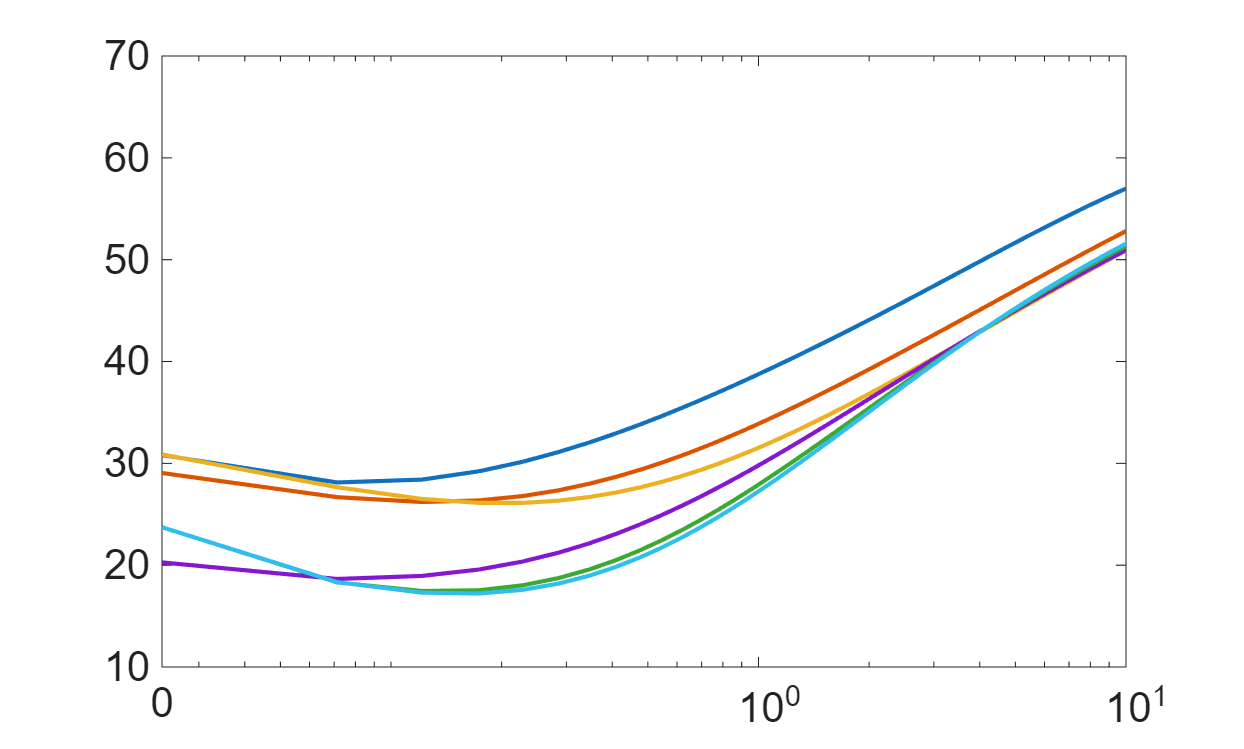} \\[-1ex]
			& {\tiny Hyperparameter $\lambda$} & {\tiny Hyperparameter $\lambda$} & {\tiny Hyperparameter $\lambda$}
		\end{tabular}
	\end{center}
	{\footnotesize {\em Notes}: On the $y$-axis we plot the asymptotic IRF estimation risk. The $x$-axis is on a logarithmic scale with zero as the left endpoint. Different lines correspond to different lag orders $p$.  }\setlength{\baselineskip}{4mm}
\end{figure}

We proceed by comparing the asymptotic risk of VAR and LP IRF estimators with $\Xi=\textup{I}_n$, using the same DGPs as before. Because a normalization is not needed in the IRF case, we present the actual asymptotic risk values instead of the differentials in Figure~\ref{fig:IRFrisk}. As before, the prior means $\underline{\Psi}$ and $\underline{\Phi}$ are chosen such that they imply identical IRFs. In turn, as $\lambda \longrightarrow \infty$, the estimation risk is identical for all $(\iota,p)$ (not visible in the figure).

\graphicspath{{ResultsMC/Figures/PriorControl/LargeSystem/GLP/Medium/AsymptoticIRFrisk/}}

The first column of Figure~\ref{fig:IRFrisk} shows the asymptotic IRF estimation risk for the no-misspecification case $\alpha=0$. At $\lambda=0$, the choice of $p=p_*=1$ is optimal. Because the VAR estimation is more efficient than multi-step regressions, the IRF estimation risk for $p=1$ is smaller for the VAR than for the LP. As the number of lags is increased, the asymptotic risk also increases, both for the VAR and the LP estimation. The LP estimation risk is identical for all $p>p_*=1$ because bias does not change (Lemma~\ref{thm:mulfe.eq.muirf}) and the variance does not increase (Proposition 3.1 in MPQW). The VAR IRF estimation risk is initially increasing in $p$. Once $p \geq p_*+h$, the risk stays constant and is identical to the LP risk (Corollary 3.2 of MPQW), i.e., $p=5$ and $p=6$ in the figure are identical. These patterns no longer hold once shrinkage is allowed. For both LP and VAR, some shrinkage is always desirable. Conditional on choosing $\hat{\lambda}(p)$, the level of risk is reduced substantially compared to $\lambda=0$, and the risks associated with the different lag lengths $p \in \{1,\ldots,6\}$ become very similar. For the LP, a single lag remains optimal, while for a VAR increasing the lag order to $p=2$ becomes desirable.

We now turn to the second and third columns of Figure~\ref{fig:IRFrisk}, which show asymptotic risk functions for the misspecified case of $\alpha = 2$. In terms of lag length selection, the choice of $p=1$ now often leads to the largest risk. This is true for the LP and the VAR IRF estimator. The LP estimator benefits from the lag augmentation effect for $p>p_*=1$. At $\lambda=0$, the risk drops substantially as the number of lags is increased from $p=1$ to 2. While there are no further risk reductions for $p>2$ at $\lambda=0$, shrinking toward the prior mean leads to additional performance improvements. In our setting, the optimal degree of shrinkage is approximately $\hat{\lambda}(p) \approx 0.3$ for $p>1$. Interestingly, in this case the equivalence under lag-augmentation continues to hold at the optimal shrinkage level.

VAR and LP IRF estimators differ with respect to the effect of lag length changes on risk. This can be seen by comparing the top and the bottom panels of Figure~\ref{fig:IRFrisk}. The desirability of choosing $p=p_*=1$ varies across Designs A1 and B1 under misspecification. While additional lags reduce the misspecification bias, they also increase the variance of the estimator. In either case, allowing for some shrinkage is optimal, and risk can be further improved by increasing $p$. At the optimal value of shrinkage, the risk is minimized at the largest $p$ considered.

In our design, the overall performance of LP and VAR IRF estimators is very similar, once an optimal lag length and an optimal shrinkage parameter $\lambda$ are chosen. IRF estimation differs from multi-step forecasting in regard to the effect of a lag length increase on the variance. The key difference is that forecasting relies on the $n \times np$ matrix of posterior mean estimates $M'\bar{\Psi}_T(\iota,\lambda,p)$, whereas the IRF analysis only depends on the $n \times n$ submatrix $M'\bar{\Psi}_T(\iota,\lambda,p)M$, which is an object that does not grow with $p$. In the forecasting application, an increase of $p$ always raises the variance of the prediction function, while simultaneously reducing the asymptotic bias, which leads to a clear trade-off. The PC and IRFC criteria are designed to balance the various trade-offs in a data-driven manner.

\section{Multi-Step Forecasting with Simulated Data}
\label{sec:MC}

We conduct a small-scale Monte Carlo experiment to illustrate the finite-sample forecasting performance of the MLE and LFE shrinkage predictors. A large-scale empirical analysis, albeit without shrinkage considerations, was done by \cite{MSW2006}.  In Section~\ref{subsec:MC.riskdifferentials.pc}, we compare the finite-sample risk differentials to the expected value of $PC_T$. We consider the joint PC-based selection of predictor, shrinkage, and lag length in Section~\ref{subsec:MC.JointSelection} and also provide a comparison of the risks associated with PC versus MDD-based model determination. The DGP is identical to the one used to compute the asymptotic risk functions in Section~\ref{sec:numerics}.

\subsection{Simulated Risk Differentials, Expected PC, and $\lambda$ Selection}
\label{subsec:MC.riskdifferentials.pc}

\begin{figure}[t!]
	\caption{\sc PC versus Finite Sample Risk, LFE, Medium Model, Design A/B1}
	\label{fig:RiskvsPC_alpha2_h4_p1}
	\setlength\tabcolsep{1pt}
	\adjustboxset{width=\linewidth,valign=c}
	\begin{center}
		\begin{tabular}{@{}lccc@{}}
			& {\footnotesize Risks}
			& {\footnotesize PC Criterion}
			& {\footnotesize $\hat{\lambda}$ Distribution} \tabularnewline
			
			\rotatebox[origin=c]{90}{\footnotesize $T=250$}
			& \includegraphics[width=.3\linewidth]{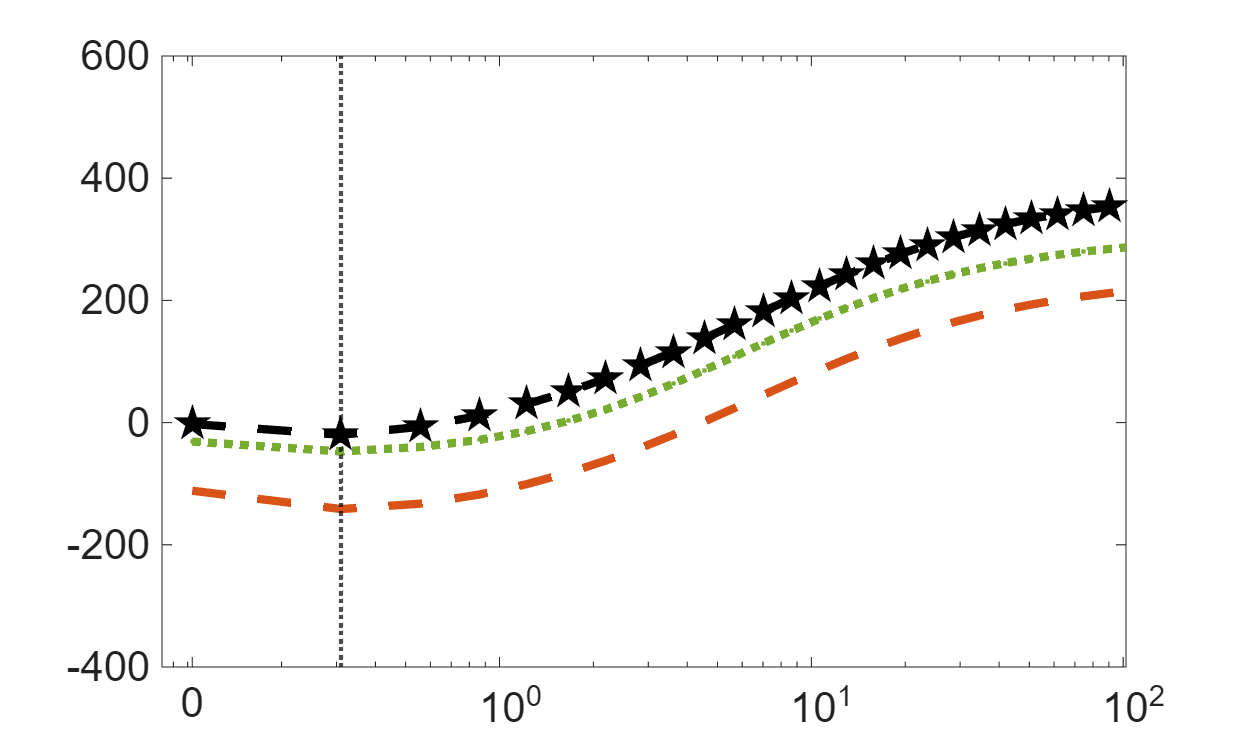} 
			& \includegraphics[width=.3\linewidth]{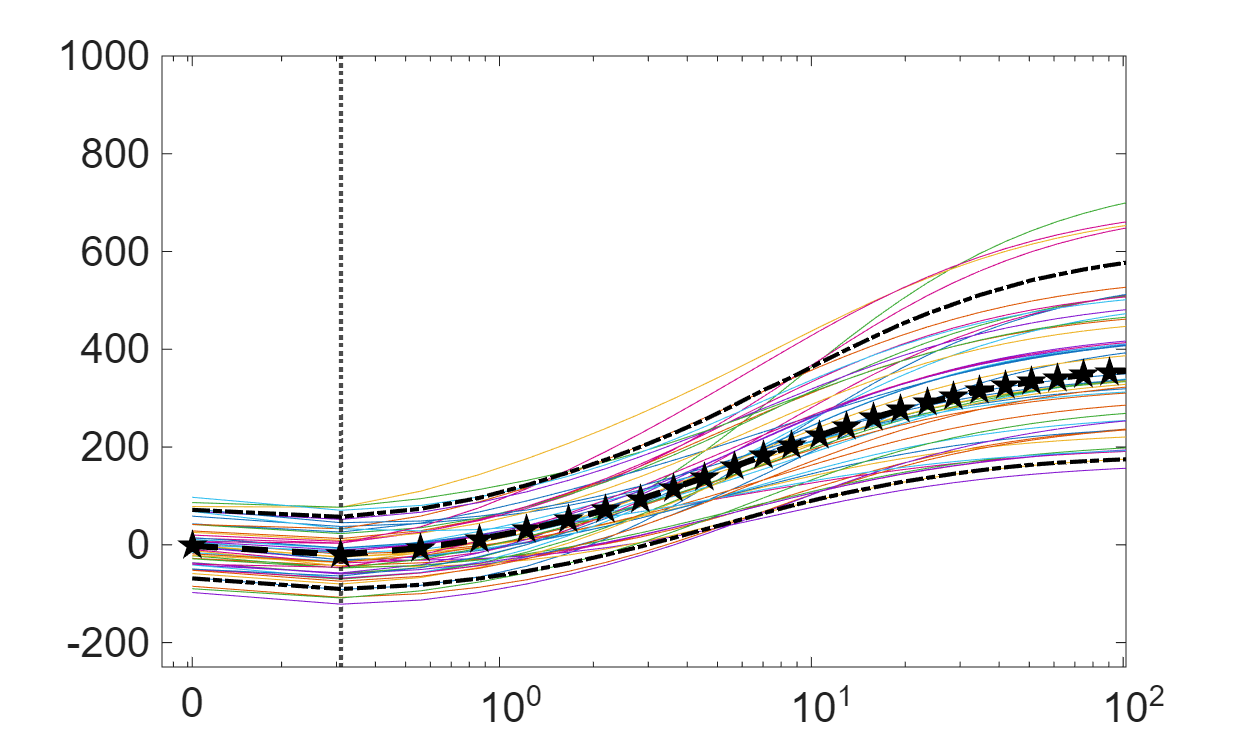} 
			& \includegraphics[width=.3\linewidth]{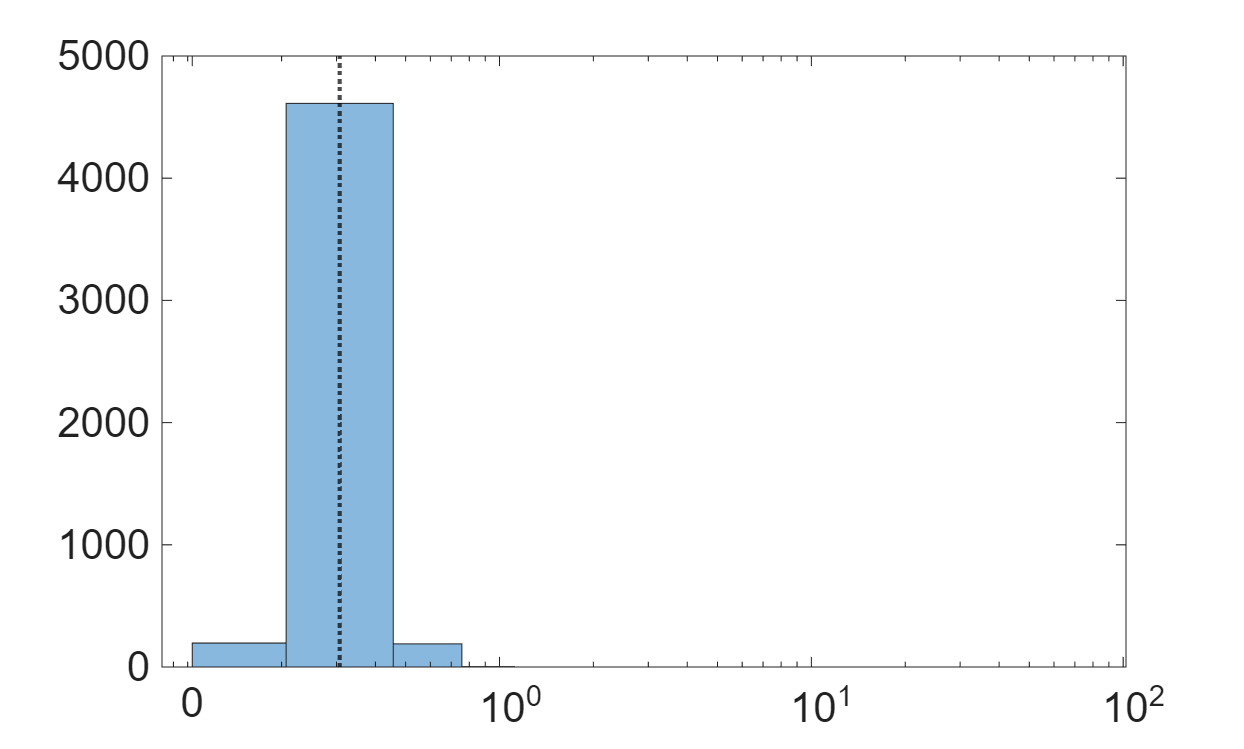} \tabularnewline
			
			\rotatebox[origin=c]{90}{\footnotesize $T=500$}
			& \includegraphics[width=.3\linewidth]{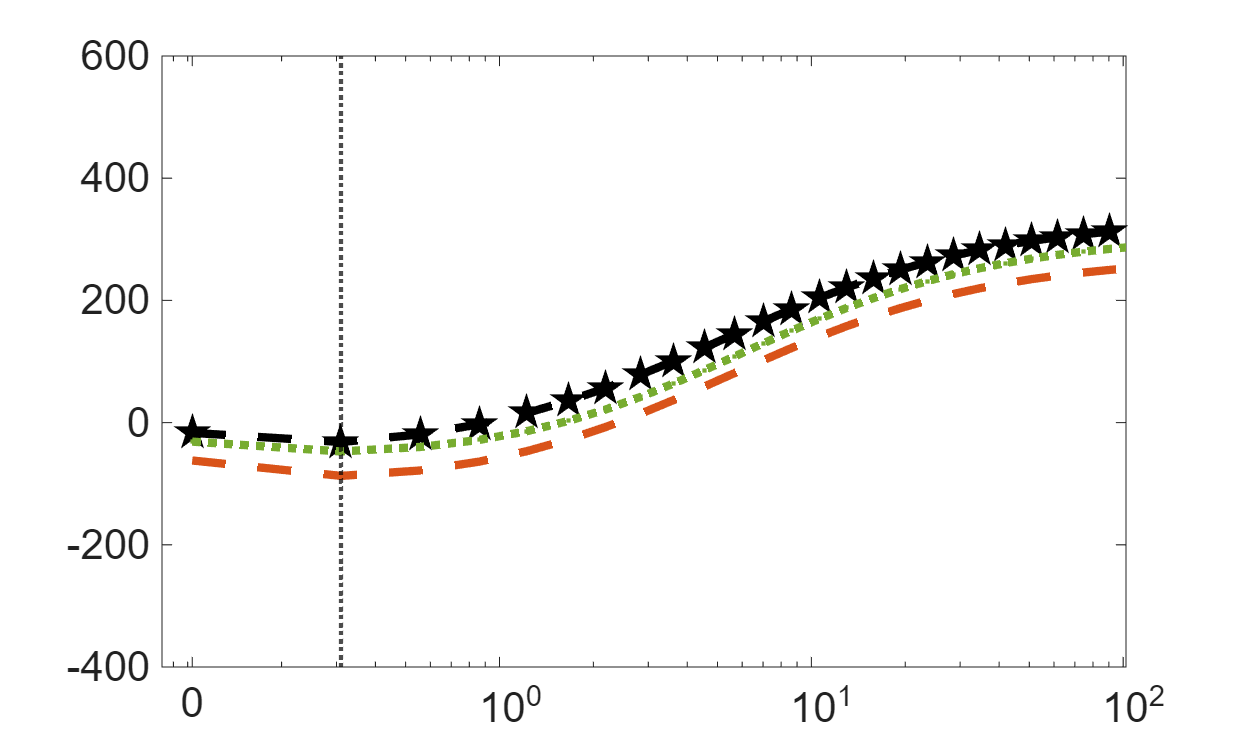} 
			& \includegraphics[width=.3\linewidth]{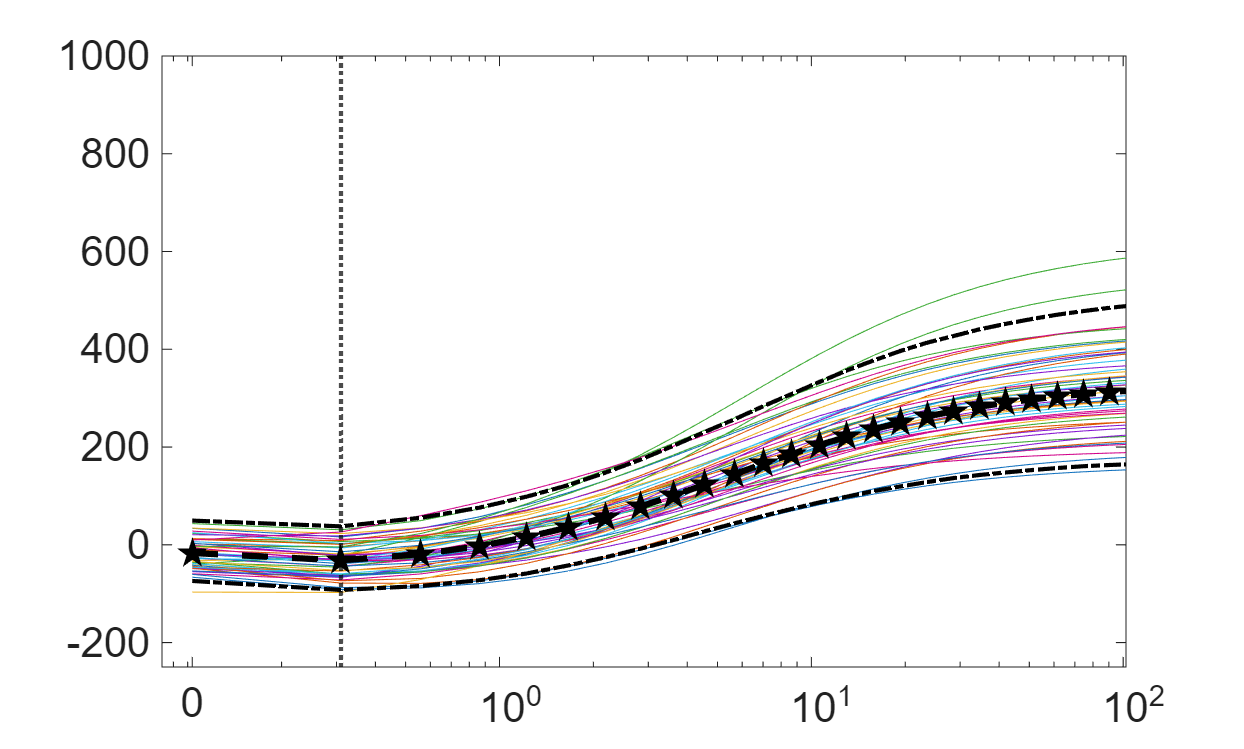} 
			& \includegraphics[width=.3\linewidth]{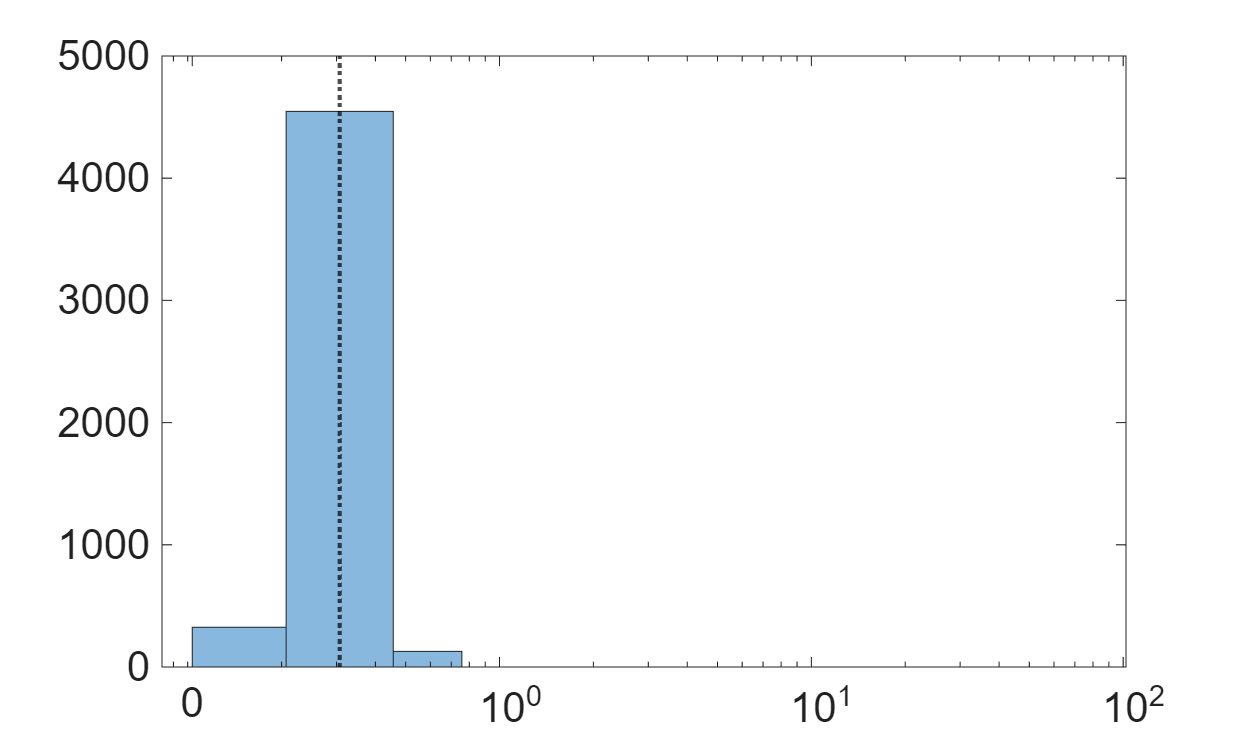} \\[-1ex]
			& {\tiny Hyperparameter $\lambda$} & {\tiny Hyperparameter $\lambda$} & {\tiny Hyperparameter $\lambda$}
		\end{tabular}
	\end{center}
	{\footnotesize {\em Notes}: $(h,\alpha)=(4,2)$ and $p=p_*=1$. The $x$-axis is on a logarithmic scale with zero as the left endpoint. Left column: asymptotic risk (dotted), Monte Carlo risk (dashed), and $\EE[PC_T]$ (starred). Center column: hairlines represent PC objective functions across Monte Carlo replications; starred black line is $\EE[PC_T]$ (same as in left column); dashed black lines are 90\% bands. Right column:  distribution of PC-selected shrinkage hyperparameter. Vertical lines indicate the value of $\lambda$ that minimizes the asymptotic risk.}\setlength{\baselineskip}{4mm}
\end{figure}

The panels in the left column of Figure~\ref{fig:RiskvsPC_alpha2_h4_p1} depict Monte Carlo risk differentials (dashed), asymptotic risk differentials (dotted), and the expected value of PC (starred). The figure shows results for $\alpha=2$, $h=4$, and $p=p_*=1$. The sample size varies across rows. For $T=250$, the average PC closely matches the asymptotic risk shape, and for $T=500$ the two lines become effectively identical. The Monte Carlo risk is slightly smaller than the asymptotic risk, but the gap narrows as $T$ increases. As a function of $\lambda$, the three curves attain their respective minima at an interior value $\hat{\lambda} \approx 0.3$. The vertical line indicates the value of $\lambda$ that minimizes the asymptotic risk. 

\graphicspath{{ResultsMC/Figures/PriorControl/LargeSystem/GLP/Medium/G0doubling/}}

In the second column of Figure~\ref{fig:RiskvsPC_alpha2_h4_p1}, we plot hairlines of the $PC_T(\iota=lfe,\lambda,p=p_*)$ objective functions, where each hairline corresponds to a particular Monte Carlo replication. The collection of hairlines illustrates the sampling variation of the PC objective function. The dashed lines represent 90\% empirical quantiles of the PC criterion across replications. The solid lines with the star symbols represent the pointwise expected values of the objective function and hence approximate $\mathbb{E}[PC_T]$ as a function of $\lambda$. They are identical to the black lines in the first column. Overall, the hairline pattern is broadly consistent with the asymptotic risk differential function. Most of the hairlines attain their minimum between $\lambda=0.1$ and $\lambda=1$ and are monotonically increasing to the right of the minimum. In the last column of Figure~\ref{fig:RiskvsPC_alpha2_h4_p1} we plot histograms of the PC-selected $\hat{\lambda}$ distribution across Monte Carlo replications. Most of the mass concentrates near the argmin of the asymptotic risk function, suggesting good selection performance by the PC criterion.

\subsection{Joint PC-Based Determination of $\iota$, $\lambda$, and $p$}
\label{subsec:MC.JointSelection}

\begin{table}[t!] 
	\caption{\sc Finite Sample Risk Differentials for $\hat{y}_{T+h}(\iota,\hat{\lambda},p)$, Medium Model}
	\label{tab:Risk.PCsel}
	\vspace*{0.2cm}
	\begin{center}
		\scalebox{0.90}{
			\begin{tabular}{@{\extracolsep{2pt}}r@{\hspace*{1.5cm}}rrrr@{\hspace*{0.8cm}}rrrr@{\hspace*{0.8cm}}rrrr} \hline \hline
				&       \multicolumn{4}{c}{A/B1, $\alpha=0$} & \multicolumn{4}{c}{A1, $\alpha=2$} & \multicolumn{4}{c}{B1, $\alpha=2$}\\  \cmidrule(l{1pt}r{20pt}){2-5}\cmidrule(l{1pt}r{20pt}){6-9}\cmidrule(l{1pt}r{1pt}){10-13}
				$p$   & LFE & MLE & $\pi$ & Joint & LFE & MLE & $\pi$ & Joint & LFE & MLE & $\pi$ & Joint \\ \hline
				\multicolumn{13}{c}{Horizon $h=2$, $T=250$} \\ \hline
				1 & -106 & -109 & {\small (18)} & -109 & -85 & -89 & {\small (10)} & -88 & -14 & -15 & {\small (20)} & -14 \\
				2 & -106 & -104 & {\small (16)} & -104 & -88 & -88 & {\small (10)} & -88 & -20 & -19 & {\small (20)} & -19 \\
				4 & -96 & -96 & {\small (24)} & -96 & -83 & -85 & {\small (13)} & -85 & -44 & -45 & {\small (29)} & -45 \\
				6 & -86 & -85 & {\small (32)} & -85 & -80 & -80 & {\small (49)} & -80 & -62 & -55 & {\small (92)} & -61 \\
				$\hat{p}$ & -98 & -100 & {\small (17)} & -100 & -80 & -80 & {\small (41)} & -80 & -62 & -54 & {\small (92)} & -61 \\
				\hline \multicolumn{13}{c}{Horizon $h = 2$, $T = 500$} \\ \hline
				1 & -88 & -91 & {\small (15)} & -91 & -60 & -64 & {\small (10)} & -64 & 12 & 8 & {\small (23)} & 9 \\
				2 & -88 & -88 & {\small (8)} & -88 & -63 & -63 & {\small (11)} & -63 & 5 & 6 & {\small (28)} & 6 \\
				4 & -79 & -80 & {\small (13)} & -80 & -60 & -62 & {\small (12)} & -62 & -22 & -21 & {\small (34)} & -21 \\
				6 & -71 & -71 & {\small (22)} & -70 & -61 & -60 & {\small (53)} & -61 & -44 & -36 & {\small (98)} & -44 \\
				$\hat{p}$ & -86 & -88 & {\small (12)} & -88 & -59 & -59 & {\small (35)} & -60 & -44 & -35 & {\small (98)} & -44 \\
				\hline \multicolumn{13}{c}{Horizon $h = 6$, $T = 250$} \\ \hline
				1 & -272 & -274 & {\small (23)} & -271 & -147 & -133 & {\small (53)} & -137 & -115 & -80 & {\small (65)} & -101 \\
				2 & -259 & -269 & {\small (15)} & -266 & -156 & -149 & {\small (46)} & -146 & -139 & -97 & {\small (77)} & -124 \\
				4 & -235 & -254 & {\small (11)} & -250 & -166 & -158 & {\small (47)} & -156 & -179 & -146 & {\small (79)} & -169 \\
				6 & -214 & -236 & {\small (16)} & -230 & -170 & -161 & {\small (62)} & -164 & -173 & -153 & {\small (79)} & -166 \\
				$\hat{p}$ & -249 & -255 & {\small (13)} & -251 & -166 & -160 & {\small (59)} & -160 & -171 & -148 & {\small (78)} & -162 \\
				\hline \multicolumn{13}{c}{Horizon $h = 6$, $T = 500$} \\ \hline
				1 & -221 & -227 & {\small (19)} & -224 & -62 & -48 & {\small (60)} & -55 & 34 & 71 & {\small (84)} & 41 \\
				2 & -209 & -225 & {\small (11)} & -222 & -73 & -62 & {\small (59)} & -63 & -5 & 54 & {\small (94)} & -1 \\
				4 & -187 & -213 & {\small (6)} & -210 & -96 & -83 & {\small (57)} & -91 & -78 & -19 & {\small (92)} & -72 \\
				6 & -169 & -200 & {\small (6)} & -197 & -106 & -97 & {\small (63)} & -103 & -87 & -47 & {\small (89)} & -83 \\
				$\hat{p}$ & -212 & -217 & {\small (9)} & -215 & -102 & -95 & {\small (61)} & -101 & -86 & -45 & {\small (91)} & -81 \\
				\hline\hline			
			\end{tabular}
		}
	\end{center}
	{\footnotesize {\em Notes}: The finite sample risk differentials are computed relative to $\hat{y}_{T+h}(lfe,0,q=6)$. $\pi$ is the percentage of times that LFE is selected by the PC criterion.}\setlength{\baselineskip}{4mm}
\end{table}

We now examine the Monte Carlo risk of LFE and MLE predictors based on data-driven hyperparameter and lag length choice. Focusing on the medium-sized model, Table~\ref{tab:Risk.PCsel} shows risk differentials between $\hat{y}_{T+h}(\iota,\hat{\lambda},p)$ and $\hat{y}_{T+h}(lfe,0,q=6)$, where $\hat{\lambda}$ is determined by minimizing $PC_T(\iota,\lambda,p)$ with respect to $\lambda$ using grid search.\footnote{We use a grid for $\lambda$ with 50 points that includes 0 and then is equally spaced on a log scale with a maximum value of $10^4$.} We report results for $\iota = lfe$, $\iota=mle$, and also use PC to choose between LFE and MLE and report the resulting risks in the column ``Joint.'' The percentage of Monte Carlo replications where PC selects LFE is reported in the column $\pi$. We consider lag lengths of 1, 2, 4, and 6. In the rows labeled $\hat{p}$, PC is also used to select the lag length. Negative numbers indicate risk reductions relative to the benchmark predictor $\hat{y}_{T+h}(lfe,0,q)$. 

Columns 2 to 5 of the table contain results for the no-misspecification case of $\alpha=0$. While for horizon $h=2$ the performance of the estimators is similar, for $h=6$, as expected, the MLE predictor attains a lower risk than the LFE predictor. This is particularly pronounced for $T=500$. The lowest risk is attained for $p=1$ because that equals the true asymptotic lag length $p_*$. The PC criterion selects the LFE predictor only in between 6\% and 32\% of the Monte Carlo repetitions. The risk under ``joint'' selection is in between the LFE and MLE risk, often closer to the latter. The risk values associated with $\hat{p}$, i.e., the case in which PC is also used to choose the lag length, are between the $p=2$ and $p=4$ values, indicating that the criterion tends to select models that are somewhat larger than $p_*$. However, stronger shrinkage to some extent compensates for the additional lags and reduces risk differentials, which was also apparent from Figure~\ref{fig:asymptotic.risk}.

Under the mild misspecification of Design A1, MLE and LFE continue to perform similarly for $h=2$. However, for $h=6$, LFE starts to dominate because it is centered at the pseudo-optimal value. It also becomes desirable to include more than $p_*=1$ lags to offset the dynamic misspecification of the VAR. The fraction $\pi$ of selecting LFE now ranges from 46\% to 63\%. The risk associated with ``joint'' selection of predictor and hyperparameter sits between the risk of the LFE and MLE predictors, often closer to the best of the two. The $\hat{p}$ risk differentials are close to the $p=6$ risk differentials, indicating that PC selects six lags with high probability. These patterns are enhanced under Design B1, which features stronger misspecification. For $h=6$ and $T=500$, LFE is selected for 91\% of the samples conditional on using $\hat{p}$ lags. Overall, the results reported here are consistent with the simulation results in S2005 and the empirical results in \cite{MSW2006} in that the use of LFE is only justified if misspecification is sufficiently large. The key takeaway is that PC helps the forecaster to adapt to the level of misspecification by choosing between LFE and MLE predictors, tuning the level of shrinkage $\lambda$, and  selecting the number of lags $p$.

\begin{table}[t!]
	\caption{\sc Finite Sample Risk Differentials for $\hat{y}_{T+h}(\iota,\hat{\lambda},\hat{p})$, $T=250$.}
	\label{tab:Risk.PCsel.250}
	\vspace*{0.2cm}
	\begin{center}
		\scalebox{0.90}{
			\begin{tabular}{l rrrrr@{\hspace*{1cm}}rrrrr}
				\hline\hline
				& \multicolumn{5}{c}{$\alpha=0$} & \multicolumn{5}{c}{$\alpha=2$} \\ \cmidrule(l{15pt}r{25pt}){2-6} \cmidrule(lr){7-11}
				Design & LFE & MLE & $\pi$ & Joint & MDD & LFE & MLE & $\pi$ & Joint & MDD \\
				\hline
				\multicolumn{11}{c}{Horizon $h=2$, Medium Model} \\ \hline 
				A1 & -98 & -100 & {\small (17)} & -100 & -107 & -80 & -80 & {\small (41)} & -80 & -88 \\
				B1 &  &  &  &  &  & -62 & -54 & {\small (92)} & -61 & -46 \\
				A2 & -503 & -532 & {\small (5)} & -531 & -539 & -493 & -461 & {\small (80)} & -483 & -415 \\
				B2 &  &  &  &  &  & -374 & -268 & {\small (98)} & -374 & -210 \\
				A3 & -44 & -44 & {\small (21)} & -44 & -49 & -36 & -37 & {\small (34)} & -37 & -43 \\
				B3 &  &  &  &  &  & -29 & -25 & {\small (93)} & -29 & -20 \\
				\hline \multicolumn{11}{c}{Horizon $h=6$, Medium Model} \\ \hline
				A1 & -249 & -255 & {\small (13)} & -251 & -274 & -166 & -160 & {\small (59)} & -160 & -135 \\
				B1 &  &  &  &  &  & -171 & -148 & {\small (78)} & -162 & -152 \\
				A2 & -1190 & -1292 & {\small (6)} & -1281 & -1373 & -682 & -307 & {\small (92)} & -650 & -128 \\
				B2 &  &  &  &  &  & -878 & -269 & {\small (98)} & -858 & -353 \\
				A3 & -108 & -113 & {\small (12)} & -110 & -123 & -65 & -72 & {\small (37)} & -69 & -80 \\
				B3 &  &  &  &  &  & -60 & -49 & {\small (74)} & -55 & -51 \\
				\hline \multicolumn{11}{c}{Horizon $h=2$, Large Model} \\ \hline
				A1 & -1493 & -1487 & {\small (89)} & -1490 & -1655 & -1622 & -1592 & {\small (99)} & -1622 & -1785 \\
				B1 &  &  &  &  &  & -1258 & -1101 & {\small (100)} & -1258 & -1594 \\
				A2 & -8742 & -8598 & {\small (100)} & -8742 & -9440 & -4263 & -5044 & {\small (99)} & -4291 & -13884 \\
				B2 &  &  &  &  &  & -374 & -2579 & {\small (100)} & -371 & -13739 \\
				A3 & -217 & -213 & {\small (89)} & -216 & -250 & -199 & -189 & {\small (94)} & -199 & -257 \\
				B3 &  &  &  &  &  & -148 & -145 & {\small (96)} & -149 & -250 \\
				\hline \multicolumn{11}{c}{Horizon $h=6$, Large Model} \\ \hline
				A1 & -3573 & -3914 & {\small (75)} & -3644 & -4197 & -3630 & -3423 & {\small (100)} & -3630 & -3754 \\
				B1 &  &  &  &  &  & -4047 & -3276 & {\small (100)} & -4047 & -3963 \\
				A2 & -20031 & -21460 & {\small (95)} & -20080 & -23180 & -21685 & -9879 & {\small (100)} & -21685 & -13435 \\
				B2 &  &  &  &  &  & -28379 & -11309 & {\small (100)} & -28379 & -16582 \\
				A3 & -563 & -606 & {\small (79)} & -569 & -653 & -526 & -478 & {\small (100)} & -526 & -540 \\
				B3 &  &  &  &  &  & -485 & -327 & {\small (100)} & -484 & -440 \\
				\hline\hline
			\end{tabular}
		}
	\end{center}
	{\footnotesize {\em Notes}: The finite sample risk differentials are computed relative to $\hat{y}_{T+h}(lfe,0,q=6)$. $\pi$ is the percentage of times that LFE is selected by the PC criterion.}\setlength{\baselineskip}{4mm}
\end{table}

In Table~\ref{tab:Risk.PCsel.250}, we report the risk differentials for PC selection of $(\lambda,p,\iota)$ for all six designs listed in Table~\ref{tab:MCDesigns}. In addition to the medium model, we now also include the large model in the analysis. The sample size is fixed at $T=250$ and we consider horizons $h=2$ and $h=6$. Because the A and B designs only differ with respect to the MA sequence $\{A_j\}_{j=1}^{10}$ that controls the misspecification, we leave the $\alpha=0$ rows for the B designs empty. Table \ref{tab:Risk.PCsel.250} further includes a column for MDD-based selection risk; see Section~\ref{subsec:modeldetermination.MDD}. 

The first important observation from Table \ref{tab:Risk.PCsel.250} is that the relative rankings of estimators, even at optimal $(\lambda,p)$, depend on the object of interest and the extent of misspecification (captured through the designs), horizon and model size. This reinforces the appeal of the task-based nature of the PC criterion if dynamic misspecification is a concern. 

Second, for the medium model, we saw in Table~\ref{tab:Risk.PCsel} that MLE was desirable under no misspecification. The PC criterion automatically acknowledges this and selects LFE only in 5\% to 21\% of the samples. Interestingly, the situation changes for the large model. While for $h=6$ MLE continues to dominate, for $h=2$ LFE is preferable. PC now tends to select LFE between 75\% and 100\% of the time. Overall, the selection problem is challenging, because the risk differentials are small. Nonetheless, selecting between the two optimized estimators using PC works well in the sense that the risk under joint selection is always between, and often closer to, the minimum of the MLE and LFE risk. Third, as we move to the case of misspecification, LFE visibly dominates for all configurations considered. Importantly, joint PC selection of $(\iota,\lambda,p)$ gets very close to the best-case scenario. In general, for the medium model, improvements appear for all horizons, while for the large model this is especially true for longer horizons. 

Fourth, we turn to the comparison between the Joint column and the MDD column. PC selection is inherently task based: for every forecast horizon and every weight matrix $W$ in each replication, we use the criterion to choose a different triplet $(\iota,\lambda,p)$. The spirit of Bayesian model selection upon which the MDD is based is intrinsically different. In our context, the goal would be to estimate one VAR specification $(mle, \hat{\lambda}, \hat{p}(\hat{\lambda}))$ -- here we pursue the model selection instead of model averaging route -- and then use the model to generate forecasts for all horizons $h$ and all weight matrices $W$. The risk associated with this workflow is reported in the MDD columns of Table~\ref{tab:Risk.PCsel.250}. By design, this workflow performs very well in the absence of misspecification ($\alpha=0$). The risk is generally lower than the joint-PC risk, but the risk differentials are modest. However, a comparison of the last two columns of Table~\ref{tab:Risk.PCsel.250} highlights that under misspecification the task-based approach works much better and leads to substantially lower prediction risk. Underlying in Table \ref{tab:Risk.PCsel.250} we found that, under misspecification, the consistent model selector MDD tends to pick $\hat{p}=p_*=1$, while the PC criterion tends to overselect the lag length to absorb some of the misspecification and reduce the bias. The overselecting behavior thus places PC closer to criteria such as the AIC.

\section{Empirical Application: IRF Estimation}
\label{sec:empirics}

We now estimate IRFs via VARs or LPs using VARs constructed from actual data. While in a pseudo-out-of-sample forecasting application it is possible to compute {\em ex-post} forecast errors from the actual data, one cannot compute estimation errors in an IRF application because the ``true'' IRFs are never observed. Thus, we proceed by documenting how often IRFC selects the VAR and LP IRF estimates, respectively, and how much shrinkage and how many lags are used.

\noindent {\bf Samples for Empirical Estimates.} We construct estimation samples by combining Hamilton-filtered time series from the FRED-QD database; see \cite{McCrackenNg2020}. We follow \cite{MSW2006} in randomly creating a large number of data sets. We do so by selecting uniformly at random 1,000 different tuples of $n=7$ series, which are demeaned and standardized.\footnote{We do discard a very small fraction of random tuples which imply explosive OLS VAR estimates.} For each of the data sets, we estimate $n$-dimensional VARs or LPs with $p\in \{ 1,2,4,6\}$ lags. This model size matches the medium model in Section \ref{sec:MC}. As in Section~\ref{sec:MC}, the prior for the VAR coefficients is centered at univariate unit-root representations (``Minnesota'' prior). Because the series are combined at random, the degree of misspecification varies across samples. Identification (or shock orthogonalization) is achieved by using the first column of the Cholesky factorization of the one-step-ahead forecast error covariance matrix (same for VAR and LP IRF estimates). We report results for estimation samples ranging from 1984:Q1 to 2019:Q4. 

\begin{figure}[t!]
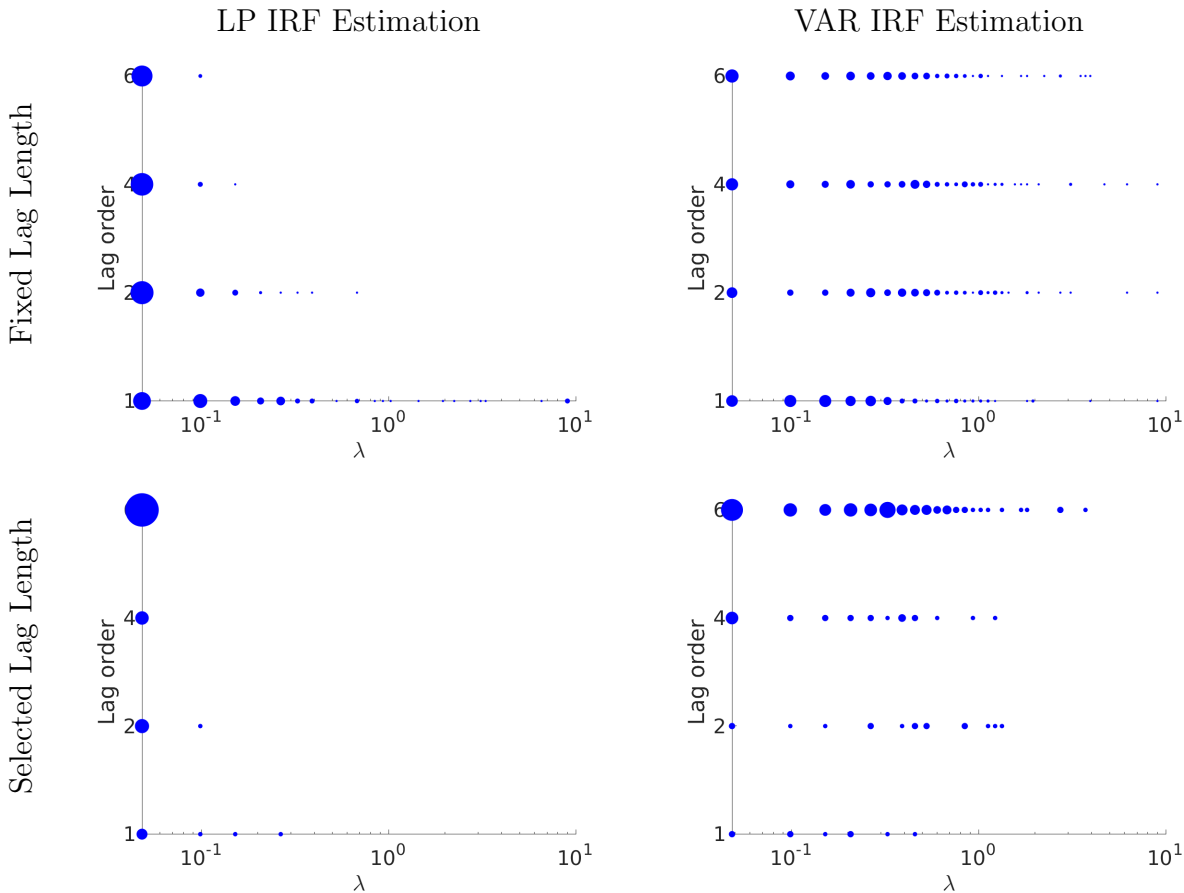

	\caption{\sc Distribution of IRFC Selected Hyperparameter $\lambda$, $h=6$.}
	\label{fig:empirics.DistLamhat.subS1}
	\vspace*{-0.5cm}
	\setlength\tabcolsep{1pt}
	\adjustboxset{width=\linewidth,valign=c}
	\begin{center} 
		\begin{tabular}{ccc}
			& {\footnotesize LP IRF Estimation} & {\footnotesize VAR IRF Estimation} \\
			\rotatebox[origin=c]{90}{\tiny Lag Length $p$ (Fixed)} &	
			\includegraphics[scale=.4]{OptimalLamP_LP_subS2_hor6.png} &
			\includegraphics[scale=.4]{OptimalLamP_VAR_subS2_hor6.png} \\
			\rotatebox[origin=c]{90}{\tiny Lag Length $\hat{p}$ (Selected)} &
			\includegraphics[scale=0.4]{OptimalLamPhat_LP_subS2_hor6.png} &
			\includegraphics[scale=0.4]{OptimalLamPhat_VAR_subS2_hor6.png} \\[-1ex]
			& {\tiny Hyperparameter $\lambda$} & {\tiny Hyperparameter $\lambda$} 
		\end{tabular}
	\end{center}
	{\footnotesize {\em Notes}: Grid values of IRFC-selected shrinkage hyperparameters for different fixed lag orders. The diameter of the dots is proportional to the frequency across samples. Fixed lag length refers to $(\hat{\lambda}(p),p)$ and each $p$-row represents 1,000 samples. Selected lag length is $(\hat{\lambda}(\hat{p}),\hat{p})$ and the number of samples across the four $\hat{p}$ rows adds up to 1,000. Estimation sample 1984:Q1-2019:Q4.}\setlength{\baselineskip}{4mm}
\end{figure}

\noindent {\bf Empirical Results for IRFC Selection.} Figure~\ref{fig:empirics.DistLamhat.subS1} provides information about the selected degree of shrinkage. The left column contains results for LP IRF estimation and the right column for VAR IRF estimation. In the top panels, we fix the lag length $p$ at the values $\{1,2,4,6\}$. For each $p$ and each of the 1,000 samples, we compute $\hat{\lambda}$. The dots represent $(p,\hat{\lambda}(p))$ and their diameters are proportional to the frequency with which this shrinkage selection occurs among the estimation samples, conditional on the four choices of $p$. In the bottom panels, for each of the 1,000 samples, we set the lag length to the selected value $\hat{p}$, providing information about the pairs $(\hat{p},\hat{\lambda}(\hat{p}))$.

\graphicspath{{ResultsEM/Figures/G0doubling/}}

For fixed lag lengths, three observations stand out. First, for the LP IRF estimator, the larger $p$, the smaller $\hat{\lambda}(p)$. This pattern is broadly consistent with the misspecification panels for the LP IRF estimation in Figure~\ref{fig:IRFrisk}. In the numerical illustration of the asymptotic risk, the shrinkage for $p=1$ under that specific DGP is substantially larger than for $p > 1$. The key mechanism is that, unlike in the case of forecasting, the variance of the LP estimator does not increase with $p$ once $p>p_*$. Thus, there is no need for more shrinkage as the number of lags is increased. Second, VAR shrinkage is generally stronger than LP shrinkage.\footnote{The hyperparameter $\lambda$ controls the relative weight of prior mean and MLE/LFE.  Comparing~\eqref{eq:barphi.mle.tildelambda} and~\eqref{eq:multistep.posterior}, notice that for $\underline{P}_{\Phi} = \underline{P}_{\Psi}$ these weights are approximately the same for MLE and LFE because $S_{T,11} \approx S_{T,hh}$. In this sense the $\hat{\lambda}$s are comparable across $\iota$.} This is also qualitatively consistent with the patterns in Figure~\ref{fig:IRFrisk}, where for $\alpha=0$ the optimal $\lambda$ for VAR IRF estimation tends to be slightly larger than for LP IRF estimation. Third, for the VAR IRF estimation the distribution of $\hat{\lambda}(p)$ does not vary as strongly with $p$ as for the LP estimation, suggesting that the bias-variance trade-off in the various samples remains relevant regardless of $p$.

The bottom row of Figure~\ref{fig:empirics.DistLamhat.subS1} replaces the fixed lag lengths $p \in \{1,2,4,6\}$ by the selected lag length $\hat{p}$. First, the most frequently selected lag length across the 1,000 samples is $\hat{p}=6$; in 88\% of the samples for LFE and 62\% for MLE. 
Second, in the vast majority of samples, the selected LP shrinkage $\hat{\lambda}(\hat{p})$ is small. In case of VAR IRF estimation, in many samples $\hat{\lambda}(\hat{p})$ is between 0.1 and 1. As we have seen previously, for LP there is no cost for increasing $\hat{p}$ beyond $p_*+1$. For the VAR, there is a potential benefit of being able to reduce the effect of misspecification
. However, additional lags are costly in terms of variance and there is a benefit to applying more shrinkage in samples in which $\hat{p}$ is large.

\begin{figure}[t!]
	\caption{\sc IRFC Selection of LP versus VAR IRF Estimate}
	\label{fig:empirics.fractionLP.Sshort}
	\vspace*{-0.5cm}
	\setlength\tabcolsep{1pt}
	\adjustboxset{width=\linewidth,valign=c}
	\begin{center} 
		\begin{tabular}{ccc}
			& $\lambda=0$ & $\lambda = \hat{\lambda}$ \\	
			\rotatebox[origin=c]{90}{\tiny Fraction LP Selected} & \includegraphics[width=0.45\textwidth]{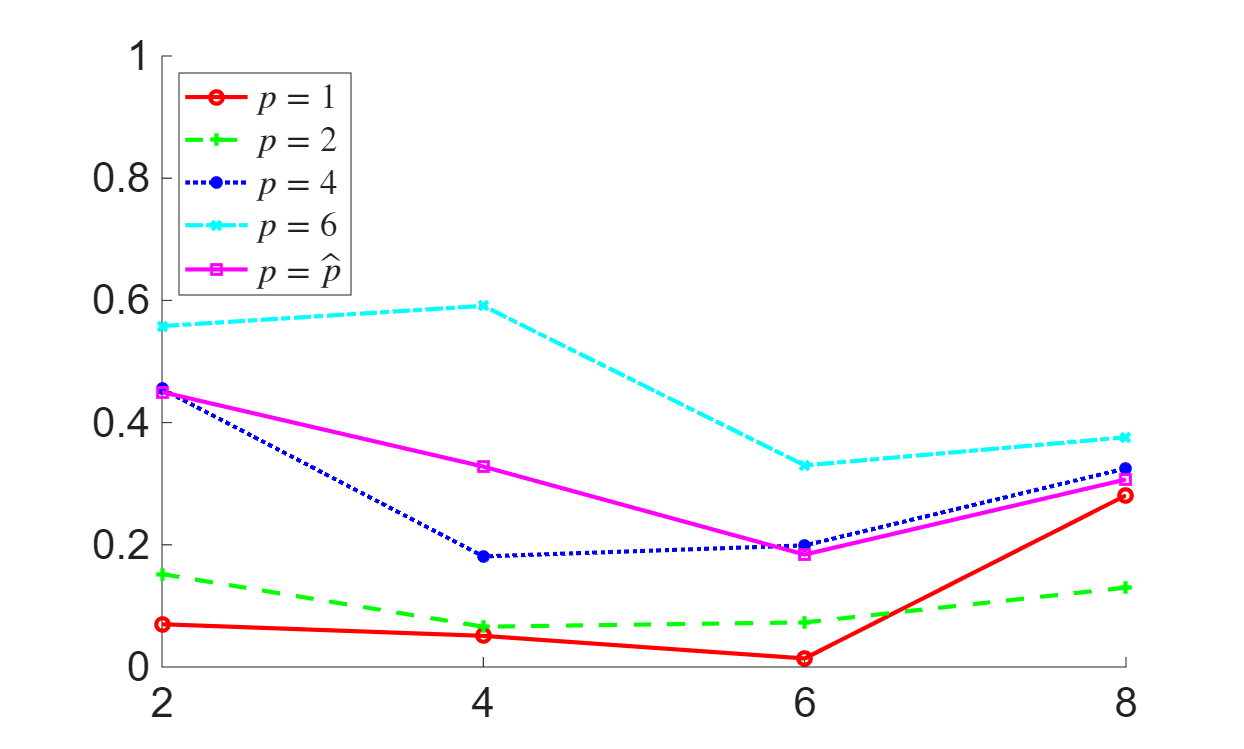} &
			\includegraphics[width=0.45\textwidth]{FrakLP_subSlong.png} \\[-0.5ex]
			& {\tiny Horizon $h$} & {\tiny Horizon $h$}
		\end{tabular}
	\end{center}
	{\footnotesize {\em Notes}: Fraction of times the IRFC selects the LP IRF estimator under different lag lengths and across different horizons. Estimation sample 1984:Q1-2019:Q4.}\setlength{\baselineskip}{4mm}
\end{figure}

Figure~\ref{fig:empirics.fractionLP.Sshort} shows the fraction of times the IRFC selects the LP IRF estimator as a function of the horizon $h$. Each line corresponds to a different lag length $p$. We also show results for the IRFC-selected lag length $\hat{p}$. The left panel corresponds to $\lambda=0$, whereas the right panel uses the IRFC-selected $\hat{\lambda}$. Conditional on $p=1$, the fraction of samples for which LP is selected is below 10\% for horizons $h=2$, 4, and 6, and rises to about 25\% for horizon 8. This pattern holds true irrespective of whether $\lambda$ is selected by IRFC or fixed at zero. For $p=2$, the fraction of LP selection remains below 20\% under no shrinkage, but increases to around 50\% once $\lambda$ is selected by IRFC.

\graphicspath{{ResultsEM/Figures/G0doubling/}}	  

For $p \in \{ 4,6\}$ the fraction of samples for which LP is selected increases conditional on $\lambda=0$. Here LP may perform relatively well because its ability to achieve correct centering outweighs increases of the variance component of the MSE criterion. If $\lambda$ is chosen optimally, then for $p \in \{ 2,4,6\}$ there is barely a difference in the number of times that LP is selected. This is consistent with the asymptotic risk lines in Figure \ref{fig:IRFrisk}. In both plots, the $p=\hat{p}$ line is very similar to the case of large $p$, which is to be expected from Figure~\ref{fig:empirics.DistLamhat.subS1}. 

The results in Figure~\ref{fig:empirics.fractionLP.Sshort} are related to the large-scale simulation experiment by LPW. Our MSE criterion corresponds to equal weights on bias and variance in their paper. The left panel in our Figure~\ref{fig:empirics.fractionLP.Sshort} could be compared to Figure~7 in LPW, except that they apply a bias correction to the VAR and LP estimators. They find that the LP estimator dominates the VAR estimator in fewer than 20\% of the samples. Our IRFC-based assessment of LP estimation is slightly more favorable, yet qualitatively in line with it. Focusing on $\hat{p}$, we select the LP IRF estimator for 20\% to 40\% of the samples. 

In their Figure~6, LPW document that, under the MSE weighting and for most of the horizons, the BVAR method performs best across their samples. Their BVAR estimator is similar to our VAR estimator with $\hat{\lambda}$.\footnote{It is similar but not the same. LPW use an MDD-based hyperparameter determination (averaging rather than selection), whereas we use a task-based selection, which is often preferable under misspecification as shown in Table~\ref{tab:Risk.PCsel.250}.}  According to the right panel in our Figure~\ref{fig:empirics.fractionLP.Sshort}, except for $p=1$, IRFC recommends selecting the LP estimator in less than 50\% of the samples, at least for horizons below $h=8$. The bottom line is that, from an MSE perspective, whether VAR or LP IRF estimation is preferable is sample-dependent: there is no clear winner. These findings discredit the widespread idea that LPs are \emph{always} preferred under misspecification. Our IRFC criterion is the first method in the literature to provide a way for empirical researchers to make a data-driven choice between estimation approaches.

\graphicspath{{ResultsEM/Figures/G0doubling/PCselection/}}	  
\begin{figure}[t!]
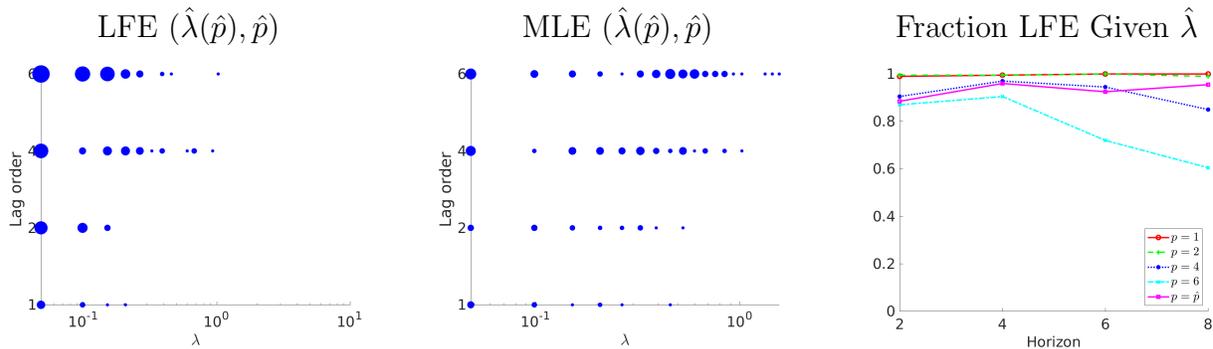

	\caption{\sc PC Selection}
	\label{fig:empirics.subS1.PC}
	\vspace*{-0.5cm}
	\setlength\tabcolsep{1pt}
	\adjustboxset{width=\linewidth,valign=c}
	\begin{center} 
		\begin{tabular}{ccccc}
			& {\footnotesize LFE $(\hat{\lambda}(\hat{p}),\hat{p})$} & {\footnotesize MLE $(\hat{\lambda}(\hat{p}),\hat{p})$} & & {\footnotesize Conditional On $\hat{\lambda}(p)$ }\\
			\rotatebox[origin=c]{90}{\tiny Lag Length $\hat{p}$} & \includegraphics[scale=.31]{OptimalLamPhat_LP_subS2_hor6.png} &
			\includegraphics[scale=.31]{OptimalLamPhat_VAR_subS2_hor6.png} & \rotatebox[origin=c]{90}{\tiny Fraction LFE Selected}
			& \includegraphics[scale=.31]{FrakLP_subSlong.png} \\[-0.5ex]
			& {\tiny Hyperparameter $\lambda$} & {\tiny Hyperparameter $\lambda$} & & {\tiny Horizon $h$}
		\end{tabular}
	\end{center}
	{\footnotesize {\em Notes}: Left and center panel: grid values of PC-selected shrinkage hyperparameters $\hat{\lambda}(\hat{p})$ for $h=6$. The diameter of the dots is proportional to the frequency of the $(\hat{\lambda}(\hat{p}),\hat{p})$ frequency. The number of samples across the four $\hat{p}$ rows adds up to 1,000. Right panel: Fraction of times the PC selects the LP IRF estimator under different lag lengths and across different horizons. Estimation sample 1984:Q1-2019:Q4.}\setlength{\baselineskip}{4mm}
\end{figure}

\noindent {\bf PC Selection.} We emphasized throughout the paper that model determination under misspecification should be based on a measure of the relevant prediction or estimation risk. We now examine how different a model selection based on multi-step forecast risk is from a choice based on the IRF estimation risk in our 1,000 samples. Suppose that a researcher uses PC instead of IRFC to determine $(\iota,\lambda,p)$. The effect on lag length choice and estimator selection is displayed in Figure~\ref{fig:empirics.subS1.PC}. The PC results can be compared to the IRFC results shown in the bottom row of Figure~\ref{fig:empirics.DistLamhat.subS1} and in the right plot of Figure~\ref{fig:empirics.fractionLP.Sshort} above. The results from the two selection criteria are readily apparent. The key differences between IRFC and PC selection are: (i) targeting multi-step prediction risk leads to more shrinkage for MLE and LFE, (ii) the lag length selected by PC tends to be smaller than the IRFC lag length for LFE, and larger for the MLE, and (iii) PC selects the LFE/LP less often than IRFC at larger $p$, including $\hat{p}$, but far more often at smaller $p$. This is consistent with the distinct features of multi-step forecasting and IRF estimation discussed previously. The $h$-step-ahead forecast risk is affected by the estimation uncertainty for $n^2p$ coefficients, whereas the IRF estimation risk is only affected by the sampling variability of the coefficient estimates for the $n\times n$ $h$th-order MA coefficient matrix. Thus, it is desirable to reduce estimation variance more strongly in the multi-step forecasting setting, by increasing the shrinkage and/or reducing the lag length and thereby the number of coefficients that need to be estimated. Hence, an important recommendation for practitioners is to use an IRF estimation risk criterion and not a criterion that measures forecast performance, when choosing the IRF estimator.

\section{Conclusion}
\label{sec:conclusion}

Multi-step forecasting and IRF estimation with VARs are challenging because the number of parameters may be large relative to the available data, and the VAR may suffer from small but consequential dynamic misspecification. The first issue can be addressed by combining likelihood information with prior information and selecting the weight in a data-driven manner. The second challenge has led econometricians to replace one-step-ahead regressions with multi-step regressions. We propose two criteria, PC and IRFC, that can be used to jointly select the estimator type, the weight on the prior, and the number of lags. Under a quadratic loss function, the criteria trade off bias and variance to minimize MSE. Our simulations and empirical evidence show that the trade-offs are non-trivial. For researchers open to task-based selection, we recommend a data-driven approach based on the proposed criteria, which can outperform rigid MDD-based selection under dynamic misspecification. In particular, if the task is IRF estimation, one should not use a criterion that measures forecast performance. Throughout, we focus on selection, but the objective functions derived here could also be used to determine averaging weights. We assume that the cross-sectional dimension $n$ of the VAR is fixed which could be relaxed in future work.

\small  \renewcommand{\baselinestretch}{1.05} \normalsize

\bibliography{VAR_PC_Ref}

@Article{Todd1984,
  author  = {Todd, Richard},
  journal = {Federal Reserve Bank of Minneapolis Quarterly Review},
  title   = {Improving Economic Forecasting with Bayesian Vector Autoregressions},
  year    = {1984},
  number  = {4},
  pages   = {18-29},
  volume  = {8},
}

@Article{Schorfheide2005,
  author  = {Schorfheide, Frank},
  journal = {Journal of Econometrics},
  title   = {VAR Forecasting Under Misspecification},
  year    = {2005},
  number  = {1},
  pages   = {99-136},
  volume  = {128},
}

@Book{Billingsley1968,
  author    = {Billingsley, Patrick},
  publisher = {John Wiley \& Sons, New York},
  title     = {Probability and Measure},
  year      = {1968},
}

@Article{MSW2006,
  author  = {Marcellino, Massimiliano and Stock, James H. and Watson, Mark W.},
  journal = {Journal of Econometrics},
  title   = {A Comparison of Direct and Iterated Multistep AR Methods for Forecasting Macroeconomic Time Series},
  year    = {2006},
  pages   = {499-526},
  volume  = {135},
}

@Article{McCrackenNg2020,
  author  = {McCracken, Michael W. and Ng, Serena},
  journal = {FRB St. Louis Working Paper},
  title   = {FRED-QD: A Quarterly Database for Macroeconomic Research},
  year    = {2020},
  volume  = {005},
}

@Article{Hamilton2018,
  author  = {Hamilton, James D.},
  journal = {Review of Economics and Statistics},
  title   = {Why You Should Never Use the Hodrick-Prescott Filter},
  year    = {2018},
  number  = {5},
  pages   = {831-843},
  volume  = {100},
}

@Article{GiannoneLenzaPrimiceri2015,
  author    = {Giannone, D. and Lenza, M. and Primiceri, G.},
  journal   = {Review of Economics and Statistics},
  title     = {Prior Selection for Vector Autoregressions},
  year      = {2015},
  number    = {2},
  pages     = {436-451},
  volume    = {97},
}

@ARTICLE{Sims1998,
  author = {Sims, Christopher A. and Zha, Tao},
  title = {Bayesian Methods for Dynamic Multivariate Models},
  journal = {International Economic Review},
  year = {1998},
  volume = {39},
  pages = {949-968},
  number = {4},
  section = {Impulse and Fluctuations},
  subsection = {VARs}
}

@Article{Litterman1986,
  author  = {Litterman, Robert B.},
  journal = {Journal of Business \& Economic Statistics},
  title   = {Forecasting with Bayesian Vector Autoregressions: Five Years of Experience},
  year    = {1986},
  number  = {1},
  pages   = {25-38},
  volume  = {4},
}

@Article{DoanLittermanSims1984,
  author  = {Doan, Thomas and Litterman, Robert and Sims, Christopher},
  journal = {Econometric Reviews},
  title   = {Forecasting and Conditional Projections Using Realistic Prior Distributions},
  year    = {1984},
  number  = {1},
  pages   = {1-100},
  volume  = {3},
}

@ARTICLE{DelNegro2004,
  author = {Del Negro, Marco and Schorfheide, Frank},
  title = {Priors from General Equilibrium Models for VARs},
  journal = {International Economic Review},
  year = {2004},
  volume = {45},
  pages = {643 -- 673},
  number = {2},
  section = {Impulse and Fluctuations},
  subsection = {DSGE-VAR}
}

@Article{Stein1981,
  author  = {Stein, Charles},
  journal = {Annals of Statistics},
  title   = {Estimation of the Mean of a Multivariate Normal Distribution},
  year    = {1981},
  number  = {6},
  pages   = {1135-1151},
  volume  = {9},
}

@Article{Bhansali1997,
  author  = {Bhansali, R. J.},
  journal = {Statistica Sinica},
  title   = {Direct Autoregressive Predictors for Multistep Prediction: Order Selection and Performance Relative to the Plug In Predictors},
  year    = {1997},
  pages   = {425-449},
  volume  = {7},
}

@Article{Ing2003,
  author  = {Ing, Ching-Kang},
  journal = {Econometric Theory},
  title   = {Multistep Prediction in Autoregressive Processes},
  year    = {2003},
  number  = {2},
  pages   = {254-279},
  volume  = {19},
}

@Article{FindleyWei2002,
  author  = {Findley, David F. and Wei, Ching-Zong},
  journal = {Journal of Multivariate Analysis},
  title   = {AIC, Overfitting Principles, and the Boundedness of Moments of Inverse Matrices for Vector Autoregressions and Related Models},
  year    = {2002},
  pages   = {415-450},
  volume  = {83},
}

@Article{IngWei2003,
  author  = {Ing, Ching-Kang and Wei, Ching-Zong},
  journal = {Journal of Multivariate Analysis},
  title   = {On Same-Realization Prediction in an Infinite-Order Autoregressive Process},
  year    = {2003},
  pages   = {130-155},
  volume  = {85},
}

@Article{LewisReinsel1985,
  author  = {Lewis, Richard and Reinsel, Gregory C.},
  journal = {Journal of Multivariate Analysis},
  title   = {Prediction of Multivariate Time Series by Autoregressive Model Fitting},
  year    = {1985},
  pages   = {393-411},
  volume  = {16},
}

@Article{LewisReinsel1988,
  author  = {Lewis, Richard A. and Reinsel, Gregory C.},
  journal = {Journal of Time Series Analysis},
  title   = {Prediction Error of Multivariate Time Series With Mis-specified Models},
  year    = {1988},
  number  = {1},
  pages   = {43-57},
  volume  = {9},
}

@Article{Reinsel1980,
  author  = {Reinsel, Gregory C.},
  journal = {Journal of the Royal Statistical Society B},
  title   = {Asymptotic Properties of Prediction Errors for the Multivariate Autoregressive Model Using Estimated Parameters},
  year    = {1980},
  number  = {3},
  pages   = {328-333},
  volume  = {42},
}

@Article{Shibata1980,
  author  = {Shibata, Ritei},
  journal = {Annals of Statistics},
  title   = {Asymptotically Efficient Selection of the Order of the Model for Estimating Parameters of a Linear Process},
  year    = {1980},
  number  = {1},
  pages   = {147-164},
  volume  = {8},
}

@Article{SpeedYu1993,
  author  = {Speed, T.P. and Yu, Bin},
  journal = {Annals of the Institute of Statistical Mathematics},
  title   = {Model Selection and Prediction: Normal Regression},
  year    = {1993},
  number  = {1},
  pages   = {35-54},
  volume  = {45},
}

@Article{Weiss1991,
  author  = {Weiss, Andrew A.},
  journal = {Journal of Econometrics},
  title   = {Multi-step Estimation and Forecasting in Dynamic Models},
  year    = {1991},
  pages   = {135-149},
  volume  = {48},
}

@Article{Bhansali1996,
  author  = {Bhansali, R. J.},
  journal = {Annals of the Institute for Statistical Mathematics},
  title   = {Asymptotically Efficient Autoregressive Model Selection for Multistep Prediction},
  year    = {1996},
  number  = {3},
  pages   = {577-602},
  volume  = {48},
}

@Book{ClementsHendry1998,
  author    = {Clements, Michael P. and Hendry, David F.},
  publisher = {Cambridge University Press},
  title     = {Forecasting Economic Time Series},
  year      = {1998},
}

@Article{Findley1983,
  author  = {Findley, David F.},
  journal = {American Statistical Association: Proceedings of Business and Economic Statistics},
  title   = {On the Use of Multiple Models for Multi-Period Forecasting},
  year    = {1983},
  pages   = {528-531},
}

@Article{Baillie1979,
  author  = {Baillie, Richard T.},
  journal = {Biometrika},
  title   = {Asymptotic Prediction Mean Squared Error for Vector Autoregressive Models},
  year    = {1979},
  number  = {3},
  pages   = {675-678},
  volume  = {66},
}

@Article{hansen2016stein,
  author  = {Hansen, Bruce E},
  journal = {Manuscript, University of Wisconsin-Madison},
  title   = {Stein Combination Shrinkage for Vector Autoregressions},
  year    = {2016},
}

@Article{Lohmeyer2018,
  author    = {Lohmeyer, Jan and Palm, Franz and Reuvers, Hanno and Urbain, Jean-Pierre},
  journal   = {Econometric Reviews},
  title     = {Focused Information Criterion for Locally Misspecified Vector Autoregressive Models},
  year      = {2018},
  issn      = {1532-4168},
  month     = feb,
  number    = {7},
  pages     = {763--792},
  volume    = {38},
  doi       = {10.1080/07474938.2017.1409410},
  publisher = {Informa UK Limited},
}

@Article{Li2022,
  author  = {Li, Dake and Plagborg-M{\o}ller, Mikkel and Wolf, Christian},
  journal = {Journal of Econometrics},
  title   = {Local Projections vs. VARs: Lessons From Thousands of DGPs},
  year    = {2024},
  number  = {2},
  pages   = {105722},
  volume  = {244},
  doi     = {10.3386/w30207},
}

@Article{PlagborgMoeller2021,
  author    = {Plagborg-M{\o}ller, Mikkel and Wolf, Christian K.},
  journal   = {Econometrica},
  title     = {Local Projections and VARs Estimate the Same Impulse Responses},
  year      = {2021},
  issn      = {0012-9682},
  number    = {2},
  pages     = {955-980},
  volume    = {89},
  doi       = {10.3982/ecta17813},
  publisher = {The Econometric Society},
}

@Article{Jorda2005,
  author    = {Jord{\`a}, {\`O}scar},
  journal   = {American Economic Review},
  title     = {Estimation and Inference of Impulse Responses by Local Projections},
  year      = {2005},
  issn      = {0002-8282},
  month     = feb,
  number    = {1},
  pages     = {161-182},
  volume    = {95},
  doi       = {10.1257/0002828053828518},
  publisher = {American Economic Association},
}

@Article{MontielOleaEtAl2024,
  author  = {Montiel Olea, Jose and Plagborg-M{\o}ller, Mikkel and Qian, Eric and Wolf, Christian},
  journal = {Econometrica},
  title   = {Double Robustness of Local Projections and Some Unpleasant VARithmetic},
  year    = {2026},
  number  = {4},
  pages   = {1313-1343},
  volume  = {94},
}

@Article{MontielOleaPlagborgMoller2021,
  author  = {Montiel Olea, Jose and Plagborg-M{\o}ller, Mikkel},
  title   = {Local Projection Inference is Easier Than You Think},
  journal = {Econometrica},
  year    = {2021},
  volume  = {89},
  number  = {4},
  pages   = {1789-1823},
}

@Article{Ludwig2024,
  author  = {Ludwig, Julian},
  journal = {SSRN Working Paper},
  title   = {Local Projections are VAR Predictions of Increasing Order},
  year    = {2026},
  volume  = {6441981},
}

@Article{Miranda-AgrippinoRicco2021,
  author  = {Miranda-Agrippino, Silvia and Ricco, Giovanni},
  journal = {Review of Economics and Statistics},
  title   = {Bayesian Local Projections},
  year    = {2025},
  number  = {5},
  pages   = {1424-1438},
  volume  = {107},
}

@Article{Muller2013,
  author  = {M\"uller, Ulrich},
  journal = {Econometrica},
  title   = {Risk of Bayesian Inference in Misspecified Models, and the Sandwich Covariance Matrix},
  year    = {2013},
  number  = {5},
  pages   = {1805-1849},
  volume  = {81},
}

@Article{IngramWhiteman1994,
  author  = {Ingram, Beth F. and Whiteman, Charles H.},
  journal = {Journal of Monetary Economics},
  title   = {Supplanting the ``Minnesota'' Prior: Forecasting Macroeconomic Time Series Using Real Business Cycle Model Priors},
  year    = {1994},
  number  = {3},
  pages   = {497-510},
  volume  = {34},
}

\small  \renewcommand{\baselinestretch}{1.3} \normalsize
		

\clearpage

\renewcommand{\thepage}{A.\arabic{page}}
\setcounter{page}{1}

\begin{appendix}
	\markright{Online Supplement -- This Version: \today }
	
	\renewcommand{\theequation}{A.\arabic{equation}}
	\setcounter{equation}{0}	
	\renewcommand*\thetable{A-\arabic{table}}
	\setcounter{table}{0}
	\renewcommand*\thefigure{A-\arabic{figure}}
	\setcounter{figure}{0}
	\renewcommand*\thetheorem{A-\arabic{theorem}}
    \setcounter{theorem}{0}
	\renewcommand*\thelemma{A-\arabic{lemma}}
\setcounter{lemma}{0}

	\begin{center}
		
		{\large {\bf Online Supplement for: Misspecification-Robust Shrinkage and Selection for VAR Forecasts and IRFs}}
		
		{\bf Oriol Gonz\'alez-Casas\'us and Frank Schorfheide}
	\end{center}

\noindent This Supplement consists of the following sections:

\begin{itemize}
	\item[A.] Proofs and Derivations
	\item[B.] Heteroskedasticity
	\item[C.] Data for Monte Carlo Design and Empirical Analysis
	\item[D.] Additional Empirical Results
\end{itemize}

\newpage

\section{Proofs and Derivations}
\label{appsec:proofs}

\subsection{Preliminaries}

The following two lemmas are useful in the proofs that follow.

\begin{lemma}\label{lmm:riskformula}
	Let $M$ be an $nq \times n$ matrix, $A$ an $nq \times nq$ matrix, $\Xi$ an $n \times n_{sh}$ matrix, where $n_{sh} \le n$, and $W$ an $n \times n$ symmetric positive definite weight matrix. Then
	\[
	\left\Vert M'AM \Xi \right\Vert_W^2
	= \tr\left\{[(MWM')\otimes (M \Xi \Xi' M')]\vecr(A)\vecr(A)'\right\}.
	\]
\end{lemma}

\noindent {\bf Proof of Lemma~\ref{lmm:riskformula}.} The result can be derived as follows:
\begin{eqnarray*}
	\left\Vert M'AM \Xi \right\Vert_W^2
	&=_{(1)}& \tr\bigg\{ W \big(M'AM \Xi\big) \big(\Xi' M'A'M \big) \bigg\} \\
	&=_{(2)}&\tr\left\{\underbrace{MWM'}_{=C}A\underbrace{M \Xi \Xi'M'}_{=B}A'\right\}\\
	&=_{(3)}&\vecr(A)'[(MWM')\otimes (M \Xi \Xi' M')]\vecr(A)\\
	&=_{(4)}&\tr\left\{[(MWM')\otimes (M \Xi \Xi' M')]\vecr(A)\vecr(A)'\right\}.
\end{eqnarray*}
The second equality uses $\tr(AB)=\tr(BA)$. The third equality is based on
\[
\tr\{CABA'\}=\vecr(A)'(C\otimes B)\vecr(A).
\]
The last equality uses $a'Ba = \tr[B a a']$. $\blacksquare$

\begin{lemma}\label{lmm:varcompanionaux_gen}
	For any $n \times 1$ vector $\nu$ and square matrix $C$ of dimension $d$, $$(\nu \otimes I_{d})C(\nu'\otimes I_{d})=(\nu\nu')\otimes C.$$
\end{lemma}
\textbf{Proof of Lemma \ref{lmm:varcompanionaux_gen}.}
By direct calculation,
\begin{align*}
	(\nu\otimes I_{d})C(\nu'\otimes I_{d})&=\begin{bmatrix}
		\nu_1I_{d}\\
		\vdots\\
		\nu_nI_{d}
	\end{bmatrix} C 
	\begin{bmatrix}
		\nu_1I_{d} &
		\cdots&
		\nu_nI_{d}
	\end{bmatrix}\\
	&=\begin{bmatrix}
		\nu_1^2C &\cdots &\nu_1\nu_nC\\
		\vdots&\ddots & \\
		\nu_n\nu_1C&&\nu_n^2C
	\end{bmatrix}\\
	&=\begin{bmatrix}
		\nu_1^2 &\cdots &\nu_1\nu_n\\
		\vdots&\ddots & \\
		\nu_n\nu_1&&\nu_n^2
	\end{bmatrix}\otimes C\\
	&=(\nu\nu')\otimes C.\quad \blacksquare
\end{align*}

\subsection{Proofs for Section~\ref{sec:VAR1}}
\label{appsubsec:proofs.sec2}

\noindent {\bf Proof of Theorem~\ref{thm:limitdis}.} The formulas for the bias terms are given by
\begin{align*}
\delta(lfe,\lambda)&=\lambda\underline{\psi}\underline{P}_\Psi(\lambda\underline{P}_\Psi+\Gamma_{yy,0})^{-1}\\
\delta(mle,\lambda)&=\lambda\sum_{j=0}^{h-1}F^j\underline{\phi}\underline{P}_\Phi(\lambda \underline{P}_{\Phi}+\Gamma_{yy,0})^{-1} F^{h-1-j}\\
\mu(lfe,\lambda)&=\sum_{j=0}^{h-1}F^j\Gamma_{zy,h-j}(\lambda\underline{P}_\Psi+\Gamma_{yy,0})^{-1}\\
\mu(mle,\lambda)&=\sum_{j=0}^{h-1}F^j\Gamma_{zy,1} (\lambda \underline{P}_{\Phi}+\Gamma_{yy,0})^{-1} F^{h-1-j}.
\end{align*}
The asymptotic covariance matrix between $(\iota,\lambda)$ and $(\iota',\lambda')$ predictors takes the form
$$\textbf V=\begin{bmatrix}
	V(mle,\lambda)&&&\\Cov(lfe,\lambda;mle,\lambda)&V(lfe,\lambda)&&\\Cov(mle,\lambda';mle,\lambda)&Cov(mle,\lambda';lfe,\lambda)&V(mle,\lambda')&\\Cov(lfe,\lambda';mle,\lambda)&Cov(lfe,\lambda';lfe,\lambda)&Cov(lfe,\lambda';mle,\lambda')&V(lfe,\lambda')
\end{bmatrix}$$ with the elements defined as
\begin{align*}
	Cov(lfe,\lambda;lfe,\lambda')&=\sum_{i=0}^{h-1}\sum_{j=0}^{h-1}(F^i\Sigma_{\epsilon\epsilon}F^{j'})\otimes((\lambda\underline{P}_\Psi'+\Gamma_{yy,0})^{-1}\Gamma_{yy,j-i}(\lambda'\underline{P}_\Psi+\Gamma_{yy,0})^{-1})\\
	Cov(mle,\lambda;mle,\lambda')&=\sum_{i=0}^{h-1}\sum_{j=0}^{h-1}(F^i\Sigma_{\epsilon\epsilon}F^{j'})\otimes (F^{h-1-i'}(\lambda \underline{P}_{\Phi}'+\Gamma_{yy,0})^{-1}\Gamma_{yy,0}(\lambda' \underline{P}_{\Phi}+\Gamma_{yy,0})^{-1}F^{h-1-j})\\
	Cov(mle,\lambda;lfe,\lambda')&=\sum_{i=0}^{h-1}\sum_{j=0}^{h-1}(F^i\Sigma_{\epsilon\epsilon}F^{j'})\otimes (F^{h-1-i'}(\lambda \underline{P}_{\Phi}'+\Gamma_{yy,0})^{-1}\Gamma_{yy,h-1-j}(\lambda' \underline{P}_{\Psi}+\Gamma_{yy,0})^{-1}),
\end{align*} and trivially $V(\iota,\lambda)=Cov(\iota,\lambda;\iota,\lambda)$.
Because $\Gamma_{yy,h-1-j} = F^{h-1-j} \Gamma_{yy,0}$, we obtain that 
\[
   Cov(mle,0;lfe,0) = \sum_{i=0}^{h-1}\sum_{j=0}^{h-1}(F^i\Sigma_{\epsilon\epsilon}F^{j'})\otimes (F^{h-1-i'}\Gamma_{yy,0}^{-1}F^{h-1-j} ) = V(mle,0).
\]

\noindent {\bf Analysis of LFE.} First, note that 
$$\Bar{\Psi}_T(lfe,\lambda)-F^h=(\underline{\Psi}_T-F^h)\lambda T\underline{P}_\Psi\Bar{P}^{-1}_\Psi+(\hat{\Psi}_T(lfe)-F^h)S_{T,hh}\Bar{P}^{-1}_\Psi.$$ Moreover, the LFE can be written as $$\hat{\Psi}_T(lfe)=F^h+\alpha T^{-1/2}\left(\sum_{j=0}^{h-1}\sum_{t=1}^TF^jz_{t-j}y_{t-h}'\right)S_{T,hh}^{-1}+\left(\sum_{j=0}^{h-1}\sum_{t=1}^TF^j\epsilon_{t-j}y_{t-h}'\right)S_{T,hh}^{-1}.$$
Therefore,
\begin{eqnarray*}
	\Bar{\Psi}_T(lfe,\lambda)-F^h&=&(\underline{\Psi}_T-F^h)\lambda T\underline{P}_\Psi\Bar{P}^{-1}_\Psi\\
	&&+\alpha T^{-1/2}\left(\sum_{j=0}^{h-1}\sum_{t=1}^TF^jz_{t-j}y_{t-h}'\right)\Bar{P}^{-1}_\Psi\\
	&&+\left(\sum_{j=0}^{h-1}\sum_{t=1}^TF^j\epsilon_{t-j}y_{t-h}'\right)\Bar{P}^{-1}_\Psi .
\end{eqnarray*}
By the same steps as in \cite{Schorfheide2005} and equations \eqref{eq:prior.mean.drift} and \eqref{eq:prior.precision.drift}, $$T^{1/2}\left(\Bar{\Psi}_T(lfe,\lambda)-F^h\right)=\delta(lfe,\lambda)+\alpha\mu(lfe,\lambda)+\zeta_T(lfe,\lambda),$$ where 
\begin{eqnarray*}      \delta(lfe,\lambda)&=&\lambda\underline{\psi}\underline{P}_\Psi(\lambda\underline{P}_\Psi+\Gamma_{yy,0})^{-1}\\
	\mu(lfe,\lambda)&=&\sum_{j=0}^{h-1}F^j\Gamma_{zy,h-j}(\lambda\underline{P}_\Psi+\Gamma_{yy,0})^{-1}\\
	\zeta_T(lfe,\lambda)&\Longrightarrow & N(0,V(lfe,\lambda))\\
	V(lfe,\lambda)&=&\sum_{i=0}^{h-1}\sum_{j=0}^{h-1}(F^i\Sigma_{\epsilon\epsilon}F^{j'})\otimes((\lambda\underline{P}_\Psi+\Gamma_{yy,0})^{-1}\Gamma_{yy,j-i}(\lambda\underline{P}_\Psi+\Gamma_{yy,0})^{-1}).
\end{eqnarray*}

\noindent {\bf Analysis of MLE.} By a first order Taylor expansion, $$\Phi^h-F^h=\sum_{j=0}^{h-1}F^j(\Phi-F)F^{h-1-j}+\mathcal{R}(\Phi-F).$$
Note that 
$$\Bar{\Phi}_T(mle,\lambda)-F=(\underline{\Phi}_T-F)\lambda T\underline{P}_\Phi\Bar{P}^{-1}_\Phi+(\hat{\Phi}_T(mle)-F)S_{T,11}\Bar{P}^{-1}_\Phi,$$ so it follows that 
\begin{eqnarray*}
	\Bar{\Psi}(mle,\lambda)-F^h&=&\lambda T\sum_{j=0}^{h-1}F^j(\underline{\Phi}_T-F)\underline{P}_\Phi\Bar{P}^{-1}_\Phi F^{h-1-j}\\
	&& +\sum_{j=0}^{h-1}F^j(\hat{\Phi}_T(mle)-F)S_{T,11}\Bar{P}^{-1}_\Phi F^{h-1-j}\\
	&& +\mathcal{R}(\Bar{\Phi}_T(mle,\lambda)-F).
\end{eqnarray*}
By \cite{Schorfheide2005} and equations \eqref{eq:prior.mean.drift} and \eqref{eq:prior.precision.drift}, $$T^{1/2}\left( \Bar{\Psi}(mle,\lambda)-F^h\right)=\delta(mle,\lambda)+\alpha\mu(mle,\lambda)+\zeta_T(mle,\lambda),$$ where
\begin{eqnarray*}
	\delta(mle,\lambda)&=&\lambda\sum_{j=0}^{h-1}F^j\underline{\phi}\underline{P}_\Phi(\lambda \underline{P}_{\Phi}+\Gamma_{yy,0})^{-1} F^{h-1-j}\\
	\mu(mle,\lambda)&=&\sum_{j=0}^{h-1}F^j\Gamma_{zy,1} (\lambda \underline{P}_{\Phi}+\Gamma_{yy,0})^{-1} F^{h-1-j}\\
	\zeta_T(mle,\lambda)&\Longrightarrow&N(0,V(mle,\lambda))\\
	V(mle,\lambda)&=&\sum_{i=0}^{h-1}\sum_{j=0}^{h-1}(F^i\Sigma_{\epsilon\epsilon}F^{j'})\otimes (F^{h-1-i'}(\lambda \underline{P}_{\Phi}'+\Gamma_{yy,0})^{-1}\Gamma_{yy,0}(\lambda \underline{P}_{\Phi}+\Gamma_{yy,0})^{-1}F^{h-1-j}).
\end{eqnarray*}
The covariance follows from the same arguments as in \cite{Schorfheide2005}. $\blacksquare$

\noindent {\bf Proof of Theorem~\ref{thm:prediction.risk}.}
The difference between the conditional
expectation of $y_{T+h}$ (omitting the tilde) and the predictor
$\hat{y}_{T+h}(\iota,\lambda)$ is given by
\begin{eqnarray*}
	T^{1/2}(\mathbb{E}_T[ y_{T+h}] - \hat{y}_{T+h}(\iota,\lambda) )
	&=& \alpha  \left(
	\sum_{j=0}^{h-1} F^j  \mathbb{E}_T[z_{T+h-j}] - \mu_*(prd) y_{T} \right) \\
	&& + \alpha  [ \mu_*(prd) - \mu(\iota,\lambda)] y_{T} -  \zeta_T(\iota,\lambda) y_{T}\\
	&& - \delta(\iota,\lambda)y_T.
\end{eqnarray*}

The normalized prediction risk can then be expressed as follows:
\begin{eqnarray}
		\lefteqn{T  \mathbb{E} \bigg[ \tr \{ W ( \mathbb{E}_T[y_{T+h}] - \hat{y}_{T+h}(\iota,\lambda) )
		(  \mathbb{E}_T[ y_{T+h}] - \hat{y}_{T+h}(\iota,\lambda) )' \} \bigg] } \label{eq:risk} \\
		&=_{(1)}&   \alpha^2 \tr \bigg\{ W( \mu_*(prd) - \mu(\iota,\lambda) )\Gamma_{YY,0}(\mu_*(prd) - \mu(\iota,\lambda))' \bigg\} \nonumber \\
		&\phantom{=}_{(2)}&  + \tr \bigg\{ W \mathbb{E} \bigg[ \zeta_T(\iota,\lambda) \Gamma_{yy,0} \zeta_T(\iota,\lambda)' \bigg] \bigg\} \nonumber \\
		&\phantom{=}_{(3)}&  + \alpha^2 \tr \left\{ W \mathbb{E} \left[ \left( \sum_{j=0}^{h-1} F^j \mathbb{E}_T[z_{T+h-j}] - \mu_*(prd) y_{T} \right) \right. \right. \nonumber  \\
		&&  \times  \left. \left. \left( \sum_{j=0}^{h-1} F^j \mathbb{E}_T[z_{T+h-j}] - \mu_*(prd) y_{T} \right)' \right] \right\}  \nonumber \\
		&\phantom{=}_{(4)}&+\tr\left\{W\delta(\iota,\lambda)\Gamma_{yy,0}\delta(\iota,\lambda)'\right\} \nonumber \\
		&\phantom{=}_{(5)}&  - 2 \alpha \tr \left\{ W  \mathbb{E}[\zeta_T(\iota,\lambda)] \mathbb{E} \left[ y_{T}  \left( \sum_{j=0}^{h-1} F^j \mathbb{E}_T[z_{T+h-j}] - \mu_*(prd) y_{T} \right)' \right] \right\}  \nonumber \\
		&\phantom{=}_{(6)}& + 2 \alpha^2 \tr \left\{ W \mathbb{E} \left[ \sum_{j=0}^{h-1} F^j \mathbb{E}_T[z_{T+h-j}]y_{T}' - \mu_*(prd) y_Ty_T' \right]
		( \mu_*(prd) - \mu(\iota,\lambda))' \right\} \nonumber \\
		&\phantom{=}_{(7)}&-2\alpha \tr\left\{W\mathbb{E}[\zeta_T(\iota,\lambda)]\Gamma_{yy,0}(\mu_*(prd) - \mu(\iota,\lambda))'\right\} \nonumber \\
		&\phantom{=}_{(8)}&- 2 \alpha \tr \left\{ W  \delta(\iota,\lambda) \mathbb{E} \left[ y_{T}  \left( \sum_{j=0}^{h-1} F^j \mathbb{E}_T[z_{T+h-j}] - \mu_*(prd) y_{T} \right)' \right] \right\} \nonumber \\
		&\phantom{=}_{(9)}& -  2 \alpha \tr \left\{ W  ( \mu_*(prd) - \mu(\iota,\lambda))\Gamma_{yy,0}\delta(\iota,\lambda)' \right\} \nonumber \\
		&\phantom{=}_{(10)}& +2\tr \left\{ W \mathbb{E} [ \zeta_T(\iota,\lambda)] \Gamma_{yy,0} \delta(\iota,\lambda)' \right\}. \nonumber
\end{eqnarray}

Since
\[ \tr[WABA'] = \vecr(A)'(W \otimes B) \vecr(A) \]
and $\tr[AB] = \tr[BA]$ we can rewrite term (2) in \eqref{eq:risk} as
$$
\tr \bigg\{ W \mathbb{E} \bigg[ \zeta_T(\iota,\lambda) \Gamma_{yy,0} \zeta_T(\iota,\lambda)' \bigg] \bigg\}
= \tr \bigg\{ ( W \otimes \Gamma_{yy,0} ) \mathbb{E}\left[\zeta_T(\iota,\lambda) \zeta_T(\iota,\lambda)'\right] \bigg\}
$$
with the understanding that on the right-hand side of the equation $\zeta_T(\iota,\lambda)$ is vectorized. 
Under Assumption~1 in S2005, the sequence $\| \zeta_T(\iota,\lambda) \|^2 $ is uniformly integrable. Hence, we can deduce that (see Theorem 3.5 of \cite{Billingsley1968}) $$\tr \bigg\{ (W \otimes \Gamma_{yy,0}) \mathbb{E} \bigg[ \zeta_T(\iota,\lambda) \zeta_T(\iota,\lambda)' \bigg] \bigg\}
\longrightarrow   \tr \bigg\{ ( W \otimes \Gamma_{yy,0} ) V( \iota,\lambda) \bigg\}.$$
Moreover, uniform integrability of $\| \zeta_T(\iota,p) \|^2 $ implies that $\mathbb{E} [ \zeta_T(\iota,\lambda)]=o(1)$, and so terms (5), (7), and (10) in \eqref{eq:risk} are $o(1)$.
Since $$ \mathbb{E} \left[ \sum_{j=0}^{h-1} F^j \mathbb{E}_T[z_{T+h-j}]y_{T}' \right]
=  \sum_{j=0}^{h-1} F^j \Gamma_{zy,h-j} = \mu_*(prd) \Gamma_{yy,0} $$
terms (6) and (8) in \eqref{eq:risk} are $o(1)$, too.
The above simplifications allow us to rewrite the normalized prediction risk as
\begin{eqnarray*}
	\lefteqn{T  \mathbb{E} \bigg[ \tr \{ W ( \mathbb{E}_T[ y_{T+h}] - \hat{y}_{T+h}(\iota,\lambda) )
	(  \mathbb{E}_T[ y_{T+h}] - \hat{y}_{T+h}(\iota,\lambda) )' \} \bigg] } \\
	&=_{(1)}&   \alpha^2 \tr \bigg\{ W( \mu_*(prd) - \mu(\iota,\lambda) )\Gamma_{yy,0}(\mu_*(prd) - \mu(\iota,\lambda))' \bigg\}\\
	&\phantom{=}_{(2)} &  + \tr \bigg\{ ( W \otimes \Gamma_{yy,0} ) V( \iota,\lambda) \bigg\} \\
	&\phantom{=}_{(3)}&  + \alpha^2 \tr \left\{ W \mathbb{E} \left[ \left( \sum_{j=0}^{h-1} F^j \mathbb{E}_T[z_{T+h-j}] - \mu_*(prd) y_{T} \right)  \left( \sum_{j=0}^{h-1} F^j \mathbb{E}_T[z_{T+h-j}] - \mu_*(prd) y_{T} \right)' \right] \right\}  \\
	&\phantom{=}_{(4)}&+\tr\left\{W\delta(\iota,\lambda)\Gamma_{yy,0}\delta(\iota,\lambda)'\right\}\\
	&\phantom{=}_{(9)}&- \;2 \alpha \tr \left\{ W  \delta(\iota,\lambda) \Gamma_{yy,0}\left(\mu_*(prd) - \mu(\iota,\lambda) \right)' \right\}\\
	&&+o(1).
\end{eqnarray*}
Hence, the desired result follows. $\blacksquare$

\noindent {\bf Proof of Theorem~\ref{thm:IRF.risk}.} Similar to the proof of Theorem~\ref{thm:prediction.risk} and hence omitted. To manipulate the IRF estimation risk the proof uses Lemma~\ref{lmm:riskformula}.

\subsection{Proofs and Derivations for Section~\ref{sec:VARp}}
\label{appsubsec:proofs.sec3}

We begin with some additional notation for the companion form representation introduced in the paper. Define
\be
\underbrace{\phi}_{n \times nq} = [\phi_1, \ldots, \phi_q], \quad
\underbrace{\Upsilon_q}_{n(q-1) \times nq}  = \left[ \begin{array}{ccccccc} 
	I_n    & \cdots & 0_n      & 0_n \\
	\vdots &  \ddots &  \vdots         & \vdots \\
	0_n    & \cdots & I_n      & 0_n 
\end{array} \right], \quad 
\underbrace{M_{\Upsilon_q}}_{nq \times n(q-1)} = \left[ \begin{array}{cccccc} 
	0_n    & \cdots & 0_n \\
	I_n    & \cdots & 0_n \\
	\vdots &  \ddots &  \vdots  \\
	0_n    & \cdots & I_n   
\end{array} \right].
\label{eq:def.upsilon}
\ee
Using this notation, the $q$-companion form matrix $\Phi$ has the following two properties
\be
M' \Phi = \phi, \quad M_{\Upsilon_q}' \Phi = \Upsilon_q.
\label{eq:Phi.properties}
\ee

The $q$-companion form matrix of a VAR($p$) takes the form 
\be
\underbrace{\Phi}_{nq \times nq} 
= \begin{bmatrix} \phi_1 \; \cdots \; \phi_{p-1} \; \phi_p & 0_{n \times n(q-p)}  \\
	\Upsilon_p & 0_{n(p-1) \times n(q-p)}  \\
	0_{n \times n} \; \cdots \; 0_{n \times n} \; I_n & 0_{n \times n(q-p)} \\
	0_{n(q-p-1) \times np}  & \Upsilon_{q-p}
\end{bmatrix}.
\label{eq:q.companionform.of.VARp}
\ee

\noindent {\bf Prior.} To construct the shrinkage estimators, we use prior means that share the restrictions of the $q$-companion form of a VAR($p$) in (\ref{eq:q.companionform.of.VARp}). For instance, in the case of the MLE, for $p>1$:
\be
\underbrace{\underline{\Phi}_T}_{nq \times nq} 
= \begin{bmatrix} \underline{\phi}_{1,T} \; \cdots \; \underline{\phi}_{p-1,T} \; \underline{\phi}_{p,T} & 0_{n \times n(q-p)}  \\
	\Upsilon_p & 0_{n(p-1) \times n(q-p)}  \\
	\cdot & \cdot \\
	\cdot & \cdot 
\end{bmatrix},
\label{eq:prior.phi}
\ee
such that
\be
M' \underline{\Phi}_T = \underline{\phi}_T, \quad \underline{\phi}_T R_p = 0.
\label{eq:prior.phi.restrictions}
\ee
For $p=1$ the second line in (\ref{eq:prior.phi}) drops out. The submatrices in the positions marked by $\cdot$ are irrelevant for the subsequent analysis. They determine the dynamics of lags $p+1,\ldots,q$, which are irrelevant in the forward iteration of a VAR($p$). We complete the prior mean specification by simply using the $q$-companion form entries in (\ref{eq:q.companionform.of.VARp}). 

As in the VAR(1) case, we use a prior covariance matrix with a Kronecker structure. Thus, one only needs to specify the $nq \times nq$ hyperparameter-scaled precision matrix $\underline{P}_{\Phi}$. In fact, it suffices to specify the $np \times np$ submatrix that corresponds to the precision of the coefficients for lags 1 through $p$. As will be shown later, the prior precision for the coefficients associated with lags $p+1$ to $q$ does not affect the first $np$ rows of the posterior mean $\Bar{\Phi}_T(mle,\lambda,p)$ and can be set to zero.

\noindent {\bf Posterior Mean.} 
If the coefficients on lags $p+1,\ldots,q$ are restricted to be zero, then one can use the formula for restricted least squares to express the posterior mean for a VAR($p$) in $q$-companion form as
\be
\Bar{\Psi}_T(lfe,\lambda,p) = \Bar{S}_{T,0h} \Bar{S}_{T,hh}^{-1} \big[ I_{nq} - R_p(R_p' \Bar{S}_{T,hh}^{-1}R_p)^{-1}R_p'\Bar{S}_{T,hh}^{-1} \big],
\label{eq:barPhi.restricted}
\ee
where 
\[
S_{T,0h} = \sum_{t=1}^T Y_t Y_{t-h}', \quad S_{T,hh} = \sum_{t=1}^T Y_{t-h} Y_{t-h}', \quad
\Bar{S}_{T,0h} = S_{T,0h} + T \lambda \underline{\Psi}_T \underline{P}_\Psi, \quad
\Bar{S}_{T,hh} = S_{T,hh} + T \lambda \underline{P}_{\Psi}.
\]
In the special case of $p=q$ we have $R_p=0$ and $\Bar{\Psi}_T(lfe,\lambda,q) = \Bar{S}_{T,0h} \Bar{S}_{T,hh}^{-1}$. The posterior mean $\Bar{\Phi}_T(mle,\lambda,p)$ is obtained by using  $\Bar{S}_{T,01}$ and $\Bar{S}_{T,11}$.

\begin{lemma} \label{lem:companion.form} (i) The companion form posterior mean has the following property:
	\[
	\bar{\Psi}_T(lfe,\lambda,p) =
	\begin{bmatrix} R'_{p \perp} \bar{S}_{T,0h} R_{p \perp} 
		\left(  R_{p\perp}'\bar{S}_{T,hh} R_{p\perp} \right)^{-1} & 0 \\
		R'_{p} \bar{S}_{T,0h} R_{p \perp} 
		\left(  R_{p\perp}'\bar{S}_{T,hh} R_{p\perp} \right)^{-1} & 0 
	\end{bmatrix}.
	\]	
	The corresponding result for $\bar{\Phi}_T(mle,\lambda,p)$ is obtained by using $\bar{S}_{T,01}$ and $\bar{S}_{T,11}$. (ii) For $p>1$, rows $n+1$ to $np$ of $\bar{\Phi}_T(mle,\lambda,p)$ take the form $\begin{bmatrix} \Upsilon_p & 0_{n(p-1) \times n(q-p)} \end{bmatrix}$; see (\ref{eq:prior.phi}).
\end{lemma}

The first part of the Lemma implies that the LFE version of the restrictions (\ref{eq:prior.phi.restrictions}) also holds for the posterior mean. Moreover, it states that $R_{p \perp}' \bar{\Psi}_T(lfe,\lambda,p) R_{p \perp}$ is identical to the posterior mean that is obtained when estimating a VAR($p$) in $p$-companion form. For the case of $p>1$, the second part of the Lemma implies the additional result that for the MLE based shrinkage estimator rows $n+1$ to $np$ contain $\Upsilon_p$ and zeros, i.e., the posterior mean inherits the form of the prior mean in (\ref{eq:prior.phi}). This is important, because the plug-in predictor will be constructed as $[\bar{\Phi}_T(mle,\lambda,p) ]^h$; see (\ref{eq:estimator.mle.plugin}).

\noindent {\bf Proof of Lemma~\ref{lem:companion.form}.} (i) Partition 
\[
\bar{S}_{T,hh}^{-1} = \begin{bmatrix} A_{11} & A_{12} \\ A_{21} & A_{22} \end{bmatrix}
\]
such that the partitions conform with the position of zeros and ones in $R_p$ and $R_{p \perp}$. Then: 
\begin{eqnarray}
	\bar{\Psi}_T(\cdot) &=& \bar{S}_{T,0h} \bar{S}_{T,hh}^{-1} \big[ I_{nq} - R_p(R_p' \bar{S}_{T,hh}^{-1}R_p)^{-1}R_p'\bar{S}_{T,hh}^{-1} \big] \label{eq:simplify.restrictions} \\
	&=& \bar{S}_{T,0h} \begin{bmatrix} A_{11} & A_{12} \\ A_{21} & A_{22} \end{bmatrix}
	\left( \begin{bmatrix} I_{11} & 0 \\ 0 & I_{22} \end{bmatrix}
	- \begin{bmatrix} 0 & 0 \\ 0 & A_{22}^{-1} \end{bmatrix} \begin{bmatrix} A_{11} & A_{12} \\ A_{21} & A_{22} \end{bmatrix} \right) \nonumber \\
	&=& \bar{S}_{T,0h} \begin{bmatrix} A_{11} & A_{12} \\ A_{21} & A_{22} \end{bmatrix}
	\left( \begin{bmatrix} I_{11} & 0 \\ 0 & I_{22} \end{bmatrix}
	- \begin{bmatrix} 0 & 0 \\A_{22}^{-1}A_{21} &  I_{22} \end{bmatrix} \right) \nonumber \\		
	&=& \bar{S}_{T,0h} \begin{bmatrix} A_{11} & A_{12} \\ A_{21} & A_{22} \end{bmatrix}
	\begin{bmatrix} I_{11} & 0 \\ -A_{22}^{-1}A_{21} & 0 \end{bmatrix} \nonumber \\
	&=& \bar{S}_{T,0h} \begin{bmatrix} A_{11} - A_{12}A_{22}^{-1}A_{21} & 0 \\ 0 &0 \end{bmatrix} \nonumber \\
	&=& \bar{S}_{T,0h} \begin{bmatrix} \left(  R_{p\perp}'\bar{S}_{T,hh} R_{p\perp} \right)^{-1} & 0 \\ 0 & 0 \end{bmatrix} \nonumber \\
	&=& \begin{bmatrix} R'_{p \perp} \bar{S}_{T,0h} R_{p \perp} & R'_{p \perp} \bar{S}_{T,0h} R_{p} \\ R'_{p} \bar{S}_{T,0h} R_{p \perp} & R'_{p } \bar{S}_{T,0h} R_{p} \end{bmatrix}  \begin{bmatrix} \left(  R_{p\perp}'\bar{S}_{T,hh} R_{p\perp} \right)^{-1} & 0 \\ 0 & 0 \end{bmatrix} \nonumber \\
	&=& \begin{bmatrix} R'_{p \perp} \bar{S}_{T,0h} R_{p \perp} 
		\left(  R_{p\perp}'\bar{S}_{T,hh} R_{p\perp} \right)^{-1} & 0 \\
		R'_{p} \bar{S}_{T,0h} R_{p \perp} 
		\left(  R_{p\perp}'\bar{S}_{T,hh} R_{p\perp} \right)^{-1} & 0 
	\end{bmatrix}. \nonumber 
\end{eqnarray}

\noindent (ii) Part (i) implies that the last $n(q-p)$ columns of the first $np$ rows of $\bar{\Phi}_T(mle,\lambda,p)$ are equal to zero as required. Now consider
\begin{eqnarray*}
	\lefteqn{R'_{p \perp} \bar{S}_{T,01} R_{p \perp} 
		\left(  R_{p\perp}'\bar{S}_{T,11} R_{p\perp} \right)^{-1} } \\
	&=& \left[ R'_{p \perp} S_{T,01} R_{p \perp} \left(R'_{p \perp} S_{T,11} R_{p \perp} \right)^{-1} \left(R'_{p \perp} S_{T,11} R_{p \perp} \right)  + \lambda T R'_{p \perp} \underline{\Phi}_T \underline{P}_\Phi R_{p \perp} \right] \left(  R_{p\perp}'\bar{S}_{T,11} R_{p\perp} \right)^{-1}.
\end{eqnarray*}
Notice that $R'_{p \perp} S_{T,01} R_{p \perp} \left(R'_{p \perp} S_{T,11} R_{p \perp} \right)^{-1}$ is the OLS estimator of a VAR($p$) written in $p$-companion form. It can be expressed as
\[
R'_{p \perp} S_{T,01} R_{p \perp} \left(R'_{p \perp} S_{T,11} R_{p \perp} \right)^{-1} = \begin{bmatrix} M' S_{T,01} R_{p \perp} \left(R'_{p \perp} S_{T,11} R_{p \perp} \right)^{-1} \\ \Upsilon_p \end{bmatrix} .
\]
The last $n(p-1)$ rows are equal to $\Upsilon_p$ because $y_{t-1}, \ldots, y_{t-p+1}$ lie in the space spanned by $y_{t-1}, \ldots, y_{t-p}$. 

For the (weighted) prior mean we obtain from (\ref{eq:prior.phi}):
\begin{eqnarray*}
	\lefteqn{R'_{p \perp} \underline{\Phi}_T \underline{P}_\Phi R_{p \perp}} \\
	&=&
	R'_{p \perp} \begin{bmatrix} \underline{\phi}_{1,T} \; \cdots \; \underline{\phi}_{p-1,T} \; \underline{\phi}_{p,T} & 0_{n \times n(q-p)}  \\
		\Upsilon_p & 0_{n(p-1) \times n(q-p)}  \\
		\cdot & \cdot \\
		\cdot & \cdot 
	\end{bmatrix} \begin{bmatrix} \underbrace{\underline{P}_{\Phi,11}}_{np \times np} & \underbrace{\underline{P}_{\Phi,12}}_{np \times n(q-p)} \\
		\underbrace{\underline{P}_{\Phi,21}}_{n(q-p) \times np} & \underbrace{\underline{P}_{\Phi,22}}_{n(q-p) \times n(q-p)} \end{bmatrix}
	R_{p \perp} \\
	&=&  \begin{bmatrix} \underline{\phi}_{1,T} \; \cdots \; \underline{\phi}_{p-1,T} \; \underline{\phi}_{p,T} \\ \Upsilon_p \end{bmatrix} \underline{P}_{\Phi,11}.
\end{eqnarray*} 
The result follows by noting that rows $n+1$ to $np$ of the posterior mean $R'_{p \perp}\bar{\Phi}_T(mle,\lambda,p)R_{p \perp}$, the weighted MLE, and the prior mean are identical, and they are equal to $\Upsilon_p$. $\blacksquare$

\noindent We now proceed with the proofs of the results in the main text and some additional derivations.

\noindent {\bf Proof of Lemma~\ref{lem:MFhRp}.} {\em (i) The ``if'' part} can be proved as follows. Because $p \ge p_*$ we can partition $F$ as follows: 
\be
F = \begin{bmatrix} \underbrace{F_{11,1}}_{np \times np} & \underbrace{0}_{np \times n(q-p)} \\
	\underbrace{F_{21,1}}_{n(q-p) \times np } & \underbrace{F_{22,1}}_{n(q-p) \times n(q-p)} \end{bmatrix}.
\ee 
The ``if'' part of the lemma follows if $F^h$ has the form
\be
F^h = \begin{bmatrix} \underbrace{F_{11,h}}_{np \times np} & \underbrace{0}_{np \times n(q-p)} \\
	\underbrace{F_{21,h}}_{n(q-p) \times np } & \underbrace{F_{22,h}}_{n(q-p) \times n(q-p)} \end{bmatrix},
\label{eq:Fh.partition}
\ee
which is true for $h=1$. For $h > 1$ we can use a proof by induction. Suppose that (\ref{eq:Fh.partition}) is true for $h$. Then
\be
F^{h+1}  =  F F^{h} 
= \begin{bmatrix} F_{11,1} & 0 \\ F_{21,1} & F_{22,1} \end{bmatrix}
\begin{bmatrix} F_{11,h} & 0 \\ F_{21,h} & F_{22,h} \end{bmatrix}
=
\begin{bmatrix} F_{11,1} F_{11,h}  & 0 \\ F_{21,1} F_{11,h} + F_{22,1} F_{21,h} & F_{22,1} F_{22,h} \end{bmatrix}.
\ee
Thus, the ``if'' part of the Lemma holds for any $h$. 

{\em (ii) The ``and only if'' part} can be proved as follows. Because $p < p_*$ we can partition $F$ as follows: 
\be
F = \begin{bmatrix} \underbrace{F_{11,1}}_{np \times np} & \underbrace{F_{12,1}}_{np \times n(p_*-p)}& \underbrace{0}_{np \times n(q-p_*)} \\
	\underbrace{F_{21,1}}_{n(p_*-p) \times np} & \underbrace{F_{22,1}}_{n(p_*-p) \times n(p_*-p)}& \underbrace{0}_{n(p_*-p) \times n(q-p_*)} \\
	\underbrace{F_{31,1}}_{n(q-p_*) \times np } & \underbrace{F_{32,1}}_{n(q-p_*) \times n(p_*-p)} & \underbrace{F_{33,1}}_{n(q-p_*) \times n(q-p_*)} \end{bmatrix}.
\ee 
The ``only if'' part of the lemma follows if $F^h$ has the form
\be
F^h = \begin{bmatrix} \underbrace{F_{11,h}}_{np \times np} & \underbrace{F_{12,h}}_{np \times n(p_*-p)}& \underbrace{0}_{np \times n(q-p_*)} \\
	\underbrace{F_{21,h}}_{n(p_*-p) \times np} & \underbrace{F_{22,h}}_{n(p_*-p) \times n(p_*-p)}& \underbrace{0}_{n(p_*-p) \times n(q-p_*)} \\
	\underbrace{F_{31,h}}_{n(q-p_*) \times np } & \underbrace{F_{32,h}}_{n(q-p_*) \times n(p_*-p)} & \underbrace{F_{33,h}}_{n(q-p_*) \times n(q-p_*)} \end{bmatrix}
\label{eq:Fh.partition.2},
\ee
where $F_{12,h} \not=0$. Note that this is true for $h=1$ because $p<p_*$. We proceed by induction. Suppose that (\ref{eq:Fh.partition.2}) is true for $h$. Then
\begin{eqnarray}
	F^{h+1}  &=&  F F^{h} 
	= \begin{bmatrix} F_{11,1} & F_{12,1} & 0 \\ F_{21,1} & F_{22,1} & 0 \\ F_{31,1} & F_{32,1} & F_{33,1} \end{bmatrix}
	\begin{bmatrix} F_{11,h} & F_{12,h} & 0 \\ F_{21,h} & F_{22,h} & 0 \\ F_{31,h} & F_{32,h} & F_{33,h} \end{bmatrix} \\
	&=&
	\begin{bmatrix} \cdot & F_{11,1}F_{12,h} + F_{12,1} F_{22,h}  & 0 \\ 
		\cdot & \cdot & 0 \\
		\cdot & \cdot & F_{33,1} F_{33,h} \end{bmatrix}.
\end{eqnarray}
Because $F_{11,1}$ and $F_{12,h}$ are both non-zero, the $(1,2)$ element of $F^{h+1}$ is non-zero. Thus, the ``only if'' part of the Lemma holds for any $h$. $\blacksquare$ 

\noindent {\bf Formulas for the Asymptotic Representation of Companion Form Posteriors.} Iterating the companion form DGP $h$ periods forward, we can write:
\[
S_{T,0h} = F^h S_{T,hh} + \alpha T^{-1/2}\left( \sum_{j=0}^{h-1} \sum_{t=1}^T F^jZ_{t-j}Y_{t-h}' \right)  
+ \left( \sum_{j=0}^{h-1} \sum_{t=1}^T F^j (M\epsilon_{t-j})Y_{t-h}' \right).
\]
It can now be shown that 
\begin{eqnarray*}
	\mu(lfe,\lambda,p)&=& \sum_{j=0}^{h-1}F^j\Gamma_{ZY,h-j}\bar{Q}_p, \quad \delta(lfe,\lambda,p) = \lambda\underline{\psi} \,\underline{P}_{\Psi} \bar{Q}_p, \\
	\zeta_T(lfe,\lambda,p) &=& \frac{1}{\sqrt{T}} \left(\sum_{j=0}^{h-1}\sum_{t=1}^TF^j(M \epsilon_{t-j})Y_{t-h}'\right) \bar{Q}_{p} \\
	\bar{Q}_p &=& \left(\Gamma_{YY,0}+\lambda\underline{P}_\Psi\right)^{-1}\left[I_{nq}-R_p(R_p'\left(\Gamma_{YY,0}+\lambda\underline{P}_\Psi\right)^{-1}R_p)^{-1}R_p'\left(\Gamma_{YY,0}+\lambda\underline{P}_\Psi\right)^{-1}\right].
\end{eqnarray*}
As for $p=1$, $\delta(\cdot)$ represents the bias induced by shrinkage, and $\zeta_T(\cdot)$ is a random variable that asymptotically has mean zero and determines the limit variance of the LP estimator. Finally, the term $\mu(\cdot)$ captures the misspecification bias. A similar calculation can be done for the MLE. 

In order to express the covariance formulas we need to make the dependence of $\bar{Q}_p$ on $(\iota,\lambda)$ explicit in the notation. We do so by writing $\bar{Q}_p(\iota,\lambda)$. For instance, for $p=q$ we obtain
\begin{eqnarray*}
	\Bar{Q}_{q}(\iota,\lambda)
	= \left\{ \begin{array}{ll} (\Gamma_{YY,0}+\lambda\underline{P}_\Psi)^{-1} & \mbox{if} \; \iota = lfe \\
		(\Gamma_{YY,0}+\lambda\underline{P}_\Phi)^{-1} & \mbox{if} \; \iota = mle \end{array} \right. \; .
\end{eqnarray*}

Following steps similar to those in the proof of Theorem \ref{thm:limitdis}, it can be shown that the covariance formulas for the companion form model are
\begin{eqnarray}
	V(mle,\lambda,p) &=&\sum_{i=0}^{h-1}\sum_{j=0}^{h-1}(F^i\Sigma_{EE}F^{j'})\otimes (F^{h-1-i'}\Bar{Q}_{p}'(mle,\lambda)\Gamma_{YY,0}\Bar{Q}_{p}(mle,\lambda) F^{h-1-j}) \nonumber  \\
	V(lfe,\lambda,p)&=& \sum_{i=0}^{h-1}\sum_{j=0}^{h-1}(F^i\Sigma_{EE}F^{j'})\otimes (\Bar{Q}_{p}'(lfe,\lambda)\Gamma_{YY,j-i}\Bar{Q}_{p}(lfe,\lambda)) \label{appeq:cov.companionform} \\
	Cov(lfe,0,q;lfe,\lambda,p) &=&\sum_{i=0}^{h-1}\sum_{j=0}^{h-1}(F^i\Sigma_{EE}F^{j'})\otimes (\Gamma_{YY,0}^{-1}\Gamma_{YY,j-i}\Bar{Q}_{p}(lfe,\lambda)) \nonumber \\
	Cov(lfe,0,q;mle,\lambda,p) &=&\sum_{i=0}^{h-1}\sum_{j=0}^{h-1}(F^i\Sigma_{EE}F^{j'})\otimes (\Gamma_{YY,0}^{-1}\Gamma_{YY,h-1-i}'\Bar{Q}_{p}(mle,\lambda) F^{h-1-j}) \nonumber. 	
\end{eqnarray}

More specifically, the covariance formulas can be derived as follows. Vectorizing $\zeta_T(\iota,\lambda,p)$ yields
\begin{eqnarray*}
	\vecr(\zeta_T(mle,\lambda,p))&=&\sum_{j=0}^{h-1}(F^j\otimes F^{h-1-j'}\bar{Q}_{p}(mle,\lambda)')\textup{vec}\left(T^{-1/2}\sum_{t=1}^TY_{t-1}E_t'\right)+o_p(1)\\
	\vecr(\zeta_T(lfe,\lambda,p))&=&\sum_{j=0}^{h-1}(F^j\otimes\bar{Q}_{p}'(lfe,\lambda))\textup{vec}\left(T^{-1/2}\sum_{t=1}^TY_{t-h+j}E_t'\right)+o_p(1).
\end{eqnarray*}
Based on Assumption~\ref{a:dgp}, the terms $$\textup{vec}\left(T^{-1/2}\sum_{t=1}^TY_{t-h+j}E_t'\right),\quad j=0,\dots,h-1,$$ jointly satisfy a central limit theorem for vector martingale difference sequences, with covariance matrix $\Sigma_{EE}\otimes\Gamma_{YY,j-i}.$ To see why, note that
\begin{eqnarray*}
	\lefteqn{ \EE\left[\textup{vec}\left(Y_{t-h+j}E_t'\right)\textup{vec}\left(Y_{t-h+i}E_t'\right)'\right]} \\
	&=_{(1)}&\EE\left[(E_t\otimes I_{nq})\textup{vec}(Y_{t-h+j})\textup{vec}(Y_{t-h+i})'(E_t'\otimes I_{nq})\right] \\
	&=_{(2)}&\EE\left[(E_t\otimes I_{nq})Y_{t-h+j}Y_{t-h+i}'(E_t'\otimes I_{nq})\right]\\
	&=_{(3)}&\EE\left[(M\epsilon_t\otimes I_{nq})Y_{t-h+j}Y_{t-h+i}'(\epsilon_t'M'\otimes I_{nq})\right]\\
	&=_{(4)}&\EE\left[(M\epsilon_t\epsilon_t'M'\otimes Y_{t-h+j}Y_{t-h+i}')\right]\\
	&=_{(5)}&\Sigma_{EE}\otimes\Gamma_{YY,j-i}+o(1).
\end{eqnarray*}
The first equality follows because $\textup{vec}(AB)=(B'\otimes I)\textup{vec}(A)$. The second and third equalities are by definition. The fourth equality follows by Lemma \ref{lmm:varcompanionaux_gen} above, by setting $\nu = M \epsilon_t$. The covariance formulas in (\ref{appeq:cov.companionform}) can now be generated by noting that $(A\otimes B)(C\otimes D)=(AC)\otimes (BD)$ and $\Bar{Q}_q(lfe,0) = \Gamma_{YY,0}^{-1}$. $\blacksquare$

\noindent {\bf Proof of Lemma~\ref{thm:mulfe.eq.muirf}.}  Write
\begin{eqnarray}
	\mu(lfe,0,p) &=& \plim_{T \longrightarrow \infty} \;  \left(\sum_{t=1}^T \left(\sum_{j=0}^{h-1} F^jZ_{t-j} \right)Y_{t-h}' \right)Q_{T,p}^{(h)} 
	\label{eq:A2.mulfe} 
\end{eqnarray}
where 
\[
Q_{T,p}^{(h)}=S_{T,hh}^{-1} \big[ I_{nq} - R_p(R_p' S_{T,hh}^{-1}R_p)^{-1}R_p'S_{T,hh}^{-1} \big].
\]
Thus, $\mu(lfe,0,p)$ can be interpreted as the probability limit of the least squares estimate obtained by regressing $\sum_{j=0}^{h-1} F^jZ_{t-j}$ onto $Y_{t-h}$ subject to the restriction that the coefficients on lags $p+1$ to $q$ are equal to zero. To facilitate the subsequent analysis, it is helpful to  define the exact VAR($p_*$) process
\be
Y_t^* = F Y_{t-1}^* + M \epsilon_t.
\label{eq:Ytstar}
\ee
The difference between $Y_t$ and $Y_t^*$ is that the latter excludes the $T^{-1/2}$ misspecification term. By construction, the $T \longrightarrow \infty$ limit autocovariances of the $Y_t$ process satisfy:
\be
\Gamma_{ZY^*,h} = \Gamma_{ZY,h} \quad \mbox{and} \quad
\Gamma_{YY^*,h} = \Gamma_{YY,h}.
\label{eq:Gamma.eq.Gammastar}
\ee
This justifies the replacement of  $Y_t$ by $Y_t^*$ in the subsequent analysis. 

The statement of the lemma concerns $M'\mu(lfe,0,p)M$, which corresponds to estimates of the coefficients that relate $M'\sum_{j=0}^{h-1} F^jZ_{t-j}$ to $y^*_{t-h}=M'Y_{t-h}^*$, after controlling for $y^*_{t-h-1},\ldots,y^*_{t-h-p+1}$. We now apply the FWL theorem to rewrite the coefficient estimates. The application of the FWL theorem involves regressing the right-hand-side variable $y_{t-h}^*$ on $y_{t-h-1}^*, \ldots, y_{t-h-p+1}^*$ and constructing residuals, which we denote by $\tilde{y}^*_{t-h}$. To represent the OLS estimator, we do not have to transform the left-hand-side variable $M'\sum_{j=0}^{h-1} F^jZ_{t-j}$ because the underlying projection is idempotent. Thus, we define
\[
\tilde{z}_{t}^* = \sum_{j=0}^{h-1} M' F^j Z_{t-j}
\]
and note that the probability limit of the least squares estimator takes the form
\be
M'\mu(lfe,0,p) M = \plim_{T \longrightarrow \infty} \;
\left( \frac{1}{T} \sum_{t=1}^T \tilde{z}_t^* \tilde{y}_{t-h}^{*'} \right) \left( \frac{1}{T} \sum_{t=1}^T \tilde{y}_{t-h}^* \tilde{y}_{t-h}^{*'}  \right)^{-1} = \Gamma_{\tilde{z} \tilde{y},h} \Gamma_{\tilde{y}\tilde{y},0}^{-1}.
\ee

Next, observe that $\tilde{y}^*_{t-h}$ is essentially the residual from a VAR($p-1$). Using the companion form notation, we can write
\be
\tilde{y}_{t-h}^* = M' \big( Y_{t-h}^* - \big( S_{T,01}^*(S_{T,11}^*)^{-1}Q_{T,p-1}^{*(1)} + O_p(T^{-1}) \big) Y_{t-h-1}^* \big). 
\ee
It is straightforward to verify using (\ref{eq:Ytstar}) and (\ref{eq:Gamma.eq.Gammastar}) that
\be
\Gamma_{\tilde{y} \tilde{y},0} = \plim_{T \rightarrow \infty} \; \frac{1}{T} \sum_{t=1}^T \tilde{y}_{t-h}^*\tilde{y}_{t-h}^{*'} \;   
= \Sigma_{\epsilon \epsilon} \quad  \text{ if and only if } p-1\geq p_*,
\ee
and that the moments and autocovariances of $\tilde{y}_{t}^*$ are asymptotically equivalent to the moments and autocovariances of the error terms $\epsilon_t$. 

\noindent {\em (i) ``If'' Part.} Recall that for $p>p_*$ $\tilde{y}_{t-h}^* \approx \epsilon_{t-h}$. Using the Ergodic Theorem, we deduce that
\[
\Gamma_{\tilde{z}\tilde{y},h} = \plim_{T \rightarrow \infty} \; \frac{1}{T} \sum_{t=1}^T \tilde{z}_t^* \tilde{y}_{t-h}^{*'} 
= \mathbb{E} \left[ \sum_{j=0}^{h-1} M'F^jZ_{t-j} \epsilon_{t-h}' \right].
\]
Plugging in the definition of $Z_{t-j}$, we obtain
\be
\Gamma_{\tilde{z}\tilde{y},h} 
=  \sum_{j=0}^{h-1} \sum_{l=1}^\infty  M'F^j M A_l \mathbb{E} \big[ \epsilon_{t-j-l} \epsilon_{t-h}' \big] .
\ee
Notice that the only terms with a non-zero expected value are the ones for which $j+l=h$. This leads to
\be
\Gamma_{\tilde{z}\tilde{y},h} 
=  \sum_{j=0}^{h-1}   M'F^j{M}A_{h-j}  \Sigma_{\epsilon \epsilon}, 
\ee
and, given that $M'M=\textup{I}_n$, we deduce that
\be
\Gamma_{\tilde{z}\tilde{y},h} \big( \Gamma_{\tilde{y} \tilde{y},0} \big)^{-1} = \sum_{j=0}^{h-1}   M'F^jMA_{h-j}=M'\left(\sum_{j=0}^{h-1} F^jMA_{h-j}M'\right)M =M'\mu_*(irf)M
\ee
as required.

\noindent {\em (ii) ``And Only If'' Part.} The equality between the LFE asymptotic bias and the IRF centering breaks down for $p-1<p_*$ because by projecting $y_{t-h}$ on fewer than $p_*$ lags in the application of the FWL theorem, the residuals are no longer asymptotically equivalent to $\epsilon_{t-h}$. Instead, because of omitted lags, they depend also on $\epsilon$s dated $t-h-1$ and earlier. Following the steps of the calculations for the ``if'' part, it is straightforward to see that the additional terms create a wedge between $M'\mu(lfe,0,p)M$ and $M'\mu_*(irf)M$. $\blacksquare$

\subsection{Proofs and Derivations for Section~\ref{sec:modeldetermination}}
\label{appsubsec:proofs.sec4}

\noindent {\bf Proof of Theorem~\ref{thm:ure} for $p=1$.} Using the asymptotic representation of $\Bar{\Psi}(\iota,\lambda)$ given
in Theorem~\ref{thm:limitdis}, the in-sample loss can be decomposed as follows
\begin{eqnarray*}
	\lefteqn{T \cdot  MSE(\iota,\lambda)}  \\
	&=&  \sum_{t=1}^T (y_t - F^hy_{t-h})(y_t - F^hy_{t-h})'  \nonumber \\
	&=& - T^{-1/2} \sum_{t=1}^T (y_ty_{t-h}' - F^hy_{t-h}y_{t-h}')
	(\delta(\iota,\lambda)+ \alpha \mu(\iota,\lambda) + \zeta_T(\iota,\lambda) + o_p(1) )' \nonumber \\
	&& - T^{-1/2} \sum_{t=1}^T (\delta(\iota,\lambda)+ \alpha \mu(\iota,\lambda) + \zeta_T(\iota,\lambda) + o_p(1) )
	(y_t y_{t-h}' - F^hy_{t-h} y_{t-h}')'  \nonumber  \\
	&& +  ( \delta(\iota,\lambda)+\alpha \mu(\iota,\lambda) + \zeta_T(\iota,\lambda) + o_p(1) )
	\left( T^{-1} \sum_{t=1}^T y_{t-h}y_{t-h}' \right) \nonumber \\
	&& \times (\delta(\iota,\lambda)+ \alpha \mu(\iota,\lambda) + \zeta_T(\iota,\lambda) + o_p(1) )' \nonumber. 
\end{eqnarray*}
From the definition of $\zeta_T(lfe,\lambda)$, it follows that
\begin{eqnarray*}
	\lefteqn{ T^{-1/2} \sum_{t=1}^T (y_ty_{t-h}' - F^hy_{t-h}y_{t-h}')}\\
	&=&\alpha \sum_{j=0}^{h-1}\left(T^{-1}\sum_{t=1}^TF^jz_{t-j}y_{t-h}'\right)+\sum_{j=0}^{h-1}\left(F^jT^{-1/2}\sum_{t=1}^T\epsilon_{t-j}y_{t-h}'\right)
	\\
	&=&\left[\zeta_T(lfe,\lambda)+\alpha\mu(lfe,\lambda)+o_p(1)\right](T\Bar{P}_\Psi^{-1})^{-1}
\end{eqnarray*} 
for any $\lambda\geq0$. Without loss of generality, take $\lambda=0$, whence $$T^{-1/2} \sum_{t=1}^T (y_ty_{t-h}' - F^hy_{t-h}y_{t-h}')=\left[\zeta_T(lfe,0)+\alpha\mu_*(prd)\right]T^{-1}S_{T,hh}.$$
Therefore,
\begin{eqnarray*}
	\lefteqn{T \cdot  \tr\left\{W\cdot MSE(\iota,\lambda)\right\} }  \\
	&=&  \tr\left\{W\sum_{t=1}^T (y_t - F^hy_{t-h})(y_t - F^hy_{t-h})'\right\}  \\
	&& -2\tr\left\{W \left[\zeta_T(lfe,0)+\alpha\mu_*(prd)\right](T^{-1}S_{T,hh})
	\left[\delta(\iota,\lambda)+ \alpha \mu(\iota,\lambda) + \zeta_T(\iota,\lambda) + o_p(1) \right]'\right\} \\
	&&+  \tr\bigg\{W\left[ \delta(\iota,\lambda)+\alpha \mu(\iota,\lambda) + \zeta_T(\iota,\lambda) + o_p(1) \right]
	\bigg( T^{-1}S_{T,hh} \bigg) \\
	&& \times 
	\left[\delta(\iota,\lambda)+ \alpha \mu(\iota,\lambda) + \zeta_T(\iota,\lambda) + o_p(1) \right]'\bigg\}.
\end{eqnarray*}
Observe that $T^{-1}S_{T,hh}=\Gamma_{yy,0}+o_p(1)$, hence
\begin{eqnarray*}
	\lefteqn{ T \bigg( \tr\left\{W\cdot MSE(\iota,\lambda)\right\}-\tr\left\{W\cdot MSE(lfe,0)\right\} \bigg) } \\
	&=&-2\tr\left\{W \left[\zeta_T(lfe,0)+\alpha\mu_*(prd)\right]\Gamma_{yy,0}
	\left[\delta(\iota,\lambda)+ \alpha \mu(\iota,\lambda) + \zeta_T(\iota,\lambda)  \right]'\right\} \\
	&&+  \tr\left\{W\left[ \delta(\iota,\lambda)+\alpha \mu(\iota,\lambda) + \zeta_T(\iota,\lambda) \right]
	\Gamma_{yy,0}
	\left[\delta(\iota,\lambda)+ \alpha \mu(\iota,\lambda) + \zeta_T(\iota,\lambda)  \right]'\right\}\\
	&&+\tr\left\{W \left[\zeta_T(lfe,0)+\alpha\mu_*(prd)\right]\Gamma_{yy,0}
	\left[ \alpha \mu_*(prd) + \zeta_T(lfe,0) \right]'\right\} +o_p(1).
\end{eqnarray*}
Statement (i) now follows from Theorem \ref{thm:limitdis}, the Continuous Mapping Theorem and a straightforward rearrangement of terms.

For statement (ii), from part (i) and uniform integrability of the in-sample loss differential it is easy to see that
\begin{eqnarray*}
	\lefteqn{\mathbb{E}\left[\Delta_{L,T} (\iota,\lambda)\right]} \\
	&\longrightarrow&  \mathbb{E}\left[\left\Vert\delta(\iota,\lambda)+\alpha \mu(\iota,\lambda) + \zeta(\iota,\lambda)\right\Vert_{W\otimes\Gamma_{yy,0}}^2\right]+\mathbb{E}\left[\left\Vert\alpha\mu_*(prd)+\zeta(lfe,0)\right\Vert_{W\otimes\Gamma_{yy,0}}^2\right]\\
	&&-2\mathbb{E}\left[\tr\left\{W \left[\alpha\mu_*(prd)+\zeta(lfe,0)\right]\Gamma_{yy,0}
	\left[\delta(\iota,\lambda)+ \alpha \mu(\iota,\lambda) + \zeta(\iota,\lambda)  \right]'\right\}\right].
\end{eqnarray*}
Working out the expected values according to Theorem \ref{thm:limitdis} yields
\begin{eqnarray*}
	\mathbb{E}\left[\Delta_{L,T} (\iota,\lambda)\right]&\longrightarrow& 
	\left\Vert \delta(\iota,\lambda)+\alpha \mu(\iota,\lambda)\right\Vert_{W\otimes\Gamma_{yy,0}}^2+tr\left\{(W\otimes\Gamma_{yy,0})V(\iota,\lambda)\right\}\\
	&&+\alpha^2\left\Vert\mu_*(prd)\right\Vert_{W\otimes\Gamma_{yy,0}}^2+\tr\left\{(W\otimes\Gamma_{yy,0})V(lfe,0)\right\}\\
	&&-2\alpha \tr\left\{W \mu_*(prd)\Gamma_{yy,0}
	\left[\delta(\iota,\lambda)+ \alpha \mu(\iota,\lambda)  \right]'\right\}\\
	&&-2 \tr\left\{(W\otimes\Gamma_{yy,0})Cov(lfe,0;\iota,\lambda)\right\}   .                      
\end{eqnarray*}
Using the definitions of $\bar{\cal R}_B(\iota,\lambda)$ and $\bar{\cal R}_V(\iota,\lambda)$ in Theorem~\ref{thm:prediction.risk} and recognizing that $\bar{\cal R}_B(lfe,0)=0$ we can write the r.h.s. as
\begin{eqnarray*}	
	\mbox{r.h.s} &=& 	\left\Vert \delta(\iota,\lambda)+\alpha \mu(\iota,\lambda)\right\Vert_{W\otimes\Gamma_{yy,0}}^2
	+\alpha^2\left\Vert\mu_*(prd)\right\Vert_{W\otimes\Gamma_{yy,0}}^2\\
	&&-2\alpha \tr\left\{W \mu_*(prd)\Gamma_{yy,0}
	\left[\delta(\iota,\lambda)+ \alpha \mu(\iota,\lambda)  \right]'\right\}\\
	&&+\bar{\cal R}_V(\iota,\lambda) +\bar{\cal R}_V(lfe,0) - 2 \tr\left\{(W\otimes\Gamma_{yy,0})Cov(lfe,0;\iota,\lambda)\right\}  \\
	&=& \bar{\cal R}_B(\iota,\lambda) +\bar{\cal R}_V(\iota,\lambda) - \big(\bar{\cal R}_B(lfe,0) +\bar{\cal R}_V(lfe,0) \big) \\
	&& + 2\bar{\cal R}_V(lfe,0) - 2 \tr\left\{(W\otimes\Gamma_{yy,0})Cov(lfe,0;\iota,\lambda)\right\}. \quad \blacksquare
\end{eqnarray*}

\noindent {\bf Proof of Theorem~\ref{thm:ure} for general $p$.} The argument parallels the $p=1$ case, using the $q$-companion form of Section~\ref{sec:VARp}; recall that predictions of $y_t = M'Y_t$ employ all coefficients $M'\bar{\Psi}_T(\iota,\lambda,p)$. Because $p_* \le p \le q$, Lemma~\ref{lem:MFhRp} gives $M'F^hR_p = 0$, so premultiplying the companion-form expansion~(\ref{eq:LFE.limit}) (and its MLE analogue) by $M'$ removes the $O_p(\sqrt{T})$ restriction term and yields
\be
\sqrt{T}\, M'\big( \bar{\Psi}_T(\iota,\lambda,p) - F^h \big) = M'\big[ \delta(\iota,\lambda,p) + \alpha \mu(\iota,\lambda,p) + \zeta_T(\iota,\lambda,p) \big] + o_p(1).
\label{eq:ure.proof.Mexpansion}
\ee
Writing the in-sample forecast error as 
\[
y_t - M'\bar{\Psi}_T(\iota,\lambda,p)Y_{t-h} = M'(Y_t - F^h Y_{t-h}) - M'(\bar{\Psi}_T(\iota,\lambda,p)-F^h)Y_{t-h}
\] 
and using~(\ref{eq:ure.proof.Mexpansion}), the in-sample loss decomposes as
\begin{eqnarray*}
	\lefteqn{T \cdot MSE(\iota,\lambda,p)}\\
	&=& \sum_{t=1}^T M'(Y_t - F^hY_{t-h})(Y_t - F^hY_{t-h})'M \\
	&& - T^{-1/2} M'\big(S_{T,0h}-F^h S_{T,hh}\big)\big[\delta(\iota,\lambda,p)+\alpha\mu(\iota,\lambda,p)+\zeta_T(\iota,\lambda,p)+o_p(1)\big]'M\\
	&& - T^{-1/2} M'\big[\delta(\iota,\lambda,p)+\alpha\mu(\iota,\lambda,p)+\zeta_T(\iota,\lambda,p)+o_p(1)\big]\big(S_{T,0h}-F^h S_{T,hh}\big)'M\\
	&& + M'\big[\delta(\iota,\lambda,p)+\alpha\mu(\iota,\lambda,p)+\zeta_T(\iota,\lambda,p)+o_p(1)\big]\big(T^{-1}S_{T,hh}\big)\\
	&& \phantom{+ M'} \times \big[\delta(\iota,\lambda,p)+\alpha\mu(\iota,\lambda,p)+\zeta_T(\iota,\lambda,p)+o_p(1)\big]'M,
\end{eqnarray*}
where $S_{T,0h}=\sum_{t=1}^T Y_t Y_{t-h}'$ and $S_{T,hh}=\sum_{t=1}^T Y_{t-h}Y_{t-h}'$. Iterating the companion-form DGP~(\ref{eq:dgp.companionform}) $h$ periods forward, the unrestricted ($q$-lag) benchmark satisfies
\[
T^{-1/2}\big(S_{T,0h}-F^h S_{T,hh}\big) = \big[\zeta_T(lfe,0,q)+\alpha\mu_*(prd,q)\big]\big(T^{-1}S_{T,hh}\big)+o_p(1),
\]
where we used $\mu(lfe,0,q)=\mu_*(prd,q)$. Substituting this identity, taking $\tr\{W \cdot\}$, subtracting the benchmark $\tr\{W\cdot MSE(lfe,0,q)\}$ (which cancels the common first term), and replacing $T^{-1}S_{T,hh}=\Gamma_{YY,0}+o_p(1)$ gives
\begin{eqnarray*}
	\lefteqn{ T\big( \tr\{W\cdot MSE(\iota,\lambda,p)\} - \tr\{W\cdot MSE(lfe,0,q)\}\big) }\\
	&=& -2\, \tr\Big\{ W M'\big[\zeta_T(lfe,0,q)+\alpha\mu_*(prd,q)\big]\Gamma_{YY,0}\big[\delta(\iota,\lambda,p)+\alpha\mu(\iota,\lambda,p)+\zeta_T(\iota,\lambda,p)\big]'M \Big\}\\
	&& +\, \tr\Big\{ W M'\big[\delta(\iota,\lambda,p)+\alpha\mu(\iota,\lambda,p)+\zeta_T(\iota,\lambda,p)\big]\Gamma_{YY,0}\big[\delta(\iota,\lambda,p)+\alpha\mu(\iota,\lambda,p)+\zeta_T(\iota,\lambda,p)\big]'M \Big\}\\
	&& +\, \tr\Big\{ W M'\big[\zeta_T(lfe,0,q)+\alpha\mu_*(prd,q)\big]\Gamma_{YY,0}\big[\alpha\mu_*(prd,q)+\zeta_T(lfe,0,q)\big]'M \Big\} + o_p(1).
\end{eqnarray*}
Because $W$ and $\Gamma_{YY,0}$ are symmetric, collecting the three traces shows the right-hand side equals
\[
\Big\| M'\big[ \delta(\iota,\lambda,p) - \alpha\big(\mu_*(prd,q)-\mu(\iota,\lambda,p)\big) - \big(\zeta_T(lfe,0,q)-\zeta_T(\iota,\lambda,p)\big)\big] \Big\|_{W\otimes\Gamma_{YY,0}}^2 + o_p(1).
\]
Statement~(i) now follows from~(\ref{eq:LFE.limit}), the Continuous Mapping Theorem, and expanding the square, using $\zeta_T(\cdot)\Longrightarrow\zeta(\cdot)$.

For statement~(ii), part~(i) together with the uniform integrability of the in-sample loss differential (guaranteed by Assumption~\ref{a:dgp}(iv)--(v), as in the $p=1$ case) implies
\begin{eqnarray*}
	\lefteqn{\mathbb{E}\big[\Delta_{L,T}(\iota,\lambda,p)\big]}\\
	&\longrightarrow& \mathbb{E}\Big[\big\| M'\big(\delta(\iota,\lambda,p)+\alpha\mu(\iota,\lambda,p)+\zeta(\iota,\lambda,p)\big)\big\|_{W\otimes\Gamma_{YY,0}}^2\Big] \\
	&& +\mathbb{E}\Big[\big\| M'\big(\alpha\mu_*(prd,q)+\zeta(lfe,0,q)\big)\big\|_{W\otimes\Gamma_{YY,0}}^2\Big]\\
	&& -2\,\mathbb{E}\Big[ \tr\big\{ W M'\big(\alpha\mu_*(prd,q)+\zeta(lfe,0,q)\big)\Gamma_{YY,0}\big(\delta(\iota,\lambda,p)+\alpha\mu(\iota,\lambda,p)+\zeta(\iota,\lambda,p)\big)'M \big\}\Big].
\end{eqnarray*}
Working out the expectations with $\mathbb{E}[\zeta(\cdot)]=0$, $\mathbb{E}[\vecr(\zeta(\iota,\lambda,p))\vecr(\zeta(\iota,\lambda,p))']=V(\iota,\lambda,p)$, and $Cov(lfe,0,q;\iota,\lambda,p)$ denoting the corresponding covariance between $\zeta(lfe,0,q)$ and $\zeta(\iota,\lambda,p)$, yields
\begin{eqnarray}
	\mathbb{E}\big[\Delta_{L,T}(\iota,\lambda,p)\big] &\longrightarrow&
	\big\| M'\big(\delta(\iota,\lambda,p)+\alpha\mu(\iota,\lambda,p)\big)\big\|_{W\otimes\Gamma_{YY,0}}^2 \nonumber \\
	&& + \tr\big\{\big((MWM')\otimes\Gamma_{YY,0}\big)V(\iota,\lambda,p)\big\} \nonumber \\
	&& +\alpha^2\big\| M'\mu_*(prd,q)\big\|_{W\otimes\Gamma_{YY,0}}^2 \nonumber \\
	&& + \tr\big\{\big((MWM')\otimes\Gamma_{YY,0}\big)V(lfe,0,q)\big\} \nonumber \\
	&& -2\alpha\, \tr\big\{ W M'\mu_*(prd,q)\Gamma_{YY,0}\big[\delta(\iota,\lambda,p)+\alpha\mu(\iota,\lambda,p)\big]'M\big\} \nonumber \\
	&& -2\, \tr\big\{\big((MWM')\otimes\Gamma_{YY,0}\big)Cov(lfe,0,q;\iota,\lambda,p)\big\}. \label{eq:ure.proof.Elimit}
\end{eqnarray}
The generalizations of the risk components in Theorem~\ref{thm:prediction.risk} to the $q$-companion form are
\begin{eqnarray*}
\bar{\cal R}_B(\iota,\lambda,p) &=& \big\| M'\big(\delta(\iota,\lambda,p)+\alpha(\mu(\iota,\lambda,p)-\mu_*(prd,q))\big)\big\|_{W\otimes\Gamma_{YY,0}}^2, \quad \\
\bar{\cal R}_V(\iota,\lambda,p) &=& \tr\big\{\big((MWM')\otimes\Gamma_{YY,0}\big)V(\iota,\lambda,p)\big\}.
\end{eqnarray*}
The first, third, and fifth terms on the right-hand side of~(\ref{eq:ure.proof.Elimit}) combine into $\bar{\cal R}_B(\iota,\lambda,p)$. Recognizing that $\delta(lfe,0,q)=0$ and $\mu(lfe,0,q)=\mu_*(prd,q)$ imply $\bar{\cal R}_B(lfe,0,q)=0$, we obtain
\begin{eqnarray*}
	\mathbb{E}\big[\Delta_{L,T}(\iota,\lambda,p)\big] &\longrightarrow&
	\bar{\cal R}_B(\iota,\lambda,p)+\bar{\cal R}_V(\iota,\lambda,p) - \big(\bar{\cal R}_B(lfe,0,q)+\bar{\cal R}_V(lfe,0,q)\big)\\
	&& + 2\bar{\cal R}_V(lfe,0,q) - 2\, \tr\big\{\big((MWM')\otimes\Gamma_{YY,0}\big)Cov(lfe,0,q;\iota,\lambda,p)\big\},
\end{eqnarray*}
which is statement~(ii). $\blacksquare$

\noindent {\bf Marginal Data Density.} To obtain the MDD formula we use the following inversion of Bayes Theorem:
\[
   p(\theta|Y) = \frac{ p(Y|\theta) p(\theta) }{p(Y)} \quad \Leftrightarrow \quad p(Y) = \frac{ p(Y|\theta) p(\theta) }{p(\theta|Y)}.
\]
Because the densities and normalization constants of $p(Y|\theta)$, $p(\theta)$ and $p(\theta|Y)$ in a linear Gaussian regression model with conjugate matrix-normal inverse Wishart prior are known, it is straightforward to show that 
\be
p(Y|\lambda) 
= (2 \pi)^{-n T/2} \frac{|\lambda T \underline{P}_\Phi|^{n/2}}{|\bar{P}_\Phi|^{n/2}} 
\frac{\underline{C}_{IW}}{\bar{C}_{IW}} ,
\ee
where 
\[
\frac{\underline{C}_{IW}}{\bar{C}_{IW}} =
\frac{|\underline{S}|^{\underline{\nu}/2}}{|\bar{S}|^{\bar{\nu}/2}}
\frac{2^{n \bar{\nu}/2} \prod_{i=1}^{n} \Gamma((\bar{\nu}+1-i)/2) }{2^{n \underline{\nu}/2} \prod_{i=1}^{n}\Gamma((\underline{\nu}+1-i)/2)}.
\] 
We included $(\lambda,p)$ as a conditioning argument for the MDD. The hyperparameter enters the formula directly and indirectly through $\bar{S}$, $\bar{\Phi}_T$, and $\bar{P}_{\Phi}$, while the lag length $p$ enters through the number of lags stacked in $X$. 

\section{Heteroskedasticity}
\label{appsec:heterosk}

To introduce heteroskedasticity, Assumption~\ref{a:dgp}(iii) could be replaced by 

\begin{assumption}\label{ass:heterosk}
	The innovation process $\{\epsilon_t\}$ takes the form $\epsilon_t = \Sigma_{t,\epsilon \epsilon}^{1/2} \eta_t$, where:
	\begin{enumerate}
		\item[(a)] $\{\eta_t\}$ is an i.i.d. sequence with $\mathbb{E}[\eta_t] = 0$, $\EE[\eta_t \eta_t'] = I_n$, and bounded fourth moments.
		\item[(b)] The volatility matrix sequence $\{\Sigma_{t,\epsilon \epsilon} \}$ is strictly stationary, ergodic, positive definite almost surely, and independent of the standardized innovations $\{\eta_t\}$ and the initial conditions, with $\mathbb{E}[\Vert\Sigma_{t,\epsilon \epsilon} \Vert]<\infty$.
	\end{enumerate}
\end{assumption}

Here we assume that the volatility process $\Sigma_{t,\epsilon \epsilon}$ evolves independently of the innovation $\eta_t$. This includes volatility paths such as structural breaks, regime switching, and stochastic volatility. Under Assumption~\ref{ass:heterosk}, the structure of the asymptotic variances is preserved. The stationarity and ergodicity of $\{ \Sigma_{t,\epsilon \epsilon} \}$ ensures that
\[
   \frac{1}{T} \sum_{t=1}^T \Sigma_{t,\epsilon \epsilon} \overset{p}{\longrightarrow} \mathbb{E}[\Sigma_{t,\epsilon \epsilon}]=:\bar{\Sigma}_{\epsilon \epsilon}.
\]
Our derivations exploit the limits of objects such as
\be 
\label{eq:termslim}
\frac{1}{T} \sum_{t=1}^T y_t y_{t-h}', \quad \frac{1}{T} \sum_{t=1}^T z_{t-j} y_{t-h}', \quad \frac{1}{\sqrt{T}} \sum_{t=1}^T \epsilon_{t} y'_{t-h+j}.
\ee
Under Assumption \ref{ass:heterosk}, the probability limits of these three objects are similar to those in the homoskedastic case, ensuring the proofs in Appendix~\ref{appsubsec:proofs.sec2} (specifically Proof of Theorem~\ref{thm:limitdis}) go through barely unchanged. Here is a breakdown for some of the important terms:

\begin{itemize}
\item The leading term in the first object in \eqref{eq:termslim} converges to $$\EE[y_ty_{t-h}']=\sum_{j=0}^{\infty}F^{h+j}\EE[\epsilon_{t-j-h}\epsilon_{t-j-h}']F^{j'},$$ and due to Assumption \ref{ass:heterosk}, $\EE[\epsilon_t \epsilon_t'] =\EE[\EE[\epsilon_t \epsilon_t' | \Sigma_{t,\epsilon \epsilon}]] = \EE[\Sigma_{t,\epsilon \epsilon}] = \bar{\Sigma}_{\epsilon \epsilon}$. Therefore, the change from the main text is as follows (see, e.g., Proof of Theorem~\ref{thm:limitdis}): $$\EE[y_ty_{t-h}']=\sum_{j=0}^{\infty}F^{h+j}\Sigma_{\epsilon\epsilon} F^{j'} \mapsto \EE[y_ty_{t-h}']=\sum_{j=0}^{\infty}F^{h+j}\Bar{\Sigma}_{\epsilon\epsilon} F^{j'}. $$

\item Similarly, we can analyze the second term in \eqref{eq:termslim}. Recall that $z_t$ captures the infinite MA term, so that the product $z_{t-j}y_{t-h}'$ involves cross-products of $\epsilon$. Because $\{\epsilon_t\}$ is a martingale difference sequence with respect to the volatility process, cross-terms $\EE[\epsilon_t \epsilon_{t-k}']$ are zero for $k \neq 0$. For matching indices, we again get $\EE[\epsilon_t \epsilon_t'] = \bar{\Sigma}_{\epsilon \epsilon}$. Therefore, the change from the main text is as follows (see, e.g., Proof of Theorem~\ref{thm:limitdis}): $$\Gamma_{zy,h}=\sum_{j=0}^{\infty}A_{h+j}\Sigma_{\epsilon\epsilon}F^{j'} \mapsto\Gamma_{zy,h}=\sum_{j=0}^{\infty}A_{h+j}\Bar{\Sigma}_{\epsilon \epsilon}F^{j'}. $$

\item The asymptotic theory relies on the variance of the third term in \eqref{eq:termslim}. By the CLT for martingale difference sequences, the asymptotic variance is given by $$
\EE[ \Sigma_{t,\epsilon \epsilon} \otimes y_{t-h+j} y_{t-h+i}' ].$$ Therefore, the change from the main text is as follows (see, e.g., the covariance formulas in Appendix~\ref{appsubsec:proofs.sec2}): $$\Sigma_{\epsilon\epsilon} \otimes \Gamma_{yy,j-i} \mapsto \EE[ \Sigma_{t,\epsilon \epsilon} \otimes y_{t-h+j} y_{t-h+i}' ],$$ where $\EE[ \Sigma_{t,\epsilon \epsilon} \otimes y_{t-h+j} y_{t-h+i}' ]$ is obtained in practice by $\frac{1}{T}\sum_{t=1}^T\hat{\epsilon}_t\hat{\epsilon}_t'\otimes y_{t-h+j} y_{t-h+i}'$. This leads to, for instance,
\begin{align*}
	V(mle,\lambda)&=\sum_{i=0}^{h-1}\sum_{j=0}^{h-1}\big(F^i\otimes F^{h-1-i'}\Bar{Q}(mle,\lambda)'\big)\EE[ \Sigma_{t,\epsilon \epsilon} \otimes y_{t-1} y_{t-1}' ] \big(F^{j'}\otimes \Bar{Q}(mle,\lambda)F^{h-1-j} \big)\\
	V(lfe,\lambda)&=\sum_{i=0}^{h-1}\sum_{j=0}^{h-1}(F^i\otimes \Bar{Q}(lfe,\lambda)')\EE[ \Sigma_{t,\epsilon \epsilon} \otimes y_{t-h+j} y_{t-h+i}' ](F^{j'}\otimes \Bar{Q}(lfe,\lambda)),
\end{align*}
where
\[
	\Bar{Q}(\iota,\lambda)
= \left\{ \begin{array}{ll} (\Gamma_{YY,0}+\lambda\underline{P}_\Psi)^{-1} & \mbox{if} \; \iota = lfe \\
	(\Gamma_{YY,0}+\lambda\underline{P}_\Phi)^{-1} & \mbox{if} \; \iota = mle \end{array} \right. \; .
\]
The covariance formulas can be adjusted in a similar manner. The objects no longer factor out as simple Kronecker products, but they still have analogous forms. Thus, adjusting the covariance formulas in the PC and IRFC criteria will be the only relevant change in practice.
\end{itemize} 

We deduce that the results in the main text still go through under some forms of heteroskedasticity with almost no changes in the structure of the proofs or implementation. Under the relaxed Assumption \ref{ass:heterosk}, the standard estimators used in the paper's criteria remain consistent for the unconditional risk objects (since $\frac{1}{T}\sum \hat{\epsilon}_t \hat{\epsilon}_t' \stackrel{p}{\longrightarrow} \bar{\Sigma}_{\epsilon \epsilon}$).

\clearpage
\section{Data for Monte Carlo DGP and Empirical Analysis}
\label{appsec:data}

This section documents the construction of the macroeconomic panel used in the Monte Carlo DGP calibration for Sections~\ref{sec:numerics} and~\ref{sec:MC} and in the empirical application of Section~\ref{sec:empirics}. The raw data are taken from a single archived vintage of FRED-QD; the series are transformed, screened for a balanced panel, and detrended with the regression filter of \citet{Hamilton2018}.

\subsection{Data Source}
\label{appsubsec:data.source}

The raw data are the FRED-QD database of \citet{McCrackenNg2020}, a quarterly companion to FRED-MD maintained by the Federal Reserve Bank of St.~Louis. We use the 2023-08 vintage throughout, taken from the historical vintage archive rather than the continuously updated ``current'' file, so that the input is fixed and reproducible. The file can be recovered by downloading \emph{Historical Vintages of FRED-QD} from the FRED-MD/FRED-QD page maintained by Michael McCracken\footnote{\texttt{https://www.stlouisfed.org/research/economists/mccracken/fred-databases}} and extracting \texttt{FRED-QD\_2023m08.csv}. The vintage contains $245$ quarterly series spanning 1959:Q1--2023:Q2 ($258$ quarters). It follows the standard FRED-QD layout: a header row of mnemonics, a \texttt{factors} row flagging membership in the St.~Louis Fed's factor model (not used here), a \texttt{transform} row giving the \citet{McCrackenNg2020} transformation code for each series, and then the observations.

\subsection{Balanced-Panel Construction}
\label{appsubsec:data.samples}

We work with two samples, and screen each for a balanced panel: any series with a missing observation anywhere in the relevant window is dropped, and no interpolation or backcasting is performed. The screen is applied to the raw series, before any transformation. Table~\ref{apptab:samples} summarizes key features of the data sets. The shorter sample retains more series ($214$ versus $189$): dropping the 1959 observations and the post-2019 period removes the windows in which many series are unavailable, so fewer series are lost to the screen. The filtered sample begins $h+p-1 = 11$ quarters after the raw sample for the reason given in Section~\ref{appsubsec:data.hamilton}.

\begin{table}[t!]
	\centering
	\caption{Sample Construction}
	\label{apptab:samples}
	\begin{tabular}{lcc}
		\hline\hline
		& Empirical Application & Monte Carlo Calibration \\
		& (Section~\ref{sec:empirics}) & (Sections \ref{sec:numerics} and \ref{sec:MC}) \\
		\hline
		Raw sample & 1959:Q1--2023:Q2 & 1960:Q1--2019:Q4 \\
		Raw observations & 258 & 240 \\
		Series available & 245 & 245 \\
		Series after balanced-panel screen & 189 & 214 \\
		Filtered sample & 1961:Q4--2023:Q2 & 1962:Q4--2019:Q4 \\
		Filtered observations & 247 & 229 \\
		\hline
	\end{tabular}
\end{table}

Each series is first put into levels or log levels according to its FRED-QD
transformation code. We depart from the FRED-QD convention in one important
respect: we do not difference. Codes 2, 3, 5 and 6 all call for
differencing, but here the trend is removed by the Hamilton regression instead, which is the point of that filter. The mapping we apply is

\begin{table}[t!]
	\centering
	\caption{Transformation Codes and the Transformations Applied}
	\label{apptab:transform}
	\begin{tabular}{clcccc}
		\hline\hline
		& & & & \multicolumn{2}{c}{Series retained} \\
		Code & FRED-QD definition & Transformation applied & Raw & Sec.~7 & Sec.~5--6 \\
		\hline
		1 & $x_t$ & none (levels) & 22 & 14 & 18 \\
		2 & $\Delta x_t$ & none (levels) & 32 & 24 & 27 \\
		3 & $\Delta^2 x_t$ & none (levels) & 0 & 0 & 0 \\
		4 & $\log x_t$ & $z_t = 100 \log x_t$ & 0 & 0 & 0 \\
		5 & $\Delta \log x_t$ & $z_t = 100 \log x_t$ & 140 & 101 & 119 \\
		6 & $\Delta^2 \log x_t$ & $z_t = 100 \log x_t$ & 50 & 49 & 49 \\
		7 & $\Delta (x_t / x_{t-1} - 1)$ & $z_t = x_t / x_{t-1}$ & 1 & 1 & 1 \\
		\hline
		& & Total & 245 & 189 & 214 \\
		\hline
	\end{tabular}
\end{table}

\subsection{The Hamilton Filter}
\label{appsubsec:data.hamilton}

The cyclical component of each series is the residual of the forecasting
regression proposed by \citet{Hamilton2018}. For each series $z_t$ we estimate by OLS
\begin{equation}
	\label{eq:hamilton}
	z_{t+h} \;=\; \beta_0 + \beta_1 z_t + \beta_2 z_{t-1} + \beta_3 z_{t-2} + \beta_4 z_{t-3} + v_{t+h},
\end{equation}
and take the residual $\hat v_{t+h}$ as the detrended series. We use
\citeauthor{Hamilton2018}'s recommended settings for quarterly data,
\begin{equation*}
	h = 8 \text{ quarters} \qquad \text{and} \qquad p = 4 \text{ lags}.
\end{equation*}
To characterize what the filter does to the persistence of the panel, we
estimate for every series $i$, before and after filtering, the AR(1) regression
\begin{equation}
	\label{eq:ar1}
	y_{it} \;=\; \alpha_{i1} + \alpha_{i2} y_{it-1} + u_{it}
\end{equation}
by OLS, and record $\hat\alpha_{i2}$. ``Before'' refers to the transformed series $z_t$ -- that is, after the level or log-level transformation but before the Hamilton regression -- so the two estimates are computed on series in identical units and are directly comparable. Table~\ref{apptab:ar1} shows that the filter induces a shift of the $\hat{\alpha}_{i2}$ distribution away from the unit root, while maintaining substantial persistence and hence predictability of the series.

\begin{table}[t!]
	\caption{\sc AR(1) Persistence Before and After Hamilton Filtering}
	\label{apptab:ar1}
	\begin{center} 
	\begin{tabular}{lccccc}
		\hline\hline
		& Min & $q_{25}$ & Median & $q_{75}$ & Max \\
		\hline
		\multicolumn{6}{c}{Empirical sample, 1961:Q4--2023:Q2 (189 series)} \\
		\hline
		$\hat\alpha_2$, before filtering & $-0.170$ & 0.982 & 0.994 & 0.997 & 1.017 \\
		$\hat\alpha_2$, after filtering  & $-0.174$ & 0.849 & 0.876 & 0.889 & 0.937 \\
		\hline
		\multicolumn{6}{c}{Monte Carlo sample, 1962:Q4--2019:Q4 (214 series)} \\
		\hline
		$\hat\alpha_2$, before filtering & $-0.175$ & 0.984 & 0.994 & 0.997 & 1.013 \\
		$\hat\alpha_2$, after filtering  & $-0.178$ & 0.848 & 0.875 & 0.890 & 0.918 \\
		\hline
	\end{tabular}
\end{center} 
\end{table}

\subsection{The Medium and Large Data Sets for DGP Calibration}
\label{appsubsec:models}

Sections~\ref{sec:numerics} and~\ref{sec:MC} report results for two VAR sizes. The two VARs use a subset of the variables in the previously constructed Monte Carlo panel. Table~\ref{apptab:models} summarizes the variable selection for the medium and large VARs. The large model adds thirteen variables to the medium model. The variable choice is based on \citet{GiannoneLenzaPrimiceri2015}, so that the calibrated data generating processes are anchored to a specification already used in the Bayesian VAR literature. ``Code'' is the FRED-QD transformation code of Table~\ref{apptab:transform}; recall that codes 5 and 6 are carried in log levels scaled by $100$ and are not differenced, while codes 1 and 2 are carried in levels.

\begin{table}[h!]
	\caption{Variables Included in the Medium and Large Models}
	\label{apptab:models}
	\small
	\begin{center}
		\setlength{\tabcolsep}{4pt}
		\begin{tabular}{llccc}
			\hline\hline
			Mnemonic & Description & Code & Medium & Large \\
			\hline
			GDPC1    & Real gross domestic product                        & 5 & $\bullet$ & $\bullet$ \\
			GDPCTPI  & GDP: chain-type price index                        & 6 & $\bullet$ & $\bullet$ \\
			FEDFUNDS & Effective federal funds rate                       & 2 & $\bullet$ & $\bullet$ \\
			\hline
			PCECC96  & Real personal consumption expenditures             & 5 & $\bullet$ & $\bullet$ \\
			GPDIC1   & Real gross private domestic investment             & 5 & $\bullet$ & $\bullet$ \\
			HOANBS   & Nonfarm business sector: hours of all persons      & 5 & $\bullet$ & $\bullet$ \\
			COMPRNFB & Nonfarm business sector: real compensation/hour    & 5 & $\bullet$ & $\bullet$ \\
			\hline
			CPIAUCSL & CPI, all urban consumers: all items                & 6 & & $\bullet$ \\
			PPIACO   & PPI: all commodities                               & 6 & & $\bullet$ \\
			INDPRO   & Industrial production: total index                 & 5 & & $\bullet$ \\
			SRVPRD   & All employees: service-providing industries        & 5 & & $\bullet$ \\
			PRFIx    & Real private fixed investment: residential         & 5 & & $\bullet$ \\
			PNFIx    & Real private fixed investment: nonresidential      & 5 & & $\bullet$ \\
			GPDICTPI & Gross private domestic investment: price index     & 6 & & $\bullet$ \\
			CUMFNS   & Capacity utilization: manufacturing                & 1 & & $\bullet$ \\
			GS1      & 1-year Treasury constant maturity rate             & 2 & & $\bullet$ \\
			GS5      & 5-year Treasury constant maturity rate             & 2 & & $\bullet$ \\
			S\&P 500 & S\&P common stock price index: composite           & 5 & & $\bullet$ \\
			M2REAL   & Real M2 money stock                                & 5 & & $\bullet$ \\
			UMCSENTx & Consumer sentiment index                           & 1 & & $\bullet$ \\
			\hline
			& \multicolumn{2}{r}{Number of variables $n$:} & 7 & 20 \\
			\hline
		\end{tabular}
	\end{center} 
\end{table}

The two data sets are built from \texttt{Data\_\allowbreak clean\_HX\_\allowbreak 1960\_2019.csv} as follows. The full filtered panel of $214$ series is first demeaned and standardized series by series, over the whole sample, so that the mixed units noted in  Section~\ref{appsubsec:data.samples} do not affect the calibration. The variables of Table~\ref{apptab:models} are then selected, giving panels of dimension $229 \times n$ with $n=7$ and $n=20$. The estimation period is 1962:Q4--2019:Q4, that is $229$ quarterly observations, of which $228$ are usable once the one-quarter lag is taken. 

On each panel we estimate a VAR(1) by OLS without an intercept (because the data have been demeaned) and take the residual covariance matrix as the innovation covariance. These two objects, together with a set of
randomly drawn drift matrices held fixed across designs, define the DGPs.

\subsection{Data Use in the Empirical Application}
\label{appsubsec:empirical}

The analysis in Section~\ref{sec:empirics} is based on the ``Empirical Application'' panel described in Table~\ref{apptab:samples}. We follow \citet{MSW2006} in constructing a large number of data sets at random. Specifically, we draw $1,000$ subsets of $n = 7$ series, uniformly without replacement within a subset, from the $189$ available. The draw is seeded, and the same $1,000$ subsets are used for the forecasting and the impulse-response exercises. We trim the time series dimension of the estimation samples as follows: observations from 1982:Q3 to 1983:Q4 are used to initialize lags (we use $q=6$) and the estimation sample ranges from 1984:Q1--2019:Q4. As in Section~\ref{appsubsec:models}, every series is demeaned and standardized before use.

We do screen the data sets for explosiveness by computing VAR OLS estimates. A candidate data set  is retained only if its unshrunken OLS estimate is non-explosive. Otherwise it is discarded and redrawn, until $1,000$ data sets are obtained. The screen is close to non-binding on this panel. Rejection is concentrated at $p=6$: $0.4\%$ of the samples generated have a dominant root that is greater than one in the long sample, essentially all of it at $p = 6$, where $42$ coefficients per equation are fitted to $144$ observations. The estimated VARs are thus persistent but stationary, with the dominant root rising in the lag order. This mirrors the series-level persistence documented in Section~\ref{appsubsec:data.hamilton}: the Hamilton filter
leaves a great deal of serial correlation in place, and the VARs fitted to the filtered data inherit it.

\clearpage
\section{Additional Empirical Results}
\label{appsec:empirics}

Similarly to Table~\ref{tab:Risk.PCsel} in the main text, we examine the Monte Carlo risk of LFE and MLE predictors based on data-driven hyperparameter and lag length choice, where now the sample size is fixed at $T=130$. Focusing on the medium-sized model, Table~\ref{tab:Risk.PCsel.130} shows risk differentials between $\hat{y}_{T+h}(\iota,\hat{\lambda},p)$ and $\hat{y}_{T+h}(lfe,0,q=4)$, where $\hat{\lambda}$ is determined by minimizing $PC_T(\iota,\lambda,p)$ with respect to $\lambda$ using grid search. We report results for $\iota = lfe$, $\iota=mle$, and also use PC to choose between LFE and MLE and report the resulting risks in the column ``Joint.'' The percentage of Monte Carlo replications where PC selects LFE is reported in the column $\pi$. We consider lag lengths of 1, 2, 3, and 4. In the rows labeled $\hat{p}$, PC is also used to select the lag length. Negative numbers indicate risk reductions relative to the benchmark predictor.

\begin{table}[h!]
	\centering
	\caption{Finite Sample Risk Differentials for $\hat{y}_{T+h}(\iota, \hat{\lambda}, p)$, Medium Model, $T=130$}
	\scalebox{0.9}{\begin{tabular}{@{\extracolsep{2pt}}r @{\hspace*{1.5cm}} rrrr @{\hspace*{0.8cm}} rrrr @{\hspace*{0.8cm}} cccc}
		\hline\hline
		& \multicolumn{4}{c}{A/B1, $\alpha = 0$} & \multicolumn{4}{c}{A1, $\alpha = 2$} & \multicolumn{4}{c}{B1, $\alpha = 2$} \\ \cmidrule(l{1pt}r{20pt}){2-5}\cmidrule(l{1pt}r{20pt}){6-9}\cmidrule(l{1pt}r{1pt}){10-13}
		$p$ & LFE & MLE & $\pi$ & Joint & LFE & MLE & $\pi$ & Joint & LFE & MLE & $\pi$ & Joint \\
		\hline
		\hline \multicolumn{13}{c}{Horizon $h = 2$} \\ \hline
		1 & -79 & -83 & {\small (16)} & -83 & -91 & -95 & {\small (11)} & -95 & -60 & -62 & {\small (19)} & -62 \\
		2 & -81 & -74 & {\small (30)} & -75 & -93 & -94 & {\small (10)} & -93 & -62 & -60 & {\small (20)} & -60 \\
		3 & -77 & -73 & {\small (41)} & -73 & -86 & -91 & {\small (14)} & -90 & -66 & -56 & {\small (91)} & -64 \\
		4 & -71 & -69 & {\small (48)} & -69 & -78 & -81 & {\small (17)} & -80 & -69 & -71 & {\small (29)} & -70 \\
		$\hat{p}$ & -73 & -74 & {\small (28)} & -74 & -79 & -83 & {\small (14)} & -82 & -68 & -71 & {\small (29)} & -70 \\
		\hline \multicolumn{13}{c}{Horizon $h = 4$} \\ \hline
		1 & -162 & -157 & {\small (35)} & -158 & -199 & -190 & {\small (41)} & -192 & -147 & -132 & {\small (44)} & -136 \\
		2 & -153 & -145 & {\small (30)} & -146 & -190 & -196 & {\small (25)} & -193 & -143 & -132 & {\small (44)} & -132 \\
		3 & -142 & -138 & {\small (28)} & -137 & -177 & -187 & {\small (30)} & -180 & -142 & -118 & {\small (72)} & -131 \\
		4 & -131 & -128 & {\small (37)} & -126 & -164 & -166 & {\small (49)} & -162 & -149 & -134 & {\small (86)} & -144 \\
		$\hat{p}$ & -141 & -139 & {\small (28)} & -137 & -167 & -173 & {\small (42)} & -168 & -148 & -132 & {\small (83)} & -142 \\
		\hline \multicolumn{13}{c}{Horizon $h = 6$} \\ \hline
		1 & -218 & -217 & {\small (30)} & -215 & -280 & -264 & {\small (46)} & -270 & -278 & -243 & {\small (49)} & -259 \\
		2 & -204 & -208 & {\small (23)} & -204 & -278 & -277 & {\small (39)} & -267 & -276 & -244 & {\small (59)} & -254 \\
		3 & -190 & -199 & {\small (21)} & -193 & -261 & -271 & {\small (38)} & -254 & -267 & -236 & {\small (70)} & -245 \\
		4 & -176 & -186 & {\small (27)} & -179 & -246 & -256 & {\small (48)} & -242 & -255 & -245 & {\small (70)} & -247 \\
		$\hat{p}$ & -193 & -198 & {\small (22)} & -193 & -248 & -261 & {\small (45)} & -244 & -252 & -240 & {\small (68)} & -240 \\
		\hline\hline
		\multicolumn{13}{l}{\small \textit{Notes:} The finite sample risk differentials are computed relative to $\hat{y}_{T+h}(lfe, 0, q = 4)$. $\pi$ is the percentage}\\ \multicolumn{13}{l}{\small  of times that LFE is selected by the PC criterion.} \\
	\end{tabular}}
	\label{tab:Risk.PCsel.130}
\end{table}

Columns 2 to 5 of the table contain results for the no-misspecification case of $\alpha=0$. In this case, and similarly as in Table \ref{tab:Risk.PCsel}, the LFE predictor dominates the MLE at shorter horizons, while the dominance is reversed as the horizon grows larger. The lowest risk is attained for $p=1$ because that equals the true asymptotic lag length $p_*$. The PC criterion selects the LFE predictor only between 16\% and 48\% of the Monte Carlo repetitions. The risk under ``joint'' selection is in between the LFE and MLE risk, often closer to the latter. The risk values associated with $\hat{p}$, i.e., the case in which PC is also used to choose the lag length, are between the $p=2$ and $p=3$ values, indicating that the criterion tends to select models that are somewhat larger than $p_*$.

Under the mild misspecification of Design A1, MLE continues to dominate LFE despite the horizon of interest and lag order used. It also becomes desirable to include more than $p_*=1$ lags to offset the dynamic misspecification of the VAR at longer horizons. The fraction $\pi$ of selecting LFE now ranges from 10\% to 49\%. The risk associated with ``joint'' selection of predictor and hyperparameter often sits between the risk of the LFE and MLE predictors, often closer to the best of the two. The $\hat{p}$ risk differentials are close to the $p=4$ risk differentials, indicating that PC selects four lags with high probability. These patterns are enhanced under Design B1, which features stronger misspecification. For horizon $h=6$, LFE is selected for 68\% of the samples conditional on using $\hat{p}$ lags. Again, the results reported here are consistent with the simulation results in S2005 and the empirical results in \cite{MSW2006} in that the use of LFE is only justified if misspecification is sufficiently large. The key takeaway is that PC helps the forecaster to adapt to the level of misspecification by choosing between LFE and MLE predictors, tuning the level of shrinkage $\lambda$, and  selecting the number of lags $p$.

\end{appendix}

\end{document}